\documentclass[12pt,aps,prd,nofootinbib,superscriptaddress, onecolumn,preprintnumbers,balancelastpage,subeqn,floatfix]{revtex4}

\usepackage{amssymb,amsmath,amsfonts,amstext,graphics, multirow}
\usepackage{graphicx}
\usepackage{epstopdf}
\usepackage{url}
\usepackage{appendix}
\usepackage{color}

\newdimen\tableauside\tableauside=1.0ex
\newdimen\tableaurule\tableaurule=0.4pt
\newdimen\tableaustep
\def\phantomhrule#1{\hbox{\vbox to0pt{\hrule height\tableaurule width#1\vss}}}
\def\phantomvrule#1{\vbox{\hbox to0pt{\vrule width\tableaurule height#1\hss}}}
\def\sqr{\vbox{%
  \phantomhrule\tableaustep
  \hbox{\phantomvrule\tableaustep\kern\tableaustep\phantomvrule\tableaustep}%
  \hbox{\vbox{\phantomhrule\tableauside}\kern-\tableaurule}}}
\def\squares#1{\hbox{\count0=#1\noindent\loop\sqr
  \advance\count0 by-1 \ifnum\count0>0\repeat}}
\def\tableau#1{\vcenter{\offinterlineskip
  \tableaustep=\tableauside\advance\tableaustep by-\tableau-rule
  \kern\normallineskip\hbox
    {\kern\normallineskip\vbox
      {\gettableau#1 0 }%
     \kern\normallineskip\kern\tableaurule}%
  \kern\normallineskip\kern\tableaurule}}
\def\gettableau#1 {\ifnum#1=0\let\next=\null\else
  \squares{#1}\let\next=\gettableau\fi\next}

\newcommand{\gsim}{\lower.7ex\hbox{$\;\stackrel{\textstyle>}{\sim}\;$}}
\newcommand{\lsim}{\lower.7ex\hbox{$\;\stackrel{\textstyle<}{\sim}\;$}}

\newcommand{\GB}{ \mathcal{G}_{\tilde{B}} }
\newcommand{\GW}{\mathcal{G}_{\tilde{W}}}
\newcommand{\TB}{\mathcal{T}_{\tilde{B}}}
\newcommand{\BB}{\mathcal{B}_{\tilde{B}}}
\newcommand{\TW}{\mathcal{T}_{\tilde{W}}}
\newcommand{\BW}{\mathcal{B}_{\tilde{W}}}
\newcommand{\refsec}[1]{Sec.~\ref{Sec: #1}}
\newcommand{\reffig}[1]{Fig.~\ref{Fig: #1}}

\def\OO{{\cal O}}
\def\LL{{\cal L}}

\newcommand{\GeV}{\,\mathrm{GeV}}

\newcommand{\ifb}{\,\mathrm{fb}^{-1}}

\newcommand{\Br}{{\text{ Br}}}
\newcommand{\eff}{{\text{eff}}}

\newcommand{\MET}{\mbox{$E_T\hspace{-0.25in}\not\hspace{0.18in}$}}

\newcommand{\be}{\begin{eqnarray}}
\newcommand{\ee}{\end{eqnarray}}
\newcommand{\bea}{\begin{eqnarray}}
\newcommand{\eea}{\end{eqnarray}}

\newcommand{\bef}{\begin{figure}[htbp]\begin{center}}
\newcommand{\eef}{\end{center}\end{figure}}

\newcommand{\go}{\tilde{g}}
\newcommand{\Q}{\tilde{q}}
\newcommand{\LSP}{{\widetilde{\chi}}}
\newcommand{\T}{\tilde{t}}
\newcommand{\B}{\tilde{b}}
\newcommand{\Eff}{\mathcal{E}}

\begin{document}

\title{Heavy Flavor Simplified Models at the LHC}

\author{Rouven Essig}
\thanks{rouven.essig@stonybrook.edu}
\affiliation{C.N.~Yang Institute for Theoretical Physics, Stony Brook University, Stony Brook, NY 11794}
\affiliation{School of Natural Sciences, Institute for Advanced Study, Einstein Drive, Princeton, NJ}
\affiliation{Theory Group, SLAC National Accelerator Laboratory, Menlo Park, CA 94025}
\author{Eder Izaguirre}
\thanks{eder@stanford.edu}
\affiliation{Theory Group, SLAC National Accelerator Laboratory, Menlo Park, CA 94025}
\affiliation{Physics Department, Stanford University, Stanford, CA 94306}
\author{Jared Kaplan}
\thanks{jaredk@slac.stanford.edu}
\affiliation{Theory Group, SLAC National Accelerator Laboratory, Menlo Park, CA 94025}
\author{Jay G. Wacker}
\thanks{jgwacker@stanford.edu}
\affiliation{Theory Group, SLAC National Accelerator Laboratory, Menlo Park, CA 94025}

\begin{abstract}

We consider a comprehensive set of simplified models  that contribute to final 
states with top and bottom quarks at the LHC. These simplified models are used to create minimal 
search strategies that ensure optimal coverage of new heavy flavor physics involving the pair production of color octets and triplets.
We provide a set of benchmarks that are representative of model space, which can be used 
by experimentalists to perform their own optimization of search strategies.  For data sets larger than $1$ fb$^{-1}$,  same-sign dilepton and 3$b$ search regions become very powerful.
Expected sensitivities from existing and optimized searches are given.

\end{abstract}
\keywords{LHC, Supersymmetry, Jets and Missing Energy}
\maketitle
\newpage
\preprint{YITP-SB-11-34,\, SLAC-PUB-14658}
\tableofcontents

\section{Introduction}
Data from the LHC has begun to inform our understanding of electroweak symmetry breaking (EWSB).  
Since naturalness of EWSB typically requires the existence of a top partner, many promising new physics scenarios  
produce final states rich in top and/or bottom quarks (heavy flavors), see e.g.~\cite{models,pheno}.  
In many cases this is accompanied by a stable neutral particle in the final state that could be the dark matter of 
the Universe. The 7 TeV LHC run has the potential to discover new colored states with masses 
well into the TeV scale that give a new contribution to events with heavy flavor content. 
In this paper, we study a comprehensive set of simplified models to aid the 7 TeV LHC search for 
new colored particles that decay to top ($t$) and/or bottom ($b$) quarks and to a new stable particle that appears as missing energy.

A generic difficulty in LHC new physics searches is that the masses of the new states produced are not known {\it a priori}. 
This means that the kinematics of the final state signatures can vary drastically, depending on the mass spectrum of the particles produced. For heavy flavor final states, an additional challenge in developing a comprehensive search strategy is that top quarks lead to different signatures than bottom quarks.  For example, a $4t$ final state is quite distinct from 
a $4b$ final state: the former produces a large multiplicity of visible particles, thus increasing (decreasing) the energy (missing energy) of the event, while the latter has a smaller multiplicity of visible states and a large missing energy from the stable particle at the bottom of the decay chain. 
Given finite resources, it is important to design a limited set of search regions that nevertheless 
retain close to optimal discovery power for all possibilities.

An effective but minimal set of search regions can be developed using ``simplified models'' 
\cite{SimplifiedModels,Alwall:2008ve,Alwall:2008va,Alves:2011sq}, which are 
effective field theories with a minimal particle content. Only the physically relevant parameters of a full model are kept, 
namely the masses of the new states, the production cross section, and the branching fractions into the available modes. 
Different simplified model parameter values can lead to distinct signatures and thus require disparate search strategies for 
discovery.  Simplified models allow one to efficiently explore all possibilities and cast a wide net for physics 
beyond the Standard Model (SM), a necessity in searches for new physics due to the LHC inverse problem \cite{ArkaniHamed:2005px}.

This article will consider 12 simplified models in which either color octets, $\go$, or triplets, $\Q$, are pair produced and decay to a combination of $t$- and/or $b$-quarks and a stable particle $\chi^0$, with possibly some light flavor jets.  
Early LHC searches are already looking for these types of spectra, which offer an optimistic path for discovery because of the highly efficient $b$-quark tagging at the ATLAS and CMS experiments. 

From the theoretical side, as alluded to above, these signatures are well-motivated because the mechanism stabilizing the mass of the Higgs boson typically relies on a new partner for the top.  Moreover, naturalness arguments favor light third generation partners while constraints from flavor changing neutral currents and CP violation typically favor much heavier light flavor partners.  A representative example is supersymmetry, in which the gluino is the color octet that can be 
pair-produced and will decay predominantly to $t$ and $b$ quarks through the light stops and sbottoms if 
these are much lighter than the first and second generation squarks. One of the advantages of simplified models is that the results can be recast into different theories. For example, universal extra dimensions (UED) theories have a similar particle content to supersymmetry, but different spin quantum numbers. This last difference can be accounted for by a re-scaling of the production cross section by the appropriate degrees of freedom, as long as spin-correlation effects are subdominant \cite{Nojiri:2011qn}.

Within each of the 12 simplified models, we consider a discrete but broad set of masses for the $\go$ (or $\Q$) 
and the $\chi^0$, giving us a total of 2762 models.  We first find the `best' search strategy for each of the 2762 models, 
which consists of the combination of cuts that optimizes the square-root signal over background.  
We consider 8064 combinations of cuts on the number of jets, b-jets, leptons, missing transverse 
energy ($\MET$), and total event transverse energy ($H_T$).  
We then find a set of `good' search strategies, namely a set of counting experiments that will discover 
all 2762 models with great efficiency.  The `efficacy' $\Eff$ of a particular search region is the ratio of the quantity of data 
needed to discover the model with this search region to the amount of data needed using the `best' cut.  
Using a genetic algorithm, we identify a small set of search regions ($\mathcal{O}(10-20)$) 
that cover all 2762 models with $\Eff \lesssim 2$.  
Noteworthy is that $3b$ and same-sign dilepton search regions (SSDL) are particularly sensitive to these models, especially with larger data sets. 
$3b$ search regions can become an effective way to discover a broad class of heavy flavor topologies from $\go$ pair production,
 while SSDL signal regions can help in multi-top topologies where the mass spectrum of new states is compressed.

As theorists, we do not have the best tools for simulating the LHC detectors, so the set of search regions found in this study may not be optimal.  Therefore the real goal of this article is to suggest some general lessons and to provide a methodology for optimizing searches over a large parameter space of new physics models.  Moreover, we identify a set of benchmarks consisting of $\mathcal{O}(100)$ simplified models that are chosen from the full set of 2762. These benchmarks can be used by LHC experimentalists to optimize their 
search strategies and still retain excellent sensitivity to the \emph{full} space of models.  
In other words, these benchmarks taken together are representative of all 2762 models, and optimizing search strategies 
on the benchmarks is to a good approximation equivalent to optimizing on all 2762 models. 
This approach significantly cuts down the work that needs to be done by the LHC experimentalists in 
picking sensible benchmark points, and also minimizes the possibility that there are regions in parameter space that might be  
missed.  
An important caveat here is that the optimization is done over a particular set of cuts (see above), and we 
cannot assure full model space coverage if different cuts are used. However, the benchmarks chosen offer vastly different final states. The fact that $\OO(10)$ search regions are needed to cover the space of 60 benchmarks is a statement that each benchmark can give quite different kinematics from the rest.  

The organization of this paper is as follows.  \refsec{SimplifiedModels} introduces the simplified spectra used in this study. The signal and background calculation is discussed in \refsec{MC}. The reach of the LHC's latest analyses on the simplified models is estimated in \refsec{CurrentLimits}. A minimal set of search regions is found and presented in \refsec{Optimization}.  Sec.~\ref{sec:discussion} summarizes our results.  Appendix \ref{Sec: CurrentLimits Plots} and 
\ref{Sec: OptLimits Plots} show the expected cross section sensitivities for our simplified models from current 
ATLAS searches and optimized searches, respectively.  Appendix \ref{Sec: BechmarkTables} presents the benchmarks.

\section{Simplified Models}
\label{Sec: SimplifiedModels}
A variety of search regions are necessary to comprehensively cover new physics scenarios that produce heavy flavor final states.  One reason for this is that different numbers of $t$ versus $b$ quark final states can significantly alter the 
optimal search strategy.  
We introduce 12 simplified models, which we divide into two broad classes: (i) gluino-like $\go$, where the $\go$ are color 
octets that are pair-produced, and each decays to two third generation quarks and a light electroweakino-like state (`LSP'); 
and (ii) squark-like $\Q$, where the $\Q$ are color triplets that are pair-produced, and each decays to one third generation quark and a light electroweakino-like state.  In the $\go$ simplified models, we always assume the $\go$ decay through heavy off-shell squark-like particles, which themselves are inaccessible at the LHC. Similarly, in the $\Q$ simplified models, the $\go$ are assumed to be much heavier than the $\Q$.  We also include light flavor decays of the $\go$ for completeness.

\subsection{Gluino-like Models}
\label{Sec: GluinoModels}

We divide the $\go$ simplified models into two further categories, characterized by the nature of the LSP.  
In the first category (referred to as $\GB$ topologies), the LSP consists of a single neutral state, $\chi^0$.  
There are abundant examples of such theories, such as supersymmetric theories where the 
lightest supersymmetric particle is a bino or singlino.  

In the second category (referred to as $\GW$ topologies), there is a charged state $\chi^\pm$ nearly degenerate with the 
neutral state $\chi^0$.  The transition $\chi^\pm \to \chi^0$, which occurs through an off-shell $W^\pm$, is effectively 
invisible at the LHC.   In all Monte Carlo (MC) calculations, we set $m_{\chi^\pm} - m_{\chi^0} = 10\GeV$, which is an 
unobservable mass splitting in jets and missing energy analyses.  
In supersymmetric theories, this spectrum frequently occurs if the LSP is a Wino and the Higgsinos are heavy or if the LSP is a Higgsino and the Wino and Bino are heavy.  More generally, a degenerate $\chi^\pm$ and $\chi^0$ are well-motivated whenever the LSP is part of an $SU(2)_L$ multiplet.  

We now map each possible decay of the $\go\go$ pair to a simplified model. 

\begin{figure}[!t]
\begin{center}
\includegraphics[width=0.8\textwidth]{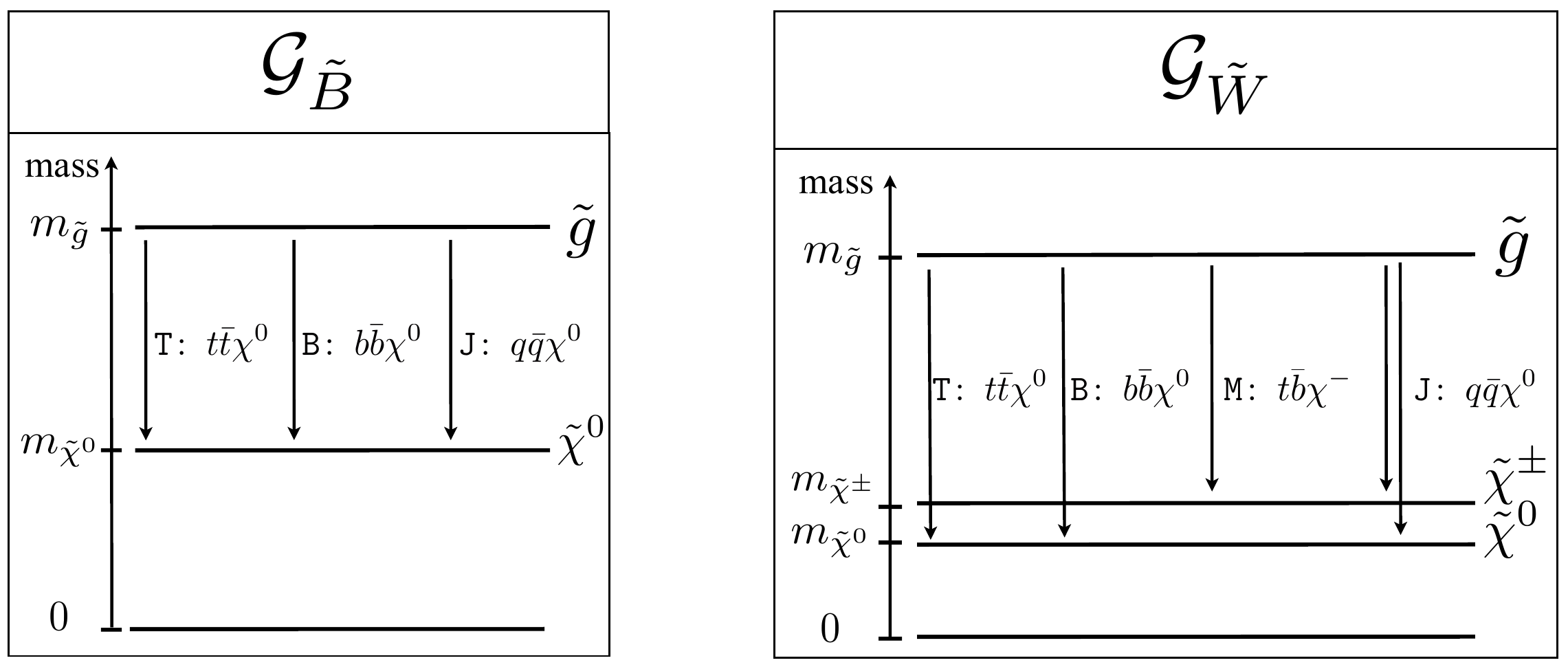}
\caption{\label{Fig: GluinoSpectra} Gluino-like simplified models used in this study.  $\GB$ models have only a light `bino'-like 
state, while $\GW$ models have an additional charged state.  The charged state is always nearly degenerate with a neutral state, so that there are not any additional visible leptons from its decay.}
\end{center}
\end{figure}

\subsubsection{$\GB$ Topologies}

The $\go$ in the $\GB$ spectra can decay in various ways, as illustrated in the left panel of \reffig{GluinoSpectra}.  The modes are
\begin{subequations}
\begin{equation}
{\tt T}: \quad \go \rightarrow \chi^0 \, t\bar{t},
\end{equation}
\begin{equation}
{\tt B}: \quad \go \rightarrow \chi^0 \, b\bar{b},
\end{equation}
\begin{equation}
{\tt J}: \quad \go \rightarrow \chi^0 \, j j .
\end{equation}
\end{subequations}
Since the $\go$ are always pair produced, the following six decay topologies are possible: 
\begin{eqnarray}\label{eq:SM1}
\GB^{\tt TT}, \quad 
\GB^{\tt BB}, \quad 
\GB^{\tt TB}, \quad 
\GB^{\tt TJ}, \quad 
\GB^{\tt BJ}, \quad 
\GB^{\tt JJ}.
\end{eqnarray} 
A thorough study of the $\GB^{\tt JJ}$ topology was performed in \cite{Alves:2011sq}; our study of light flavor jets here will be 
cursory and only done to allow comparisons with the heavy flavor searches.

The $\GB^{\tt TT}$, $\GB^{\tt BB}$ and $\GB^{\tt JJ}$ will be referred to as {\em pure topologies}, and in principle these decay topologies could be the only signal of new physics.   This is in contrast to the {\em hybrid topologies} 
$\GB^{\tt TB}$, $\GB^{\tt TJ}$, $\GB^{\tt BJ}$, which each necessarily will be accompanied by another topology. 
For example, if the $\GB^{\tt TJ}$ topology is present, the $\GB^{\tt TT}$ and $\GB^{\tt JJ}$ topologies must also be 
present in the data. 
Although hybrid topologies never appear as the only topology, they may have the largest branching ratio and can frequently be the most visible.  For instance, if $\Br(\go \rightarrow \tt{T} ) = 10\%$ and $\Br(\go\rightarrow \tt{J}) = 90\%$, then the topologies appear in the ratio 
\begin{eqnarray}
\GB^{\tt TT} : \GB^{\tt TJ} : \GB^{\tt JJ}  = 1 \%: 18\% : 81\% .
\end{eqnarray}
However, the appearance of final state top quarks can make $\GB^{\tt TJ}$ much more visible than  $\GB^{\tt JJ}$, compensating for the smaller rate.    
We will develop optimized strategies for both pure and hybrid topologies, regardless of the branching ratios into particular final states.  

\subsubsection{$\GW$ Topologies}

Models that have a $\GW$ spectrum have a new decay mode
\begin{subequations}
\begin{equation}
{\tt M}: \quad \go \rightarrow \chi^- \, t\bar{b} \rightarrow   \chi^0  \, t\bar{b}\; W^{+\,*}
\end{equation}
\begin{equation}
\bar{{\tt M}}: \quad \go \rightarrow \chi^+ \, b\bar{t} \rightarrow   \chi^0  \,b\bar{t} \; W^{-\,*}\,,
\end{equation}
\end{subequations}
where in all cases the $W^\pm$ boson is so far off-shell that it is effectively invisible (as mentioned above, for definiteness, 
we choose it to have $10$ GeV of energy).  Note that $m_{\chi^+}\gtrsim 100$ GeV from chargino searches at LEP-II 
\cite{Abbiendi:2003sc}, although in plots we will allow for much smaller $m_{\chi^+}$.  
We always take ${\tt M}$ and $\bar{\tt M}$ in equal admixtures and drop the distinction between the two decays, since they look very similar in detectors.
There are still the direct decays of $\go$ to $\tilde{\chi}^0$ which can give rise to ${\tt T}$, ${\tt B}$ and ${\tt J}$ decay modes.
Constructing the ten possible decay topologies from these four decay modes, six of them are identical to $\GB$ decay topologies. 
The four new decay topologies are
\begin{eqnarray}\label{eq:SM2}
\GW^{\tt MM}, \quad\GW^{\tt TM}, \quad
\GW^{\tt BM}, \quad
\GW^{\tt MJ}.
\end{eqnarray}

Eqs.~(\ref{eq:SM1}) and (\ref{eq:SM2}) thus give the ten possible decay topologies in $\go$ simplified models, when cascade decays are either suppressed or inaccessible, that we will study.   In supersymmetry, we could see various specific admixtures of these topologies, depending on the spectrum of squarks and the identity of the LSP.  

The decay width of a gluino into a fermion species is
$\Gamma_{\go\rightarrow \tilde{B} q\bar{q}} \propto  { Y_q^2}/{m_{\tilde{q}}^4}$,
where $Y_q$ is the quark hypercharge. 
Of course, due to the large $t$-quark mass, phase space considerations are important and can significantly modify the 
branching ratios.  Let us consider the branching ratios in several examples of plausible supersymmetric topologies, ignoring phase space factors:
\begin{itemize}
\item Bino-like LSP, lighter right handed third generation squarks. Assuming that $m_{\tilde{b}^c} = m_{\tilde{t}^c}$, this gives a ratio of $\go$ decays of ${\tt T} : {\tt B} = 4 :1$ resulting in
\begin{eqnarray}
\GB^{\tt TT} : \GB^{\tt TB} : \GB^{\tt BB} = 64\% : 32\% : 4\% .
\end{eqnarray}
\item Wino-like LSP with ${\tt TT}$ decay mode kinematically accessible
\begin{eqnarray}
\GB^{\tt TT}: \GB^{\tt TB}: \GB^{\tt BB}: \GW^{\tt MM}: \GW^{\tt TM}: \GW^{\tt BM} = 2.8\%: 5.6\%: 2.8\%: 44\%: 22\%: 22\%
\end{eqnarray}
\item Wino-like LSP with ${\tt TT}$ decay mode kinematically inaccessible
\begin{eqnarray}
 \GB^{\tt BB}: \GW^{\tt BM}: \GW^{\tt MM} =11\%:44\%: 44\%
\end{eqnarray}
\end{itemize}

\subsection{Heavy Flavor Squark-like Models}
\label{Sec: SquarkModels}

\begin{figure}[!t]
\begin{center}
\includegraphics[height=2.0in]{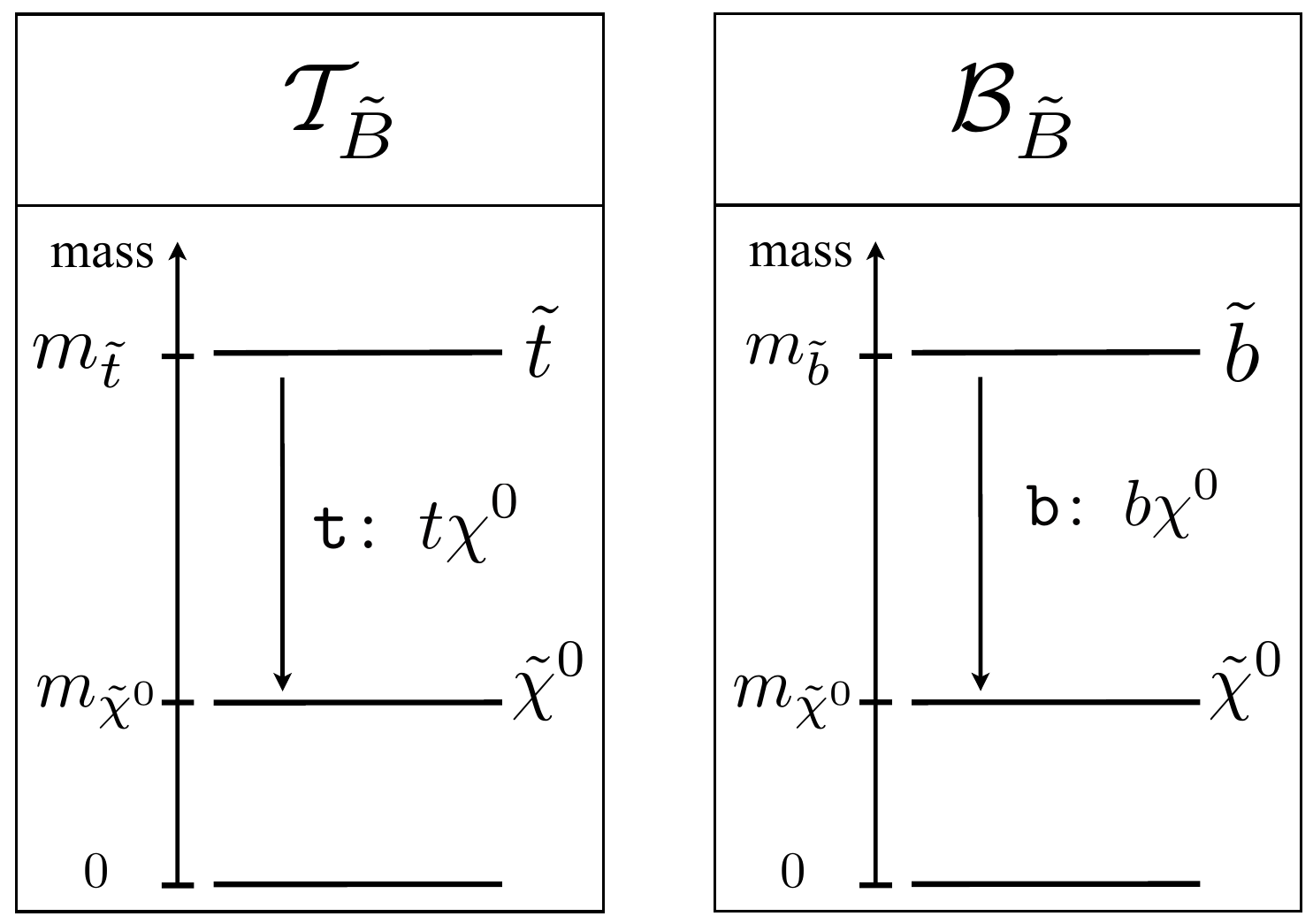}
\caption{\label{Fig:Spectra} Squark-like simplified models used in this study.}
\end{center}
\end{figure}

The second class of simplified models consists of a single color-triplet squark-like state, either $\T$ or $\B$, which are 
pair-produced.  
There are again two categories characterized by the nature of the LSP.  
A single neutral stable particle, $\chi^0$, allows the decays 
\begin{eqnarray}\label{eq:t}
\nonumber
&{\tt t}:&\quad \T  \rightarrow \chi^0 \, t,\\
&{\tt b}:& \quad \B  \rightarrow \chi^0 \, b,
\end{eqnarray}
giving rise to simplified models that we call $\TB$ and $\BB$, respectively, with topologies $\TB^{\tt t}$ and $\BB^{\tt b}$.  
Supersymmetric theories with comparable $\T$ and $\B$ masses can give $\mathcal{O}(1)$ admixtures of these 
two processes, but we will not consider these scenarios here.  

In $\TW$ and $\BW$ simplified models, a chargino-like state $\chi^\pm$ nearly degenerate with the $\chi^0$ LSP allows the decays
\begin{eqnarray}
\nonumber
&{\tt t}:&\quad \B  \rightarrow \chi^- \, t \rightarrow \chi^0 \,t\; W^-{}^*,\\
&{\tt b}:& \quad \T  \rightarrow \chi^+ \, b\rightarrow \chi^0\, b \;W^+{}^*,
\end{eqnarray}
where we assume the $W^{\pm}$ decay products are soft and unimportant, and we 
therefore again label the decays as ${\tt t}$ and ${\tt b}$ as in Eq.~(\ref{eq:t}).  
Various decay topologies are possible, but they are again effectively the same as $\TB^{\tt t}$ and $\BB^{\tt b}$ 
considered above.

\section{Backgrounds and Signal Simulation}
\label{Sec: MC}

The dominant SM backgrounds to jets and $\MET$ signatures are $t\bar{t}+ \text{ jets}$, $W^{\pm}+ \text{ jets}$, $Z^0+ \text{ jets}$, $t~+\text{jets}$, $VV +$ jets, $tV+\text{jets}$ and QCD, where $V=Z^0, W^\pm$.  The matrix elements for parton level events were computed in \texttt{MadGraph 4.4.32} \cite{Alwall:2007st} with CTEQ6L1 parton distribution functions \cite{Pumplin:2002vw}.   Variable renormalization and factorization scales are set to the transverse energy of the event \cite{Bauer:2009km}. The SM parton level processes generated are 
\begin{eqnarray}
\label{Eq: Samples}
pp  \rightarrow W^{\pm} + nj  &\quad& 1\le n\le 3\\
\nonumber
pp  \rightarrow Z^{0} + nj &\quad & 1\le n \le 3 \\
\nonumber
pp  \rightarrow t\bar{t} + nj &\quad & 0\le n \le 2\\
\nonumber
pp \rightarrow VV+ nj & \quad & 0\le n \le 2\\
\nonumber
pp\rightarrow t/\bar{t}+nj & \quad & 1\le n \le 3\\
\nonumber
pp\rightarrow t/\bar{t}+W^{\mp}+nj & \quad & 0\le n \le 2\\
\nonumber
\end{eqnarray}
where $j\in\{g,u,\bar{u}, d,\bar{d}, c,\bar{c}, s,\bar{s}, b,\bar{b}\} $ .

\begin{figure}[!t]
\begin{center}
\includegraphics[width=0.53\textwidth]{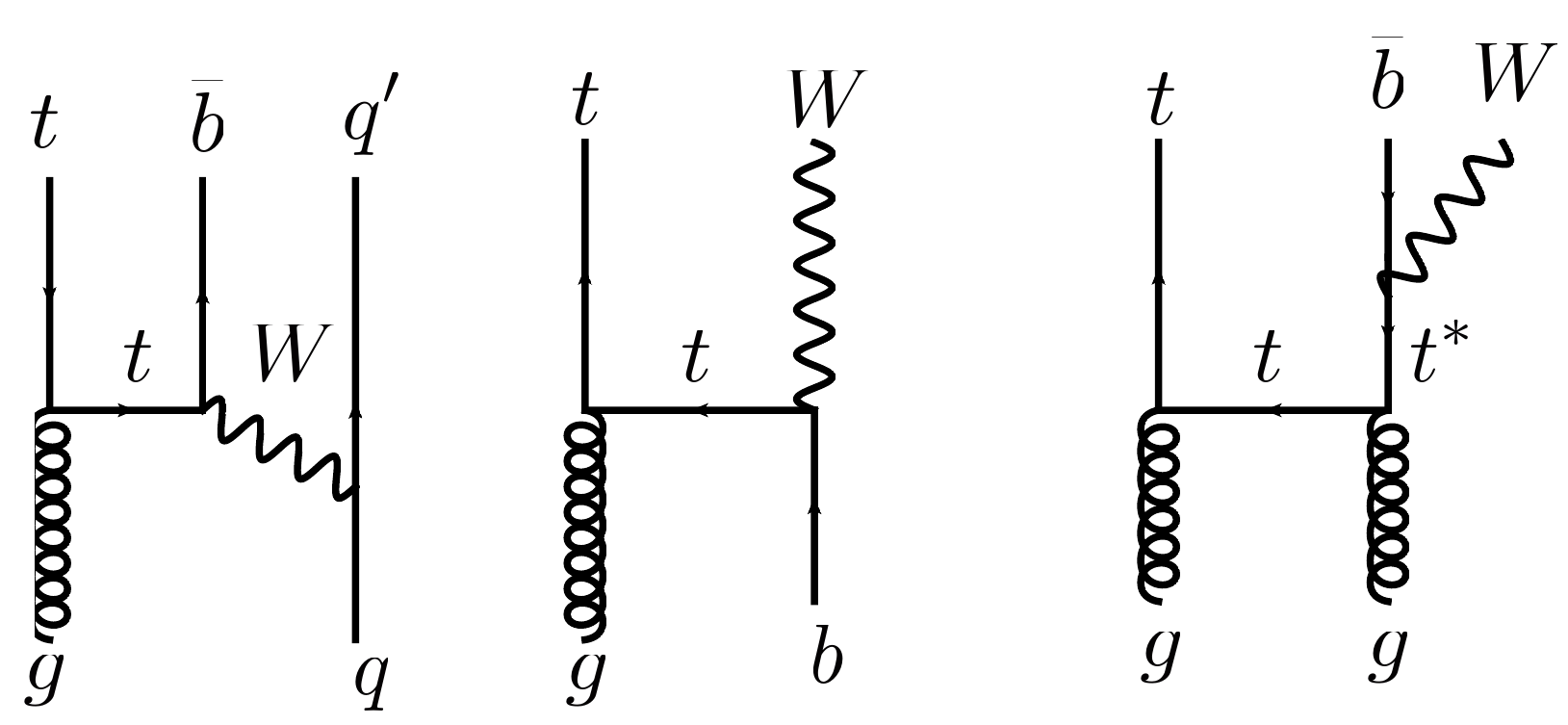}
\caption{\label{Fig: FeynmanDiagrams} Feynman Diagrams for  $t+nj$  and $t+W+nj$.  Notice the need to subtract the on-shell top contribution from  $t W \bar{b}$ sample. }
\end{center}
\end{figure}

The SM contribution to $\MET$~distributions peaks at low energies, whereas many new physics signatures produce events with large $\MET$.   Therefore it is important to have sufficient MC statistics on the tail of the $\MET$ distribution.  
To achieve sufficient statistics, different samples are generated for each SM process, where each sample has the $p_T$ of the massive particle lying in a given interval.
For instance, $Z^{0}+\text{jets}$ parton level events were divided into three samples with
\begin{eqnarray*}
0~\GeV\le &p_{T,Z^0}&\le200\GeV\\
200~\GeV < &p_{T,Z^0}&\le300\GeV\\
300~\GeV< &p_{T,Z^0}&.
\end{eqnarray*}
 In the case of two massive particles produced, such as $t\bar{t}$, the samples are divided by the larger $p_{T}$ of either massive particle in the event. 

Contributions from QCD to jets and $\MET$  can come from either detector effects and jet energy mis-measurement, or neutrinos appearing in the decay of heavy flavor hadrons. To estimate the QCD contribution to jets and $\MET$ signatures, the following subprocesses were generated
\begin{eqnarray}
\nonumber
pp\rightarrow nj'&\quad & 2\le n \le 4\\
pp\rightarrow b\bar{b}+nj &\quad & 0\le n \le 2.
\end{eqnarray}
Here $j'$ refer to light flavor jets and gluons only. To achieve sufficient statistics, the QCD and $b\bar{b}$ backgrounds were subdivided in exclusive bins delimited by the $p_T$ of the leading jet.

The signals, $\go$ and $\Q$ pair-production, were generated in association with up to two jets at parton level,
\begin{eqnarray}
pp\rightarrow \go\go + nj \quad 0\le n \le 2\\
\nonumber
pp\rightarrow \Q\bar{\Q} + nj \quad 0\le n \le 2
\end{eqnarray}
The effects of including additional radiation in signal processes have been documented in several 
studies \cite{Alwall:2008ve, Alwall:2008va, SUSYPSME,PSME}.  
We generated MC for 2762 different points in the $m_{\go} - m_{\tilde{B}}$, $m_{\T} - m_{\tilde{B}}$, and 
$m_{\B} - m_{\tilde{B}}$ planes. 

For both signal and backgrounds, the showering, hadronization, particle decays, and matching of parton 
showers to matrix element partons are done in \texttt{PYTHIA 6.4} \cite{Sjostrand:2006za}.  
We use the MLM parton shower/matrix element matching scheme with a shower-$k_{\perp}$ scheme introduced 
in \cite{SUSYPSME}. A fixed 5-flavor matching scheme is used. The matrix elements better describe hard radiation, while the parton shower generates 
softer radiation that fills out jets \cite{PSME}.  
The matching scales used here are: 
 \begin{eqnarray}
\begin{array}{|c|c|}
\hline
\text{Sample}& Q_{\text{Match}}\\
\hline\hline
t\bar{t}+\text{jets}& 100 \GeV\\
V+ \text{jets}& 40\GeV\\
t+\text{jets}& 100\GeV\\
tV+ \text{jets}& 100\GeV\\
QCD& 50\GeV\\
\go\go +\text{jets}& 100\GeV\\
\Q\bar{\Q} +\text{jets}& 100\GeV\\
\hline
\end{array}
 \end{eqnarray}
 
Hard jets  beyond the multiplicities listed in Eq.~\eqref{Eq: Samples} must be generated by the parton shower.    
In particular, for $W^\pm +\text{ jets}$ and $Z^0+\text{ jets}$, the fourth jet and beyond are generated through the 
parton shower.
This approximation has been validated by several studies  \cite{PSME}. As a cross check, for the $Z^0+\text{jets}$ background generated for our study, the 
discrepancy in the inclusive rate for four jets and $\MET$ from matching up to three jets versus matching up to 
four jets is $\OO(15\%)$, assuming a selection requirement of 50 GeV on the fourth leading jet. 
For consistency, in the analysis only samples 
with up to $2\rightarrow4$ partons are used.

\begin{figure}[!t]
\begin{center}
\includegraphics[width=6in]{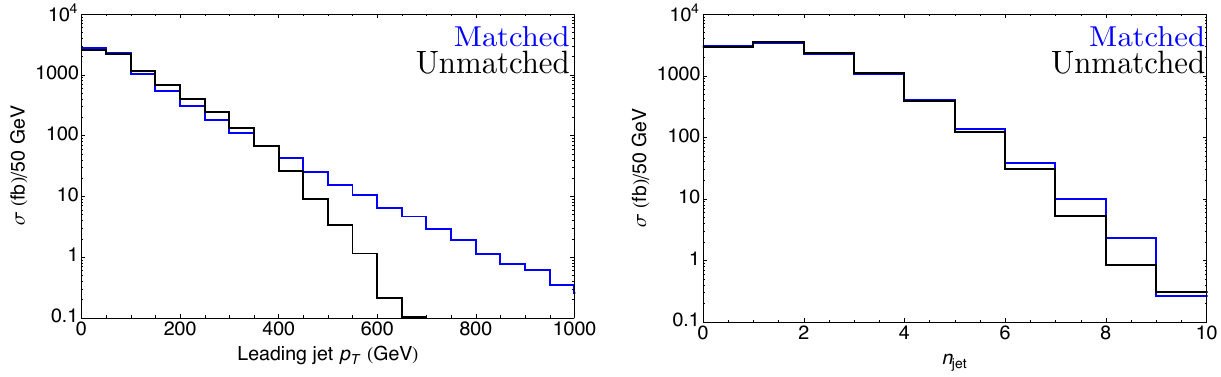}
\caption{\label{Fig: Signal Matching}  The effect of matching is shown for a signal with a $\go$ at 400 GeV and a $\chi^0$ at 390 GeV in the $\GB^{\tt BB}$ simplified model.  The left plot shows the $p_T$ spectrum of the leading jet.  The right plot shows the effect on the jet multiplicity.}
\end{center}
\end{figure}

The matching scheme is particularly important for the signal when the spectrum of the new states is compressed. 
In this signature, the final state jets from the signal can be soft and mimic QCD events. 
The signal can be enhanced by requiring hard jets, which will come from ISR or FSR. 
Fig.~\ref{Fig: Signal Matching} considers the $p_T$ spectrum of the leading jet and the jet 
multiplicity from a 400 GeV $\go$ that decays to a degenerate 390 GeV $\chi^0$ in the $\GB^{\tt BB}$ topology.  Significant 
differences are seen, generally with the unmatched sample underestimating the tail of the distributions.

Next-to-leading-order (NLO) corrections alter the predictions of both signal and background.  With parton shower/matrix element matching, the shapes of differential distributions are accurately described by tree level predictions.  The largest corrections are to the inclusive production cross section and can be absorbed into $K$-factors. The leading order cross sections of the signal are normalized to the NLO cross sections calculated in \texttt{Prospino 2.0} \cite{Beenakker:1996ch}. \reffig{NLOSigma} shows the NLO cross sections for both $\go$ and $\Q$ pair production. The $t\bar{t}+X$,$W^{\pm}+\text{jets}$, and $Z^0+\text{jets}$ leading order production cross sections are scaled to the NLO ones from \cite{Campbell:2010ff}. 

\begin{figure}[!t]
\begin{center}
\includegraphics[width=0.6\textwidth]{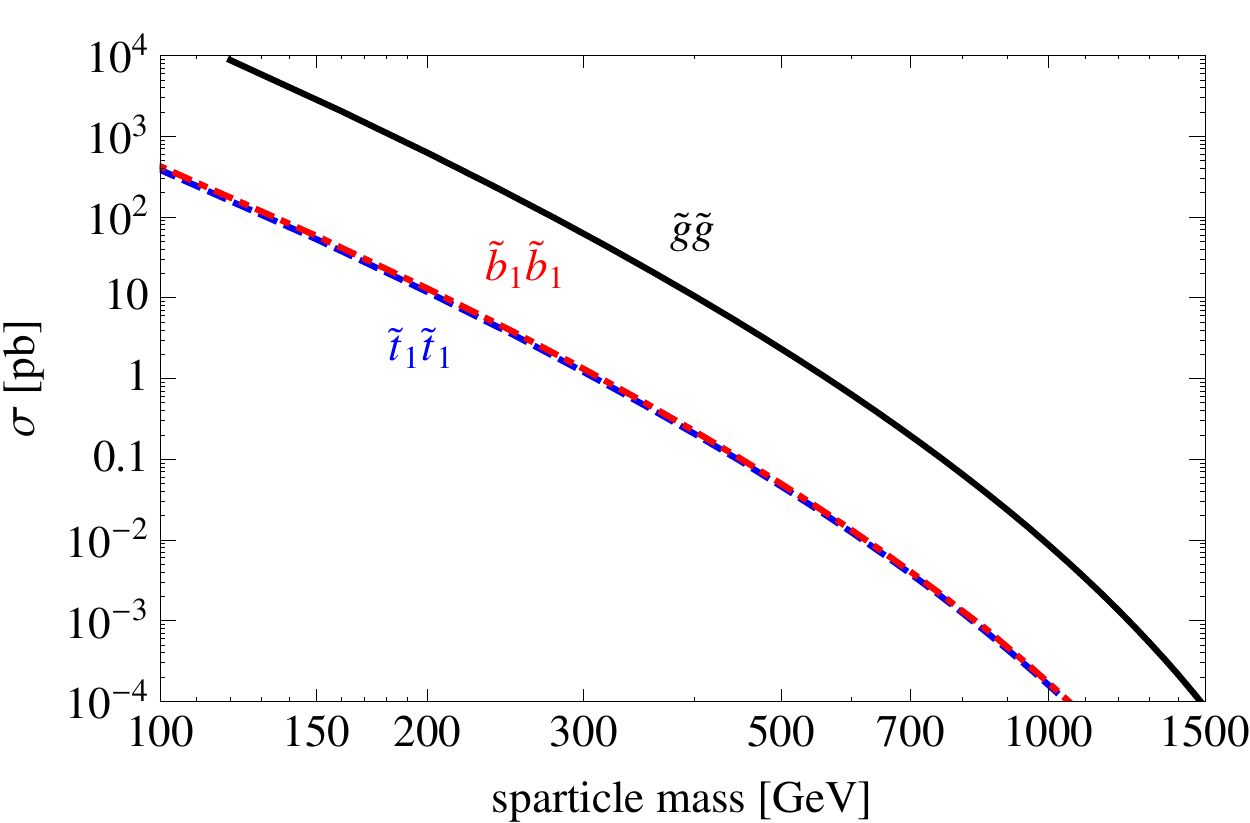}
\caption{\label{Fig: NLOSigma}  NLO cross sections for gluinos and third generation squarks, with other squarks decoupled. }
\end{center}
\end{figure}

\texttt{PGS 4} is used as a detector simulator \cite{PGS}. 
We use the {\tt PGS 4}  ATLAS card, which has been shown to reproduce results to $\OO(20\%)$ accuracy.
 
One of the drawbacks of {\tt PGS 4} is that it uses  a cone jet algorithm with $\Delta R =0.7$.  This is an infrared 
unsafe jet algorithm, but better represents the anti-$k_T$ algorithms used by the experiments than the $k_T$ algorithm.   
The SM backgrounds change by at most $\OO(10\%)$ when varying the cone size to $\Delta R=0.4$.
  
The signal offers a more varied testing ground for the effects of changing the jet algorithm.   Two competing 
effects are found. The first is that there is more out-of-cone energy  for smaller cones, resulting in less energetic jets.  
The second effect is that smaller cone jet algorithms find more jets.   
The  dependence of the kinematic cut efficiencies on $\Delta R$ varies with mass splitting between the $\go$  and $\LSP$.  
For compressed spectra, when the $p_T$ of the jets is reduced,  the efficiencies for the smaller cone size decrease 
because jets fall below the minimum jet $p_T$ requirement. 
For widely spaced spectra, where jets are energetic, more jets are found with a smaller cone size and the efficiency to have 
multiple jets passing the minimum jet requirement increases.  Altogether, the efficiencies differ by at most $\OO(20\%)$ and 
are consistent with other studies \cite{Krohn:2009zg}.
This effect is not included, and we simply use a fixed $\Delta R=0.7$.

We have modified the publicly available {\tt PGS 4} code to include a different algorithm for identifying $b$- and charm-jets.  
The algorithm now matches $B$-mesons at the MC truth level in the {\tt PYTHIA} output to the 
jet found by {\tt PGS 4} that lies closest to it in $\Delta R$.  In this way, all $B$-mesons will be 
matched to some jet.  This jet is then preliminarily identified as a $b$-jet if it lies within $\Delta R \le 0.4$ of 
the $B$-meson.  Note that different $B$-mesons (assuming there is more than one in the {\tt PYTHIA} 
event record) can be matched to the same jet, so that the number of $b$-jets in an event can be lower 
than the number of $b$-quarks produced in the event at the generator level.  
We checked the efficiency for finding $b$-jets in this way in two different samples.  
For a sample of pair-produced sbottoms with a mass of 400 GeV, where each sbottom decays to 
a bottom quark and a neutralino, we find an efficiency of $\sim 86\%$ (89\%, 90\%) when requiring 
both $b$-jets to have $p_T > 20$ GeV (10 GeV, 0 GeV).  Note that the efficiency increases by 1\% if we 
increase $\Delta R$ between the jet and the $B$-meson to 0.7 from 0.4.  
The efficiency of the original {\tt PGS 4} algorithm is 83-84\%.  
For a sample of $W^\pm b\bar{b}$ events, we find an efficiency of 75\% (78\%, 79\%) when requiring both 
$b$-jets to have $p_T > 20$ GeV (10 GeV, 0 GeV).  

After the $b$-jets are identified, we identify $c$-jets in a similar way, except now $D$-mesons are matched to jets that are 
not already $b$-jets.  

Having identified the $b$- and $c$-jets at the MC truth level, we randomly turn a 
$b$-jet into an ordinary jet without a $b$-tag with a probability given by the $p_T$ and $\eta$-dependent 
$b$-tag efficiencies found in \cite{CMSBtag}.  
We also turn ordinary jets and $c$-jets into $b$-jets with a $p_T$-dependent mistag 
rate given in \cite{CMSBtag} (for the mistag rate, we average over the small $\eta$ dependence). 

\begin{figure}[!t]
\begin{center}
\includegraphics[width=0.48\textwidth]{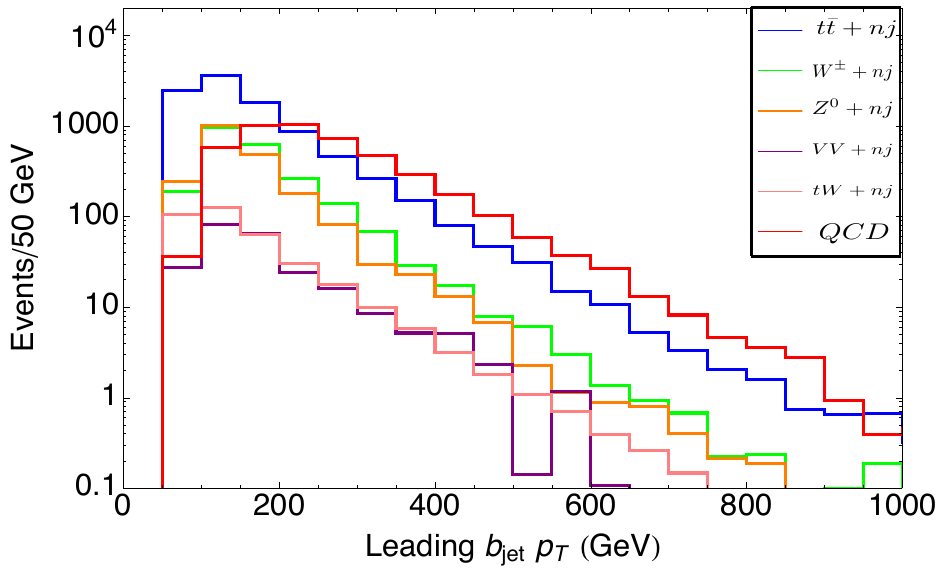}
\includegraphics[width=0.48\textwidth]{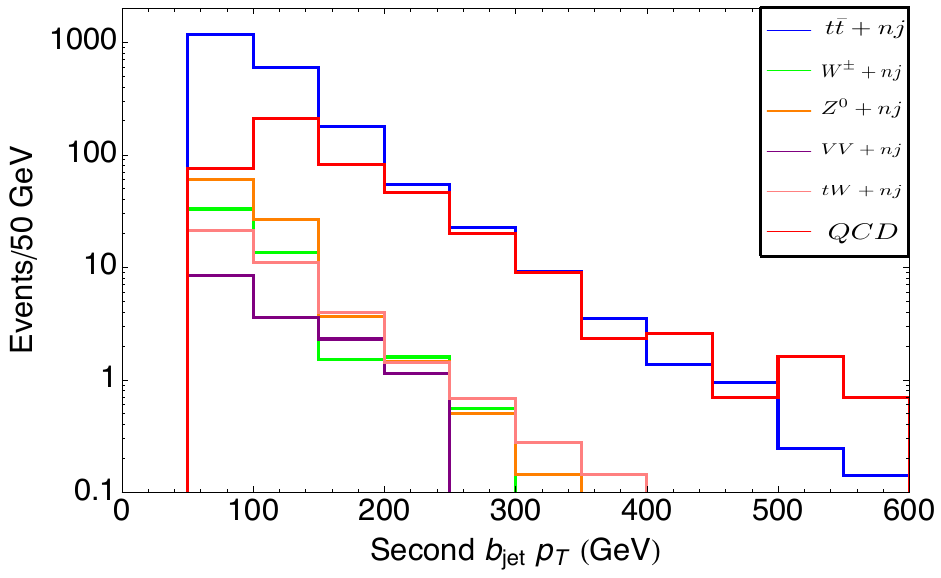}\\
\vskip 5mm
\includegraphics[width=0.48\textwidth]{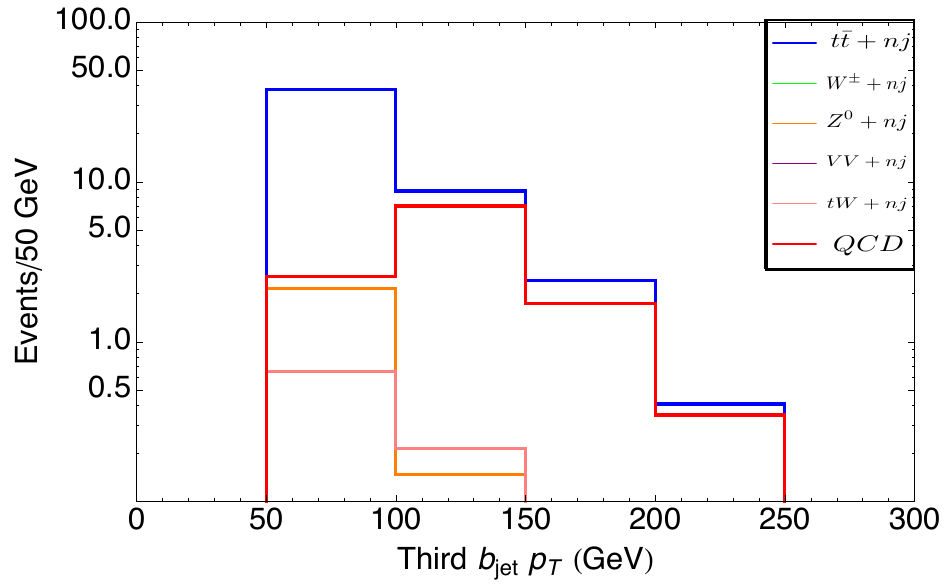}
\caption{\label{Fig: Bjet1pt}  The $p_T$ of the 1st, 2nd, and 3rd hardest $b$-jet in MC background events with at least that many $b$-jets, with a total luminosity of $1$ fb$^{-1}$, after requiring $\MET\ge100$ GeV.}
\end{center}
\end{figure}

\begin{figure}[!t]
\begin{center}
\includegraphics[width=0.48\textwidth]{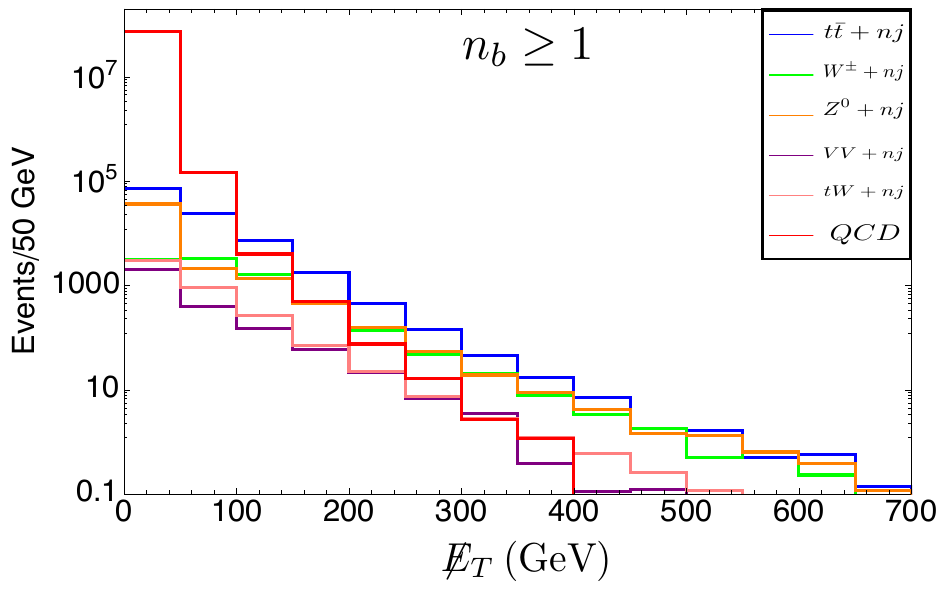}
\includegraphics[width=0.48\textwidth]{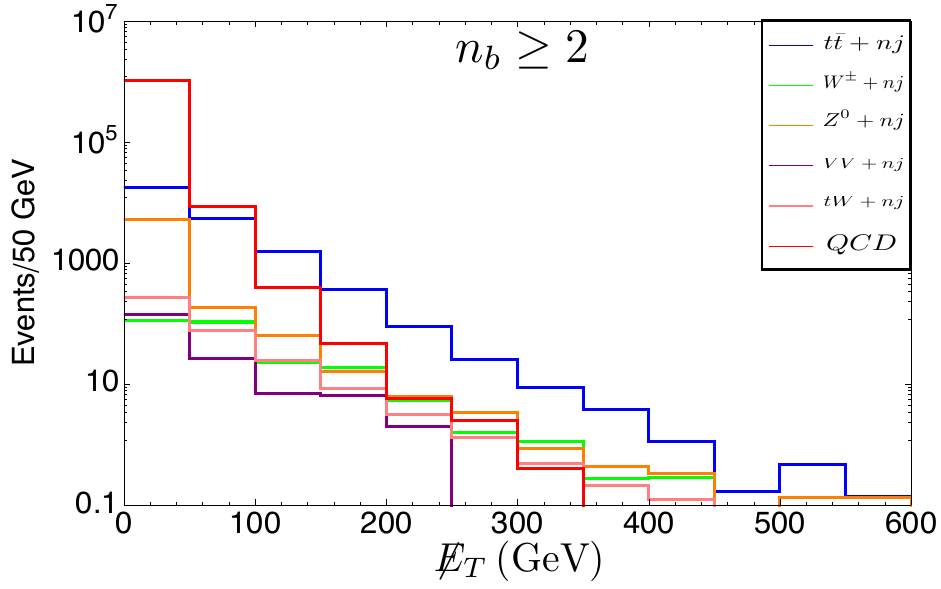}
\caption{\label{Fig: METbjetge1} \label{Fig: METbjetge2}  $\MET$ of MC background events with greater than or equal to 1 or  2 $b$-jets, with a total luminosity of $1$ fb$^{-1}$.}
\end{center}
\end{figure}

\begin{figure}[!t]
\begin{center}
\includegraphics[width=6.0in]{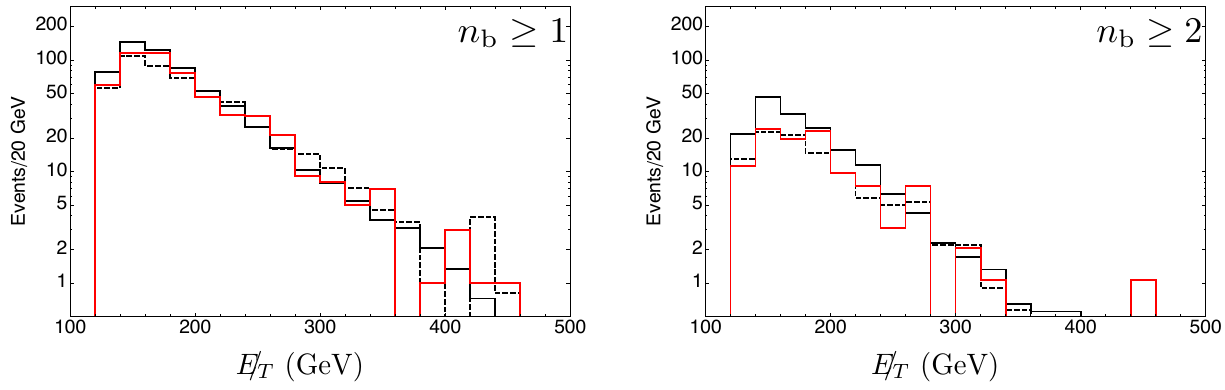}\\
\includegraphics[width=6.0in]{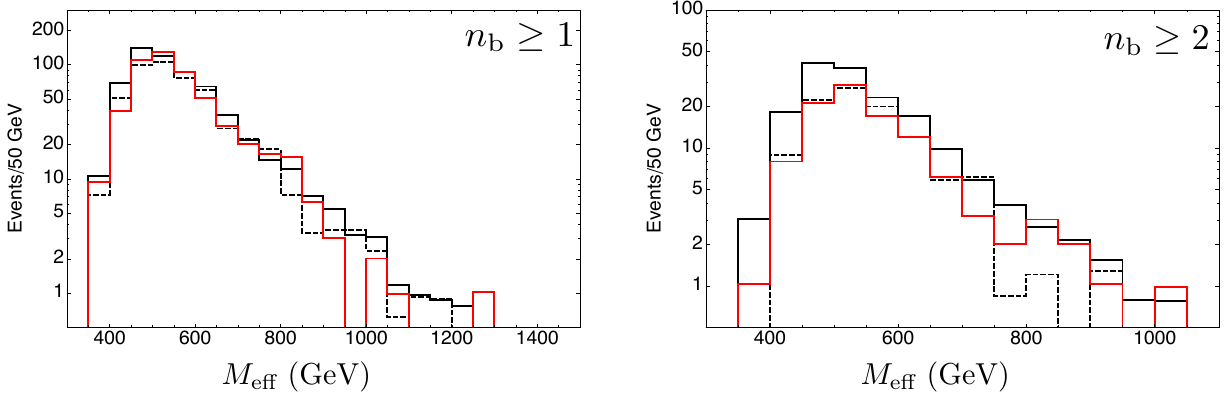}
\caption{\label{Fig: MCCheckMET} \label{Fig: MCCheckMeff}  Comparison of the backgrounds generated for this study against the ATLAS backgrounds from  \cite{Aad:5}. The top (bottom) panel shows the $\MET$ ($M_{\text{eff}}$) distributions in the $n_b\ge1$ and $n_b\ge2$ channels, respectively. The backgrounds generated in this study with \texttt{PGS} are shown in solid black, the ATLAS backgrounds in dashed black, and the data in solid red. The figures are shown assuming $\mathcal{L}=0.83$ fb$^{-1}$ of integrated luminosity. }
\end{center}
\end{figure}

With these modifications to the \texttt{PGS 4} code, the main standard model backgrounds were calculated. 
Fig.~\ref{Fig: Bjet1pt}
shows the $p_T$ spectrum of the leading, second, and third $b$-jet in the event for each of the SM backgrounds that were calculated. Fig.~\ref{Fig: METbjetge1} shows the $\MET$ distributions for backgrounds after requiring $n_b\ge1$ and $n_b\ge2$, respectively. Furthermore, Fig.~\ref{Fig: MCCheckMET} shows a comparison between the main SM backgrounds generated for this study in \texttt{PGS} vs the simulated backgrounds used by ATLAS in  \cite{Aad:5}. Reasonable agreement between distributions is found for $n_b\ge1$. For $n_b\ge2$ the distributions obtained from \texttt{PGS} over-predict the number of events in the signal region.  ATLAS currently does not use a separate jet energy calibration for $b$-jets which are systematically lower reconstructed at lower $p_T$ than their true value.   To model this effect, the $p_T$ requirement on the $b$-jets is increased from the quoted value by ATLAS of 50 GeV to 60 GeV and gives a better agreement between our backgrounds and those from ATLAS.

\section{Expected limits from existing LHC searches}
\label{Sec: CurrentLimits}

The LHC experiments have performed analyses in the jets and $\MET$ channel using 
nearly 1 fb$^{-1}$ of data from the 2010 and 2011 runs.   
It is important to study and compare the search regions in these studies to our optimized search strategies in 
Sec.~\ref{Sec: Optimization}.  
We consider several ATLAS studies, three with $35$ pb$^{-1}$ of 2010 data that remain relevant for constraining the 
low-mass regions, and three that include more data: 
\begin{enumerate}
\item A study \cite{daCosta:2011qk} with  $35$ pb$^{-1}$ of $2$ or $ 3^+$ light flavor jets and $\MET \geq 100$ GeV, with various $m_{\eff}$ and jet $p_T$ cuts, including cuts on the ratio of $\MET/m_{\eff}$
\item A study \cite{Aad:2011ks} with  $35$ pb$^{-1}$ demanding at least one $b$-jet with $0$ or $1^+$ lepton, and $\MET > 80$ or $100$ GeV, respectively 
\item A study \cite{Aad:2011xm} with $35$ pb$^{-1}$ using opposite-sign or same-sign dileptons, with a missing energy cut $\MET > 150$ or $100$ GeV, respectively
\item An updated study \cite{Aad:2011hh} of jets, leptons, and missing energy with $165$ pb$^{-1}$ requiring $\MET > 0.25 m_{\eff}$ and $m_{\eff} > 500$ GeV
\item An ATLAS study \cite{Aad:5} with $0.83$ fb$^{-1}$ requiring $\MET/m_{\eff} > 0.25$ and  $\MET > 130$ GeV, defining four signal regions with $1$ or $2$ $b$-jets and $m_{\eff} > 500$ or $700$ GeV
\item An updated study \cite{Vivarelli} with $1.03$ fb$^{-1}$ demanding $m_{\eff} > 1000$ GeV and $\MET/m_{\eff} > 0.25$ with four signal regions with various jet $p_T$ cuts
\end{enumerate}
There are also many similar relevant CMS analyses looking for jets and missing energy with or without leptons and photons,  analyses using MT2 and $\alpha_T$, and also a CMS analysis using same-sign dileptons (e.g. \cite{Kiesenhofer:2011wv,Khachatryan:2011tk,Chatrchyan:2011wb}).  While this paper was in preparation several new interesting LHC searches have emerged \cite{UNUSEDLHC}. 
We will not consider these here. 

In Appendix \ref{Sec: CurrentLimits Plots}, in Fig.~\ref{Fig:Limits}, we show the \emph{expected} 95 C.L. limits for 
$\LL=1$ fb$^{-1}$ from the above ATLAS studies on all simplified models from \refsec{SimplifiedModels} for different 
choices of the production cross section.  
Note that we use our own MC for the background modeling. 
In the next section, we find a set of minimal search regions that are needed to cover the space of simplified models, and estimate the 7 TeV LHC's sensitivity to the simplified models using this minimal set of search regions. 

\section{Optimal Search Regions}\label{Sec: Optimization}

The simplified models introduced in \refsec{SimplifiedModels} can be used to develop broad search strategies that cover the model space. Despite the reduction in the number of relevant free parameters in simplified models compared to complete theories, a multi-signal-region strategy is needed to efficiently cover all kinematic possibilities. In this section, we find a minimal number of signal regions necessary to cover the entire space of simplified models.  Then in \refsec{Benchmarks} we will propose a set of benchmark models that span the full parameter space of simplified models, in the sense that a search strategy that is sensitive to all benchmarks will also be sensitive to all of the simplified models. 

The terminology used throughout the rest of the article is introduced in what follows. We assign all events to a signal region defined by the number of 
jets ($N_{\text{jet}}$), $b$-jets ($N_{b\text{jet}}$), and leptons ($N_\ell$), as well as the missing transverse 
($\MET$) and visible energies ($H_{T}$): 
\be\label{eq:cuts}
(N_{\text{jet}}, N_{\ell}, N_{\text{bjet}}, \MET, H_{T}).
\ee
In addition to a cut of 120 GeV on the transverse momentum ($p_T$) of the 
hardest jet, we consider the following set of cuts: 
\begin{eqnarray*}
N_{\text{jet}} &\in& \{2^+, 3^+, 4^+\}\\
N_{b\text{jet}} &\in& \{ 0^+, 1^+ ,2^+, 3^+\}\\
N_{\ell} &\in& \{ 0, 1^+, 2^+, 3^+, \text{SSDL}^+, \text{OSDL}^+\}\\
\MET{}_{\text{ min}} &\in& \{ 0, 50, 100,\ldots, 500\}\GeV\\
H_{T\text{ min}} &\in& \{ 200, 300, \ldots, 1200\}\GeV\,.
\end{eqnarray*}
Here labels without a ``$+$'' are exclusive cuts, {\it e.g.} exactly zero isolated leptons are required for $N_\ell = 0$;  
the superscript ``$+$'' indicates that the search regions are inclusive, {\it e.g.} $N_{\text{jet}} = 2^+$ requires two or 
more jets; 
SSDL refers to ``same-sign di-lepton'' and OSDL refers to ``opposite sign di-lepton''; 
leptons are electrons and muons (not taus, which are treated as jets) and are required to be isolated; 
all jets (including $b$-jets) beyond the hardest one are required to have $p_T>50$ GeV and $|\eta_j|<2.5$. 
This gives rise to a set of 8,712 possible search regions.  Not all search regions are physical due to the overlap 
between the $H_T$ cut and other $p_T$ and/or $\MET$ requirements
and this reduces the number of search regions to 8064. 
Note that requiring even higher jet multiplicities may be useful \cite{Lisanti:2011tm}, but we do not consider this here. 

A given signal region or cut, $C_i$, will yield an expected limit on the cross section times branching ratio, $\sigma\times\mathcal{B}$, for a given simplified model at the 95 \% C.L. given by
\be
(\sigma\times\mathcal{B})_i &=& \frac{\Delta(B)_{i}}{\mathcal{L} \times\epsilon(M)_i}.
\ee
where $\epsilon(M)_i$ is the efficiency of $C_i$ on the model point $M$. $\Delta(B)_i$ is the allowed number of events in the signal at the 95 \% C.L. if $B$ background 
events are expected and in fact fit the data. We take 
\be
\Delta(B) &=& 2 \times \sqrt{\text{Stat}(B)^2 + (\epsilon(n_{\text{bjet}})_{\text{syst}}B)^2},
\ee
where Stat$(B)$ is the Poisson limit on $B$. We also include a systematic error, $\epsilon(n_{\text{bjet}})_{\text{syst}}$,  
as a function of $N_{b\text{jet}} $ in the signal region \footnote{The choices of systematic errors used were made following private communication with the SLAC ATLAS group; however, these are also consistent with  \cite{Aad:5}.}:
\begin{eqnarray}
\epsilon(0b)_{\text{syst}}= 20\%\quad
\epsilon(1b)_{\text{syst}}= 20\%\quad
\epsilon(2b)_{\text{syst}}= 40\%\quad
\epsilon(3b)_{\text{syst}}= 60\%.
\end{eqnarray}

The optimal cross section limit on a model $M$ is given by
\be
(\sigma\times\mathcal{B})_{\text{opt}} &=& \{\text{min}((\sigma\times\mathcal{B})_i): i\in\{1,N_{\text{cuts}}\}\}\,, 
\ee
where the number of search regions is $N_{\text{cuts}}=8064$.  
It is natural to quantify the ``goodness'' of a cut $C_i$ by the amount of LHC data needed to make a discovery 
or exclusion using that cut.  For this purpose, we introduce the efficacy of a cut
\be
\Eff(C_i) &=&\frac{(\sigma\times\mathcal{B})_i}{(\sigma\times\mathcal{B})_{\text{opt}}}. 
\ee
In words, this is just the ratio of the necessary production cross section for discovery using a cut $C_i$ divided by the cross section necessary for discovery using the optimal set of cuts.  An efficacy of $1$ is `perfect', and otherwise smaller efficacies are better.  Thus the best search strategy for all model points $M$ will be a combination of cuts $\{C_i\}$ such that $\Eff$ is close to one for every model using at least one of the cuts in the search strategy.

\begin{figure}[!t]
\begin{center}
\includegraphics[width=0.32\textwidth]{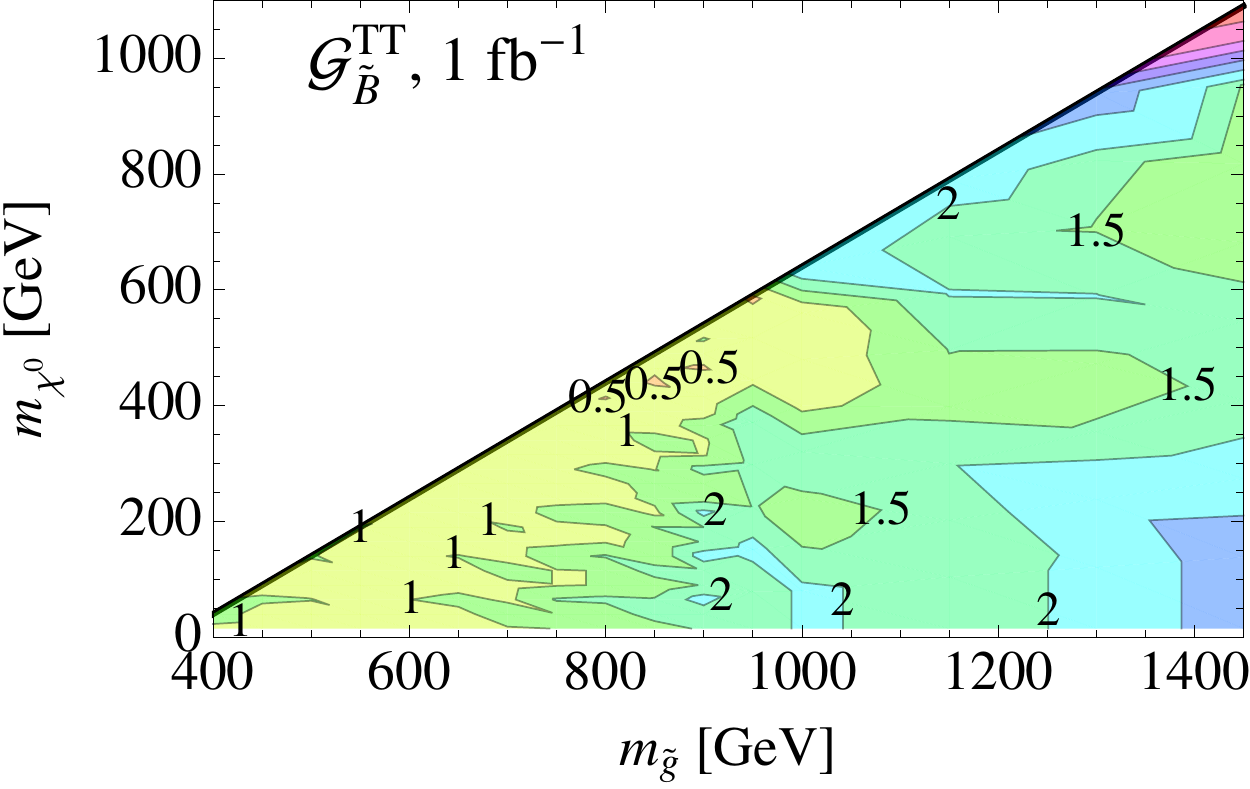}
\includegraphics[width=0.32\textwidth]{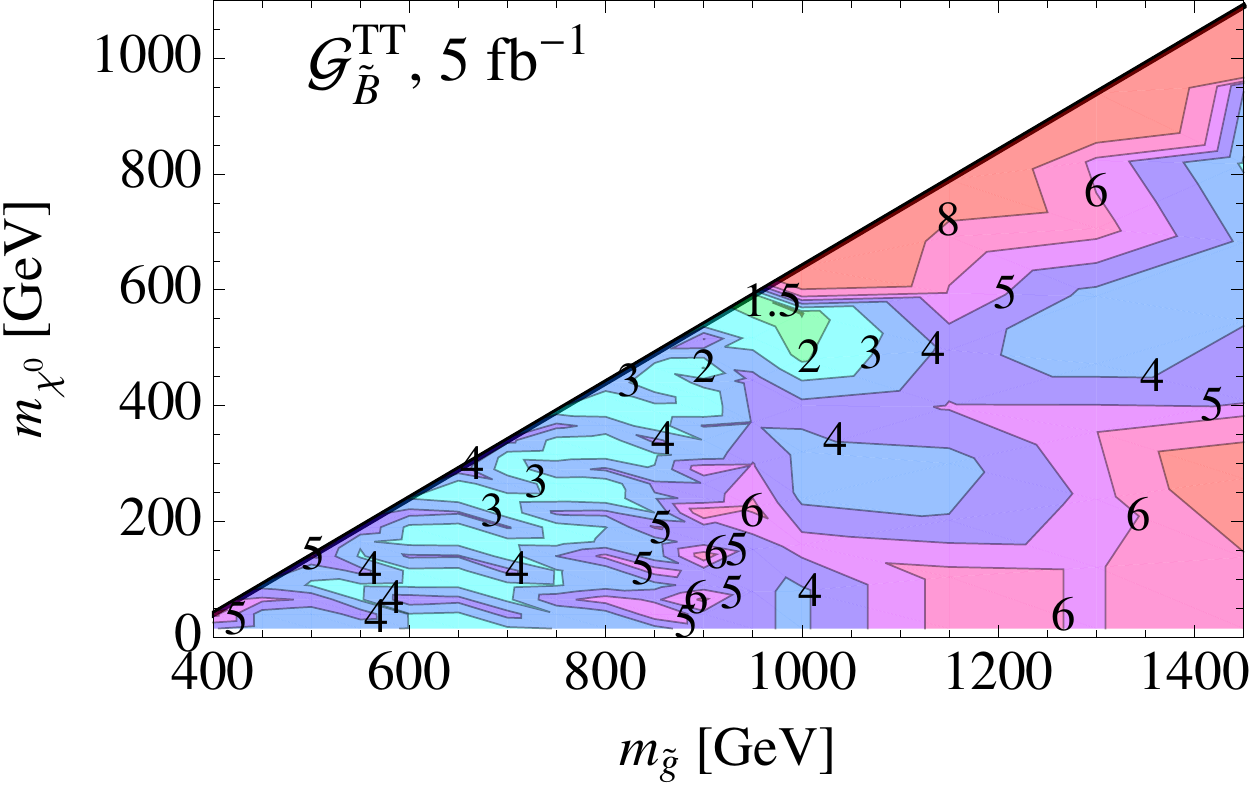}
\includegraphics[width=0.32\textwidth]{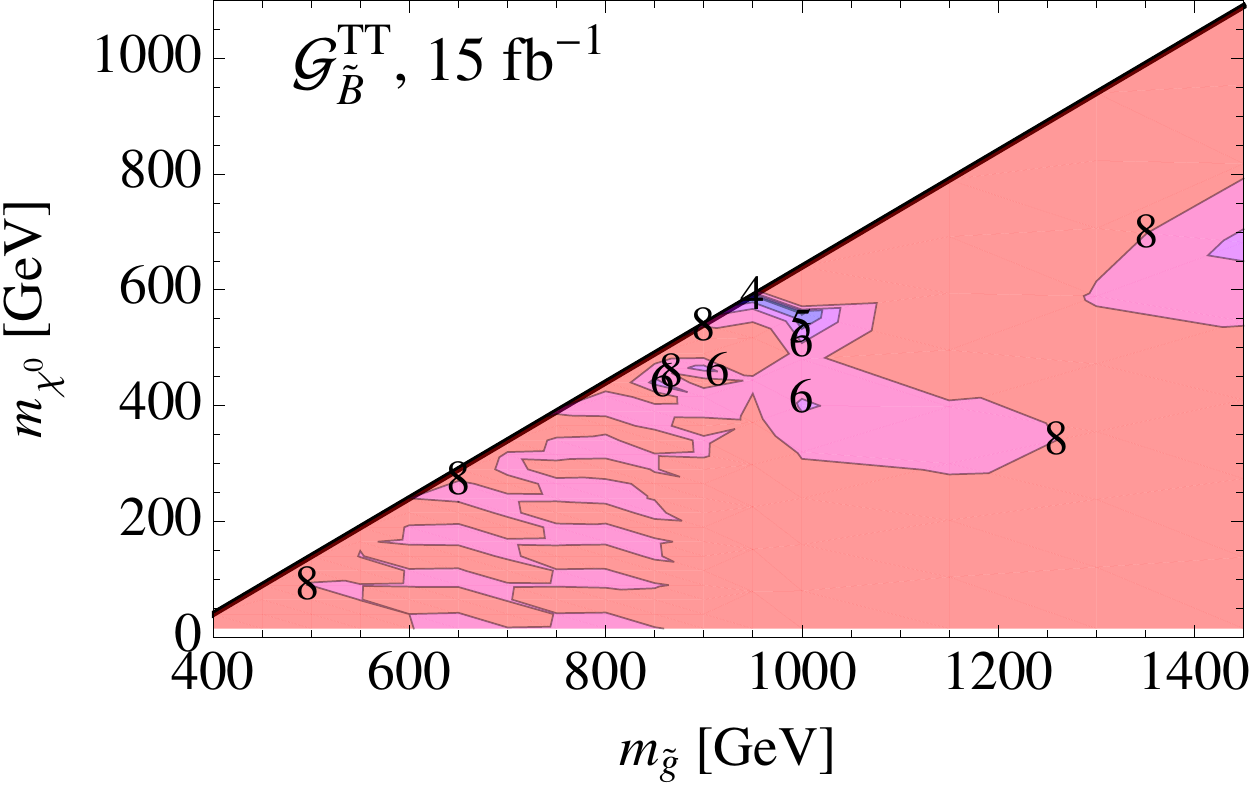}\\
\includegraphics[width=0.32\textwidth]{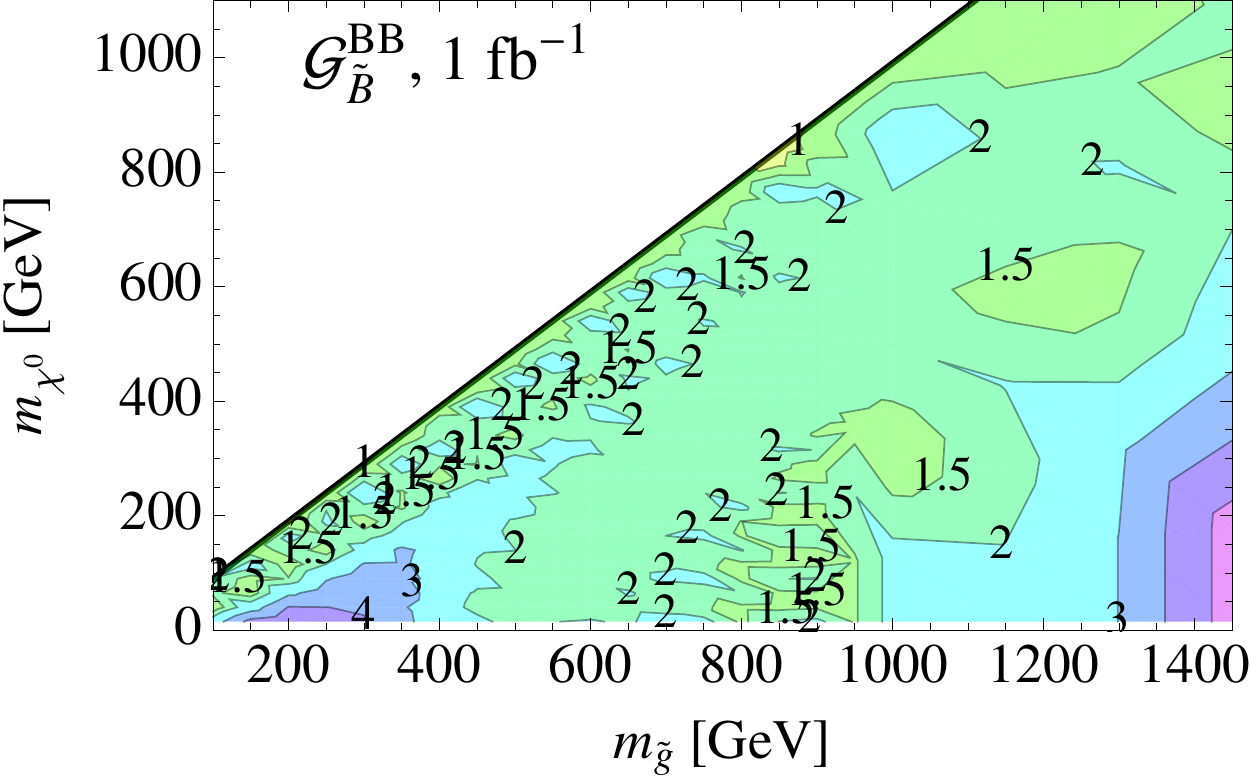}
\includegraphics[width=0.32\textwidth]{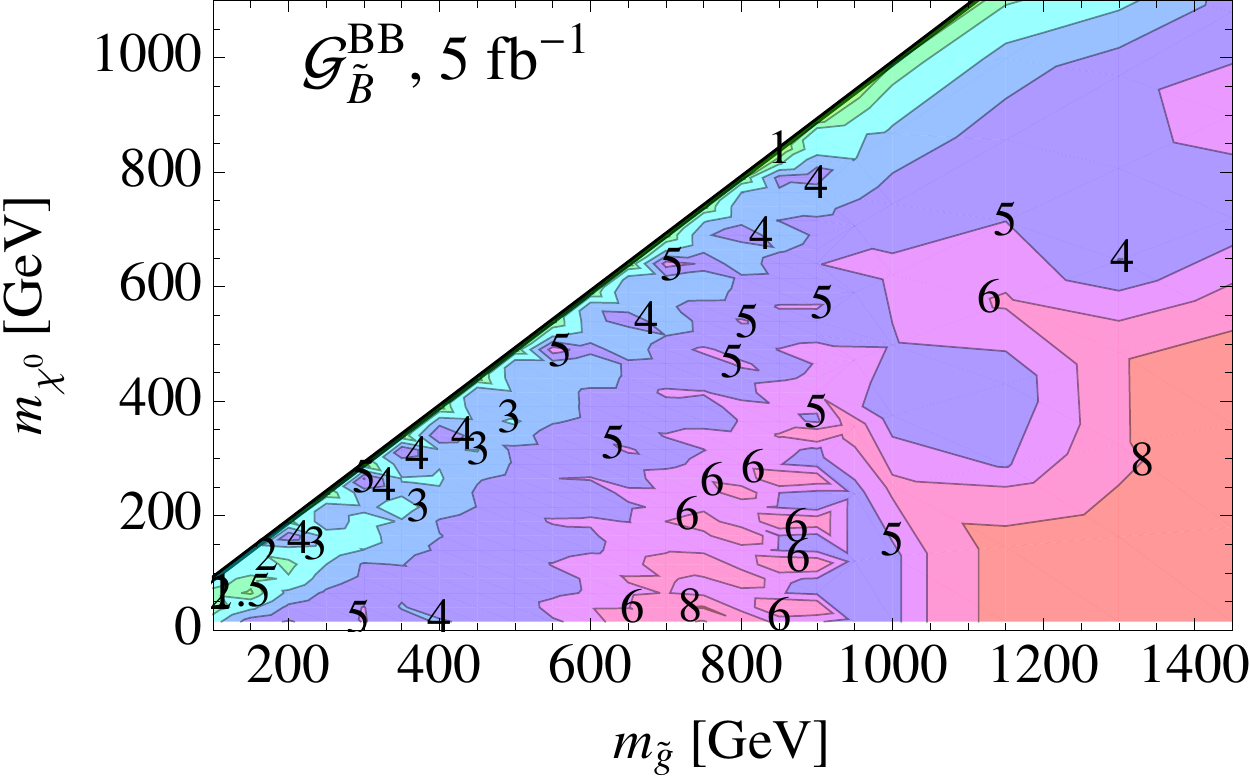}
\includegraphics[width=0.32\textwidth]{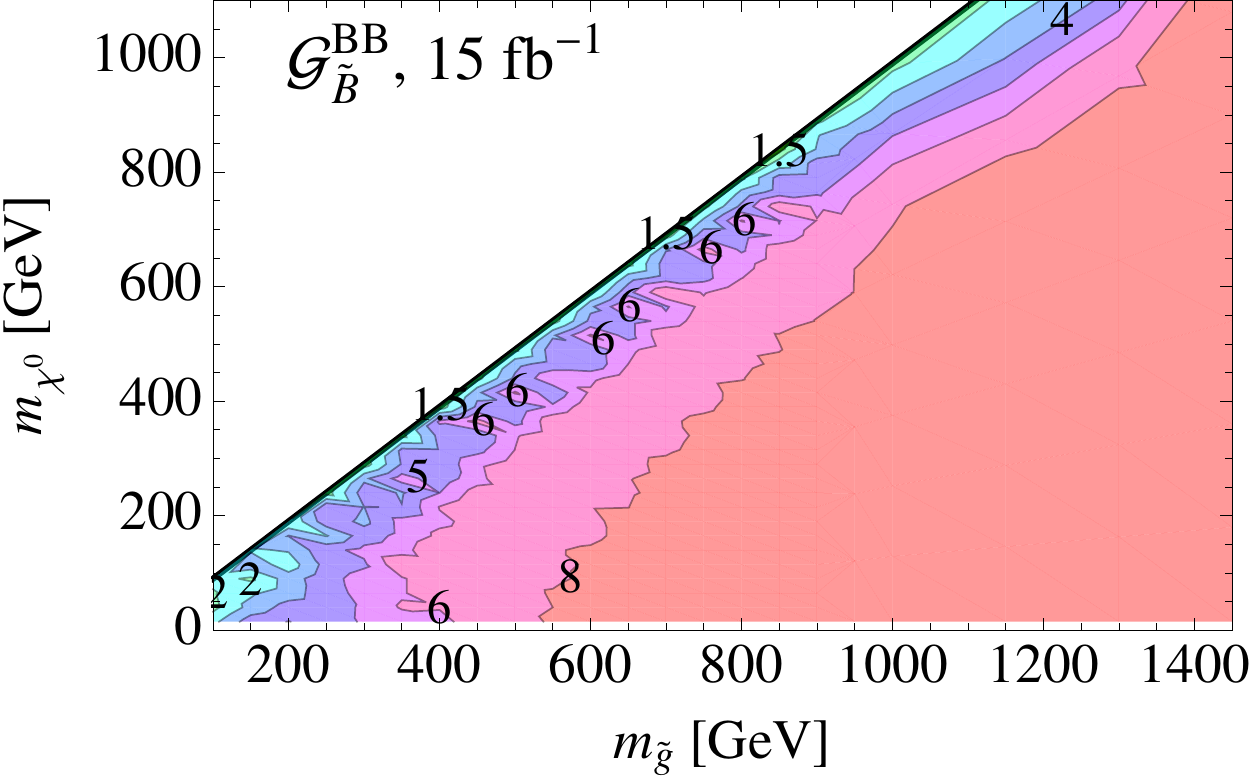}
\caption{\label{Fig:ATLASEfficacies} Expected efficacies of the ATLAS search regions on the $\GB^{\tt TT}$ (top panel) 
and $\GB^{\tt BB}$ (bottom panel) simplified models, for integrated luminosities $\LL=1$ fb$^{-1}$, $5$ fb$^{-1}$, 
and $15$ fb$^{-1}$ (first, second, and third column, respectively).  Our goal is to find a set of search regions that 
have efficacies close to $1$ for all simplified models and for each luminosity. While current ATLAS 
searches give excellent coverage, other strategies will be required for higher luminosities data sets, especially for events with many final state $b$-jets and tops.}
\end{center}
\end{figure}

It is interesting to compute the expected efficacies of the search regions 
used by public LHC analyses  (see Sec.~\ref{Sec: CurrentLimits}) for the heavy flavor simplified models.  Fig.~\ref{Fig:ATLASEfficacies} shows the efficacies for  the $\GB^{\tt TT}$ (top panel) and $\GB^{\tt BB}$ (bottom panel) topologies for $\LL=1$ fb$^{-1}$, $5$ fb$^{-1}$, and 
$15$ fb$^{-1}$ (first, second, and third column, respectively).  For the current low-luminosity searches, the LHC 
analyses have very good efficacies, and there are even a few small isolated regions where the LHC 
searches have been slightly better than the search regions considered here, due to the looser triggers allowed by lower luminosity analyses.
At higher luminosities, it is of course possible to greatly improve the search strategies. 

\subsection{Optimizing Search Strategies}
\label{sec:OptimizingSearchStrategies}

It is not feasible, nor necessary, to look at all 8064 search regions to have a reasonable efficacy over the entire 
set of simplified models.  One of our principle goals is to find a minimal 
comprehensive set of signal regions that cover the space of simplified models spanned by the topologies from 
\refsec{SimplifiedModels}. Each solution to this problem will be a set of cuts $\{ C_i \}$, where at least one $C_i$ in the set gives $\Eff(C_i)\le \Eff_{\text{crit}}$, for some ``critical'' efficacy $\Eff_{\text{crit}}$, for every point $M$ in the space of simplified models. Different solutions will be found for benchmark luminosities $\mathcal{L}=$ 1 fb$^{-1}$, 5 fb$^{-1}$, or 15 fb$^{-1}$. 

A genetic algorithm was used to construct the minimal set of search regions to cover 
the entire space of simplified models.   The configuration space for the genetic algorithm is a binary string of 8712 
bits that signify whether a particular search region is used or not (recall that there are 8712 search regions before 
imposing constraints);  thus the configuration space is $2^{8712}$ states.  A set of random cuts were created by 
turning on a weighted selection of 40 of the 8712 search regions.   The weight of each search region is proportional 
to the number of models to which the search region is sensitive (i.e.~the number of models for which the cut $C_i$ 
gives $\Eff(C_i)\le \Eff_{\text{crit}}$).   The sets of search regions are evaluated to see how 
many models they cover within the desired efficacy, and a ``fitness'' is assigned to them with the formula
\begin{eqnarray} \label{eq:FitnessFunction}
f(C, M) = \frac{1}{M_{\text{max}}^2 - (M^2 - C)}\,,
\end{eqnarray}
where $M$ is the number of models covered, $C$ is the number of search regions, and $M_{\text{max}}$ is the total number of models.  
This fitness function strongly penalizes search strategies that do not cover all models, followed by a penalty 
for having too many search regions.  

After evaluating the fitness of the search strategies, the  least fit  50\% are removed.  Pairs of fit search strategies are then selected and a new search strategy is created by taking a randomly determined fraction of each search strategy's used search regions.  For instance, if the two selected search strategies had $N_1$ and $N_2$ search regions, then a uniform random number on the unit line segment, $x$, would determine that $x N_1 $ search regions would be taken from the first search strategy and $(1-x) N_2$ would be taken from the second search strategy. So if $N_1=20$ and $N_2=30$ and $x=0.20$, 4 search regions would be taken from the first search strategy and 24 would be taken from the second.  If duplicate signal regions are selected, the duplicate is removed, reducing the number of search regions.  
After creating a new search strategy, the search is mutated to guarantee that the population of search strategies had sufficient diversity.   Each used search region has a probability of being changed to another random search region.  We 
use 6\% for this probability known as the ``mutation rate''.  Thus with the 28 search regions in the example, 1.5 changes would be made on average. 

If after ten consecutive generations no progress has been made, i.e.~if no solution has been found that covers the entire model space, then a solution is manually created by forcing every model to be covered by some search region.  This can be done by increasing the number of search regions in the search strategies until full coverage is achieved.   Finally, if every model is covered 
and no further progress is achieved for seven generations, search strategies are scoured to see if any search region can be removed without reducing coverage.  Either way, the genetic algorithm is restarted.  If no progress in reducing the number of search regions in a search strategy has been made in twenty generations, the program ends.  

Typically, the program terminated after 50 to 70 generations, and 20 to 300 distinct optimized search strategies were found each time.  While the termination of the program does not guarantee that the optimal solution has been found, re-running the program multiple times usually results in the same number of required search regions.   The distinct search strategies typically have similar features even if they differ slightly in detail.   

The program can easily identify optimized search strategies that cover all models with an efficacy less than the critical efficacy, and we found that $\Eff_{\mathrm{crit}} = 1.75$ results in just a small number ($< 10$) of signal regions that can cover all models.  We have chosen to present our benchmark optimized search regions, which will be discussed in more detail in the next section. Table \ref{tab: SearchRegions BM} presents the search regions found after optimizing over all 8712 search regions that together cover all benchmark models with an efficacy of 1.75 or better (i.e.~$\Eff_{\text{crit}}=1.75$), for $\mathcal{L}=$ 1 fb$^{-1}$, 5 fb$^{-1}$, or 15 fb$^{-1}$.  

 The optimized search strategy from Table \ref{tab: SearchRegions BM} has three distinct sets of search regions. First, it achieves sensitivity to generic light flavor jets and missing energy 
signals with a small number of regions that involve significant $\MET$ and large $H_T$ cuts.  
Second, to uncover heavy flavor physics, there are several search regions involving $b$-jets with more modest 
cuts on $H_T$.  This second category of search regions is not that different from standard signal regions used by the LHC experiments thus far in searches for heavy flavor models. 
Third, a set of non-standard cuts are found. Since many of the simplified models produce $4b$ jets in every event, we find that $3b$-jet search regions
with a very modest $\MET$ cut and a minimal $H_T$ cut can be very effective (see also \cite{pheno}).  
Furthermore, the signal region involving 
same-sign dileptons in events with $2$ or more $b$-jets achieves great sensitivity to multi-top events, particularly for points in model space where $m_{\tilde g} \approx 2 m_t + m_{\chi^0}$.

\subsection{Benchmark Models and Re-optimized Search Strategies}
\label{Sec: Benchmarks}

\begin{table}[!t]
\begin{center}
{\large{\textrm{1 fb}$^{-1}$:}} $\qquad$
\begin{tabular}{|c||c||c|c|c|c|c|}
\hline
&\text{Search Region} & $N_j$ & $N_{\ell}$ & $N_{\text{bjet}}$ & $\MET$ & $H_T$\\
\hline\hline
\text{0 $b$, High $H_T$} &1 & $4^+$ & 0 & 0 & 300 & 1000\\
\text{0 $b$, High MET}  &2 & $4^+$ & 0 & 0 & 450 & 600\\ 
\hline
\text{1 $b$, Low $H_T$} &3& $2^+$ & 0 & $1^+$ & 300 & 400\\
\text{1 $b$, High $H_T$}&4 & $3^+$ & 0 & $1^+$ & 300 & 600\\
\hline
\text{3 $b$}&5 & $4^+$ & 0 & $3^+$ & 150 & 400\\
\hline
\text{1 $b$, SSDL}&6 & $3^+$ & $\text{SSDL}^+$ & $1^+$ & 0 & 200\\
\hline 
\end{tabular}\\
\vskip 0.5cm
{\large{\textrm{5 fb}$^{-1}$:}} $\qquad$
\begin{tabular}{|c||c||c|c|c|c|c|}
\hline
&\text{Search Region} & $N_j$ & $N_{\ell}$ & $N_{\text{bjet}}$ & $\MET$ & $H_T$\\
\hline\hline
\text{0 $b$} &1 & $4^+$ & 0 & 0 & 400 & 900\\
\hline
\text{1 $b$} &2& $3^+$ & 0 & $1^+$ & 450 & 500\\
\text{1 $b$} &3& $4^+$ & 0 & $1^+$ & 350 & 500\\
\hline
\text{2 $b$}&4 & $2^+$ & 0 & $2^+$ & 400 & 400\\
\hline
\text{3 $b$, low $H_T$}&5 & $3^+$ & 0 & $3^+$ & 100 & 200\\
\text{3 $b$, high $H_T$}&6 & $4^+$ & 0 & $3^+$ & 250 & 400\\
\hline
\text{1 $b$, SSDL}&7 & $3^+$ & $\text{SSDL}^+$ & $1^+$ & 0 & 300\\
\hline 
\end{tabular}\\
\vskip 0.5cm
{\large{\textrm{15 fb}$^{-1}$:}} $\qquad$
\begin{tabular}{|c||c||c|c|c|c|c|}
\hline
&\text{Search Region} & $N_j$ & $N_{\ell}$ & $N_{\text{bjet}}$ & $\MET$ & $H_T$\\
\hline\hline
\text{0 $b$} &1 & $4^+$ & 0 & 0 & 450 & 1100\\
\hline
\text{1 $b$, High $H_T$} &2& $4^+$ & 0 & $1^+$ & 350 & 900\\
\text{1 $b$, High $\MET$} &3& $4^+$ & 0 & $1^+$ & 450 & 500\\
\hline
\text{2 $b$}&4 & $2^+$ & 0 & $2^+$ & 400 & 600\\
\hline
\text{3 $b$}&5 & $4^+$ & 0 & $3^+$ & 250 & 600\\
\hline
\text{1 $b$, SSDL}&6 & $3^+$ & $\text{SSDL}^+$ & $1^+$ & 0 & 300\\
\hline 
\end{tabular}
\vskip 0.5cm
\caption{Search regions that were optimized on the benchmark models. Together they cover virtually all models 
with an efficacy of 1.75 or better, 
for $\mathcal{L}=$ 1 fb$^{-1}$, 5 fb$^{-1}$, and 15 fb$^{-1}$.  
}
\label{tab: SearchRegions BM}
\end{center}
 \end{table}

The optimized set of signal regions discussed above cannot be obtained without sampling over a very large model space. There may be practical limitations in doing such fine sampling, especially for the experimental groups which must use full detector simulation in their analyses. In this section, we find a set of models, or benchmarks, so that if a search strategy that covers this set of models with a given efficacy, $\Eff_{\text{crit}}$, is found, it will also cover the {\it entire} space of simplified models.
An important caveat here is that this assumes that they optimize over the set of cuts in Eq.~(\ref{eq:cuts}).  Additional 
cuts, or a different set of cuts, may require slightly different benchmarks.

To make the set of benchmarks as intuitive as possible, we began with five benchmarks per topology, 
spaced to effectively span both the massless and degenerate LSP regions.  However, we found that 
these benchmarks alone fell far short of our goal.  A search optimized only for these benchmarks will 
miss roughly one third of the simplified model parameter space.  To improve the benchmark list, we 
found that the most important additions were in the simplified model topologies with many b-quarks 
but without top quarks and leptons, such as the $\GB^{\tt BB}$ and $\BB$ models.  
The set of benchmarks found are listed in Appendix \ref{Sec: BechmarkTables}.  
Note that the set is not unique, and it may also be possible to create a list with a slightly 
smaller number of benchmarks; however, the list does present a useful solution to covering the 
simplified model parameter space.  

Table \ref{tab: SearchRegions BM} presents the search regions optimized on the \emph{benchmark models}
that together cover virtually all models with an efficacy of 1.75 or better 
(i.e.~$\Eff_{\text{crit}}=1.75$), for $\mathcal{L}=$ 1 fb$^{-1}$, 5 fb$^{-1}$, or 15 fb$^{-1}$.  Fig.~\ref{Fig: BackgroundHistograms} shows the individual background contributions to the optimal search regions found for $\mathcal{L}=5$ fb$^{-1}$.
In Fig.~\ref{fig:coverage1}, we show for $\mathcal{L}=$ 5 fb$^{-1}$ the region covered by each of 
the seven signal regions for the $\GB^{\tt TT}$ and $\GB^{\tt BB}$ simplified models.  
Appendix \ref{Sec: OptLimits Plots} shows the \emph{expected} 95 C.L. limits cross section times branching 
ratio sensitivity for all simplified models in \refsec{SimplifiedModels} 
from the search regions in Table \ref{tab: SearchRegions BM} that have been optimized on the benchmarks 
in Appendix \ref{Sec: BechmarkTables}, for $\LL= 1\ifb$, $5\ifb$, and $15\ifb$.  

\begin{figure}[!t]
\begin{center}
\includegraphics[width=0.6\textwidth]{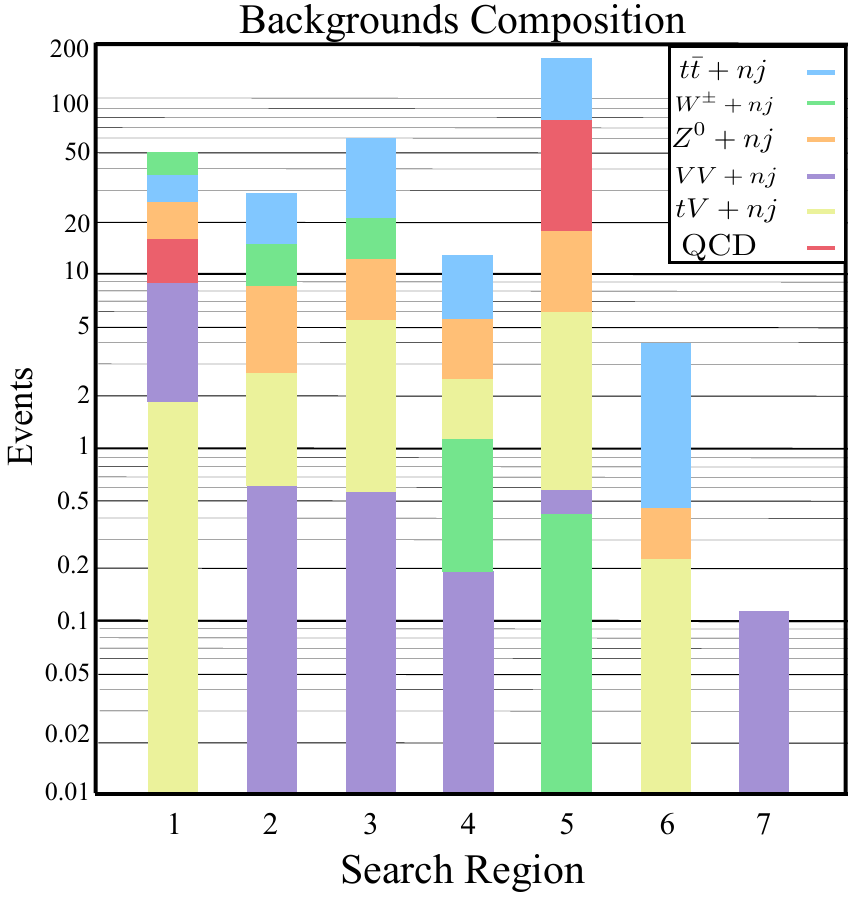}
\caption{\label{Fig: BackgroundHistograms}  Individual background contributions to each of the seven search regions in 
Table \ref{tab: SearchRegions BM} for $\mathcal{L} = 5$ fb$^{-1}$. The dominant background is usually $t\bar{t}+n j$.  
The SSDL search region is essentially background-free.}
\end{center}
\end{figure}

\begin{figure}[!t]
\begin{center}
\includegraphics[width=0.85\textwidth]{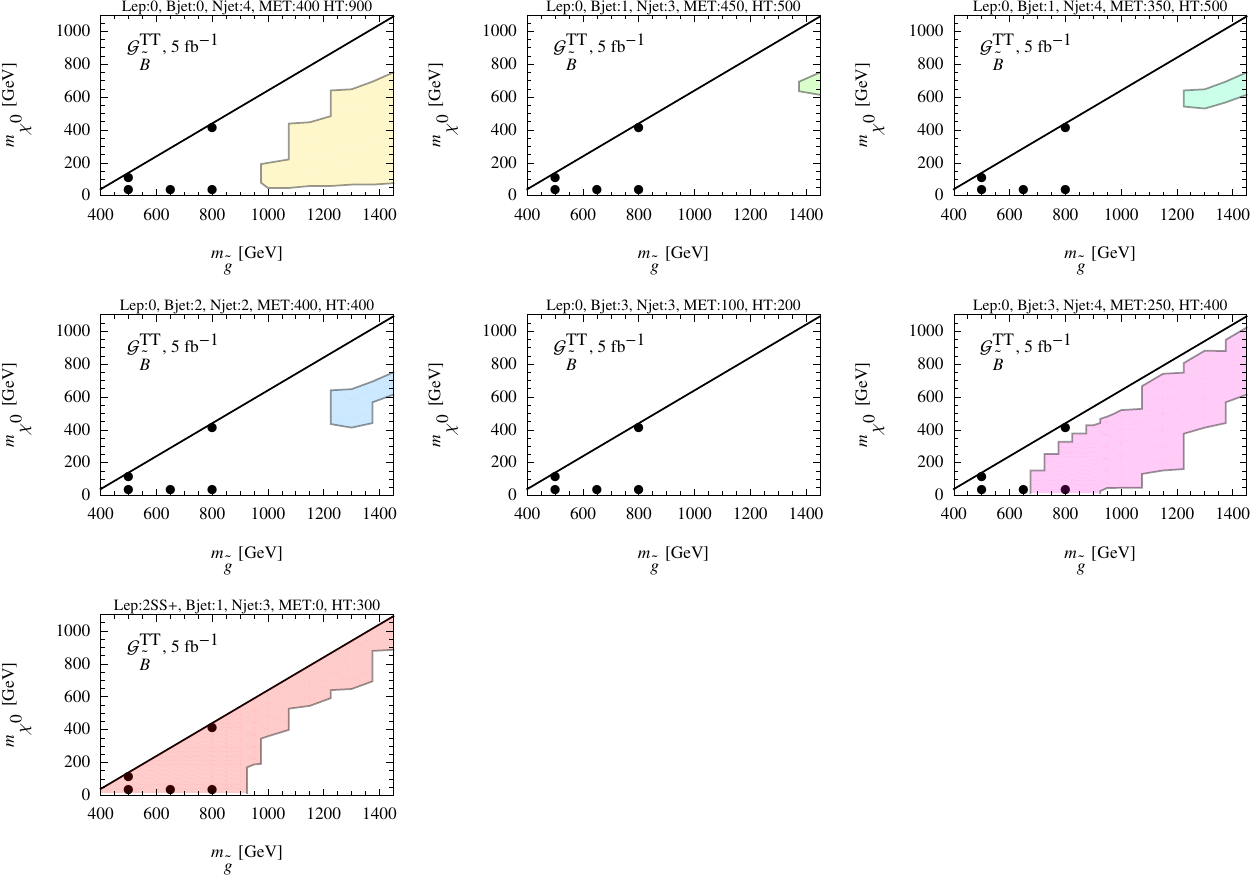} \\
\includegraphics[width=0.85\textwidth]{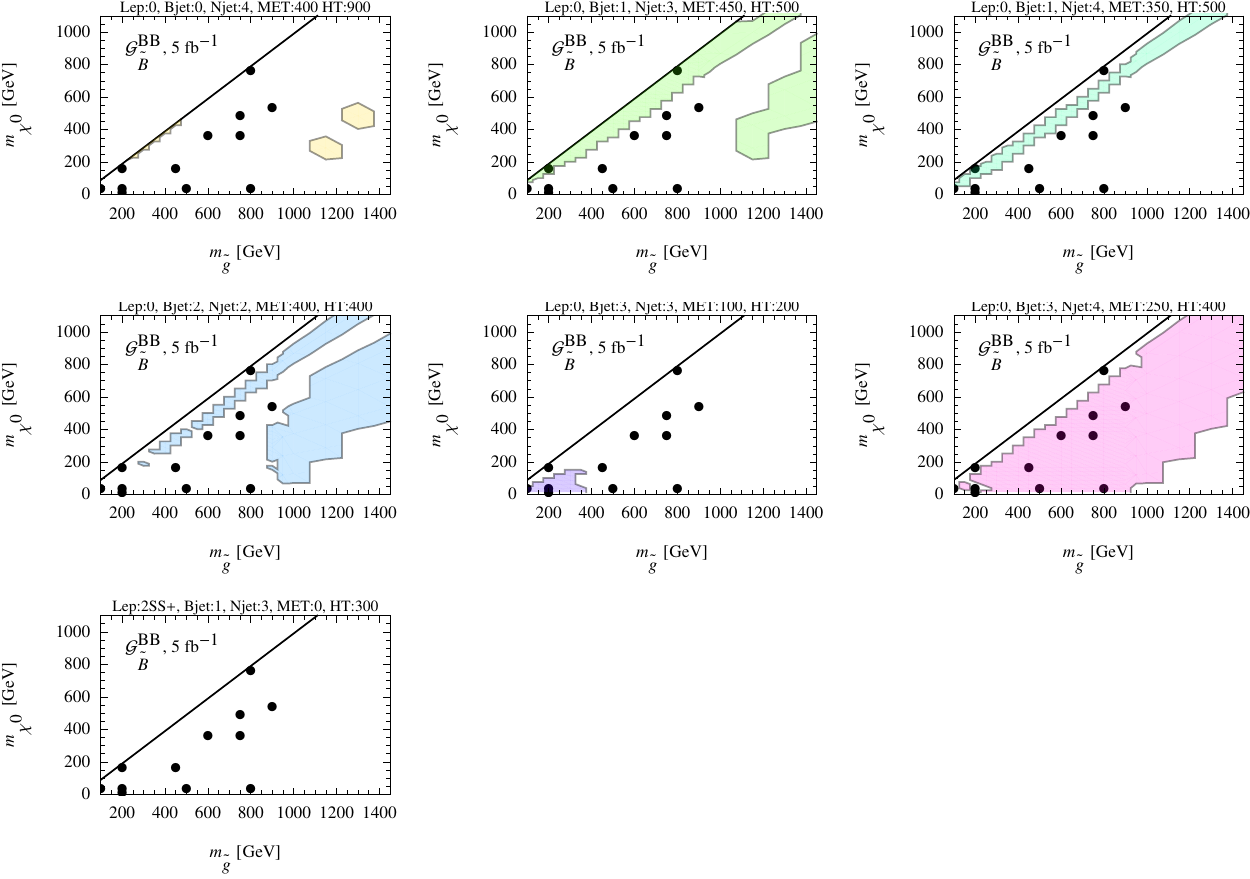}
\caption{\label{fig:coverage1} 
The regions covered by each of the seven searches in Table \ref{tab: SearchRegions BM} 
for the $\GB^{\tt TT}$ (top seven plots) and $\GB^{\tt BB}$ (bottom seven plots) topologies for $\mathcal{L}=$ 5 fb$^{-1}$.  
Note that the 3b search is particularly effective for these two topologies, and the SSDL search is also important for 
$\GB^{\tt TT}$. 
The black dots are benchmark points for these topologies (see Appendix \ref{Sec: BechmarkTables}).}
\end{center}
\end{figure}

An important observation is that the $3b$ and SSDL channels  should be utilized.  
Experimental studies, which currently only go up to $2b$ and $1 \ell$, could be better optimized by 
considering these channels once data sets with greater than 1 fb$^{-1}$ begin to be studied.  

Fig.~\ref{fig:coverage1} shows that the $3b$ and SSDL cuts are by far the most sensitive for top-rich samples.  
The SSDL cut plays an increasingly pivotal role as the size of the data sample increases.  The SSDL cut is especially useful for reducing background without requiring significant amounts of missing energy.  The combination of $b$-jets and SSDL is crucial in order to achieve optimal sensitivity to the $\GB^{\tt TT}$ simplified model near the degeneracy line where $m_{\tilde g} \approx 2 m_t + m_{\tilde \chi^0}$.  Furthermore, even in the case of topologies with mixtures of tops and bottoms, such as the $\GB^{\tt MM}$ simplified model, SSDL will be one of the dominant discovery channels.  
The $3b$ cuts are also useful for discovering simplified models with many $b$-jets and missing energy, such as the  $\GB^{\tt BB}$ model.  

To further explore the utility of $3b$ and SSDL channels, we constructed optimized searches without either or both, and the results are striking.  Without the SSDL channel, the sensitivity to top-rich samples, especially those with low missing energy, will be significantly suboptimal.  Searches without $3b$ channels seem to be even more problematic, as our sensitivity to virtually all heavy flavor color octet decays is degraded.  In fact, we were unable to find a set of search regions that are able to cover a majority of the parameter space with an efficacy better than 1.75. 

The solutions to the optimization problem depend on the choice of systematic errors made. It is natural to ask if a larger systematic error in the $3b$ region would decrease its utility, but we have found that even with $\epsilon(n_{\text{bjet}})_{\text{syst}} = 1$, the $3b$ signal region will still be useful for discovering these simplified models. 

Altogether, the inclusion of $3b$ and SSDL channels leads to a great improvement in efficacy for the majority of heavy-flavor simplified model parameter space.  Although we cannot draw firm conclusions without a full LHC detector study, our results suggest that the LHC's sensitivity to new physics will be greatly improved by including these channels. 

\section{Discussion}\label{sec:discussion}

This work presented a framework for constructing optimal search strategies sensitive to heavy flavor and missing energy signatures at the LHC. We used a set of simplified models, with each model or topology parametrized by only a small number of parameters. This model space offers a wide range of kinematics and can only be probed with a broad and 
flexible search strategy. A search strategy consists of a list of counting experiments, each to be performed in a given search region where a particular set of kinematic and selection cuts have been applied to the data.  Theorists cannot determine the optimal search strategy because we do not have access to realistic LHC detector simulations.  However, we can sidestep this issue by instead providing a more robust and useful piece of information: a set of benchmark models with the property that any search strategy sensitive to all of them will also be sensitive to the entire parameter space of heavy flavor simplified models.  

The benchmarks in Appendix \ref{Sec: BechmarkTables} have been designed to span the parameter space of simplified models involving the pair production of color octets or triplets decaying to all plausible combinations of tops, $b$-quarks, light flavor quarks, and missing energy.  Since the color octets can decay to any combination of third generation quarks, or a pair of light flavor quarks, there are a large number of possible topologies, which differ qualitatively in terms of the number and momenta of jets, leptons, and $b$-jets, and also the amount of missing energy.  Thus there is a large parameter space of models, and it is rather non-trivial that a small number of signal regions can cover all of these models very effectively.  

The notion of efficacy, $\Eff$, was used to get a quantitative handle on the optimization.  The efficacy of a search region applied to a given model is defined by the ratio of the amount of data needed to discover the model to the amount of data needed to discover the model using the optimal search region.  Our results suggest that it is possible to obtain $\Eff < 1.75$ over the entire parameter space of heavy flavor simplified models with a search strategy consisting of only $6$ or $7$ search regions, depending on the integrated luminosity.

The search regions specified here are powerful enough to cover many models simultaneously, 
but one can do better for particular 
models with a dedicated search.  For example, to gain sensitivity for $\T \to t+\chi^0$, other methods may be required for 
the 7 TeV LHC to separate the signal from the dominant $t\bar{t}$ background, which is at 
least six times larger than the signal. Fig.~\ref{Fig: StopvsBckg} shows the $\MET$ and $H_T$ distributions for a 350 GeV stop that decays to a 50 GeV LSP. Separating the signal from background is quite challenging for this topology because the signal peaks at low $\MET$ and is lost amidst the SM backgrounds. Another challenging topology is $\GB^{\tt TT}$. Fig.~\ref{Fig: 4Tvs4B} shows the $\MET$ and $N_j$ distributions for a 900 GeV gluino that decays to a 150 GeV LSP for both the $\GB^{\tt TT}$ and $\GB^{\tt BB}$ topologies.
Because the highest jet-mutliplicity search that we consider is $4^+$ jets, the cuts we propose for 
$\go\to t\bar{t} + \chi^0$ are suboptimal, as more jets would help to reduce background versus signal 
(see e.g.~\cite{Lisanti:2011tm,Aad:2011qa}).

\begin{figure}[!t]
\begin{center}
\includegraphics[width=6.0in]{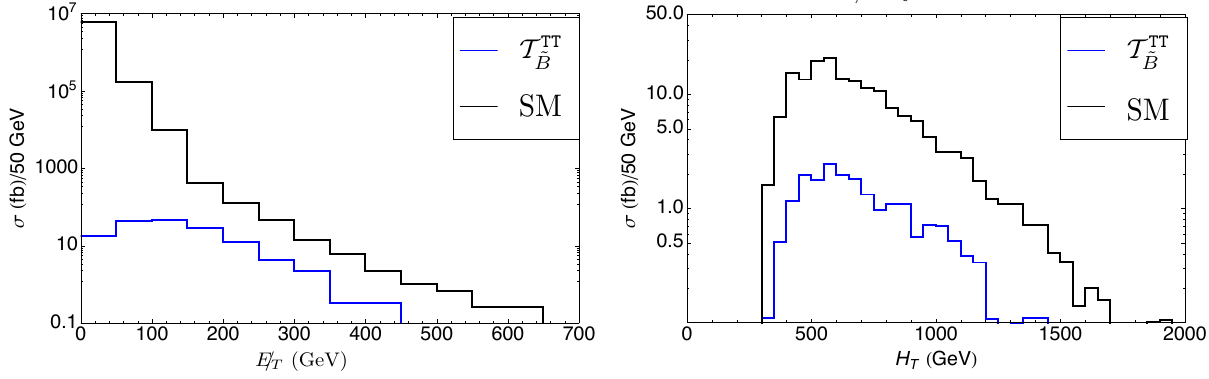}
\caption{\label{Fig: StopvsBckg}   $\MET$ (left panel) and $H_T$ (right panel) distributions for a 300 GeV stop that decays to a 50 GeV LSP from the $\TB$ topology vs SM backgrounds. The $\MET$ distribution is plotted after requiring $N_j\ge4$, $N_b\ge1$ and $H_T\ge400$ GeV. The $H_T$ distribution is shown after requiring $N_j\ge4$, $N_b\ge1$ and $\MET\ge200$ GeV.}
\end{center}
\end{figure}

\begin{figure}[!t]
\begin{center}
\includegraphics[width=6.0in]{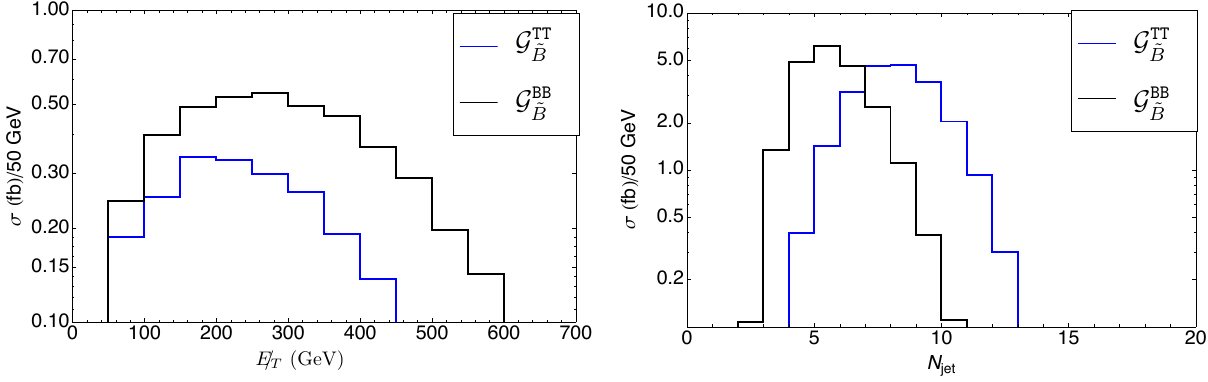}
\caption{\label{Fig: 4Tvs4B}  $\MET$ (left panel) and $N_j$ (right panel) distributions for a 900 GeV gluino that decays to a 150 GeV LSP for the $\GB^{\tt TT}$ and $\GB^{\tt BB}$ topologies. The $\MET$ distribution is plotted after requiring $N_j\ge4$ and $N_b\ge3$.}
\end{center}
\end{figure}

One lesson that emerged from this study is that the character of the optimal search strategies will change significantly as 
larger amounts of data are analyzed.  In particular, while searches involving $3b$ jets and same-sign dileptons may not be particularly useful with less than $1$ fb$^{-1}$ of data, it seems that with more data these channels will be crucial for obtaining optimal sensitivity to heavy flavor simplified models involving color octets.  This conclusion appears to be robust:  even if the systematic uncertainties on the $3b$ backgrounds are taken to be as large as $100 \%$,  the $3b$ searches will still be important for fully utilizing the LHC data.  
We look forward to an exciting year as the LHC completes its $7$ TeV run, taking an order of magnitude more data.

\acknowledgments
We thank B.~Butler, A.~Haas, P.~Hansson, P.~Schuster, A.~Schwartzman, and N.~Toro for 
many useful discussions.  
We acknowledge support from the US DOE under contract no.~DE-AC02-76SF00515. EI is supported by 
an LHCTI graduate fellowship under grant NSF-PHY-0969510.  RE also acknowledges support by the National Science Foundation under Grant No.~PHY-0969739 and PHY05-51164. JGW is partially supported by the DOE's Outstanding Junior Investigator Award and the Sloan Fellowship.
At the completion of this work, we discovered several similar but distinct studies on heavy flavor and missing energy \cite{KMRS,BKLS, PRW}.

\appendix


\section{Expected limits from existing LHC searches: Plots}
\label{Sec: CurrentLimits Plots}

In Fig.~\ref{Fig:Limits}, we show the \emph{expected} $95 \%$ C.L. limits for 
$\LL=1$ fb$^{-1}$ from the ATLAS studies in Sec.~\ref{Sec: CurrentLimits} 
on all simplified models from \refsec{SimplifiedModels} for different choices of the production cross section.  

\begin{figure}[!h]
\begin{center}
\includegraphics[width=0.32\textwidth]{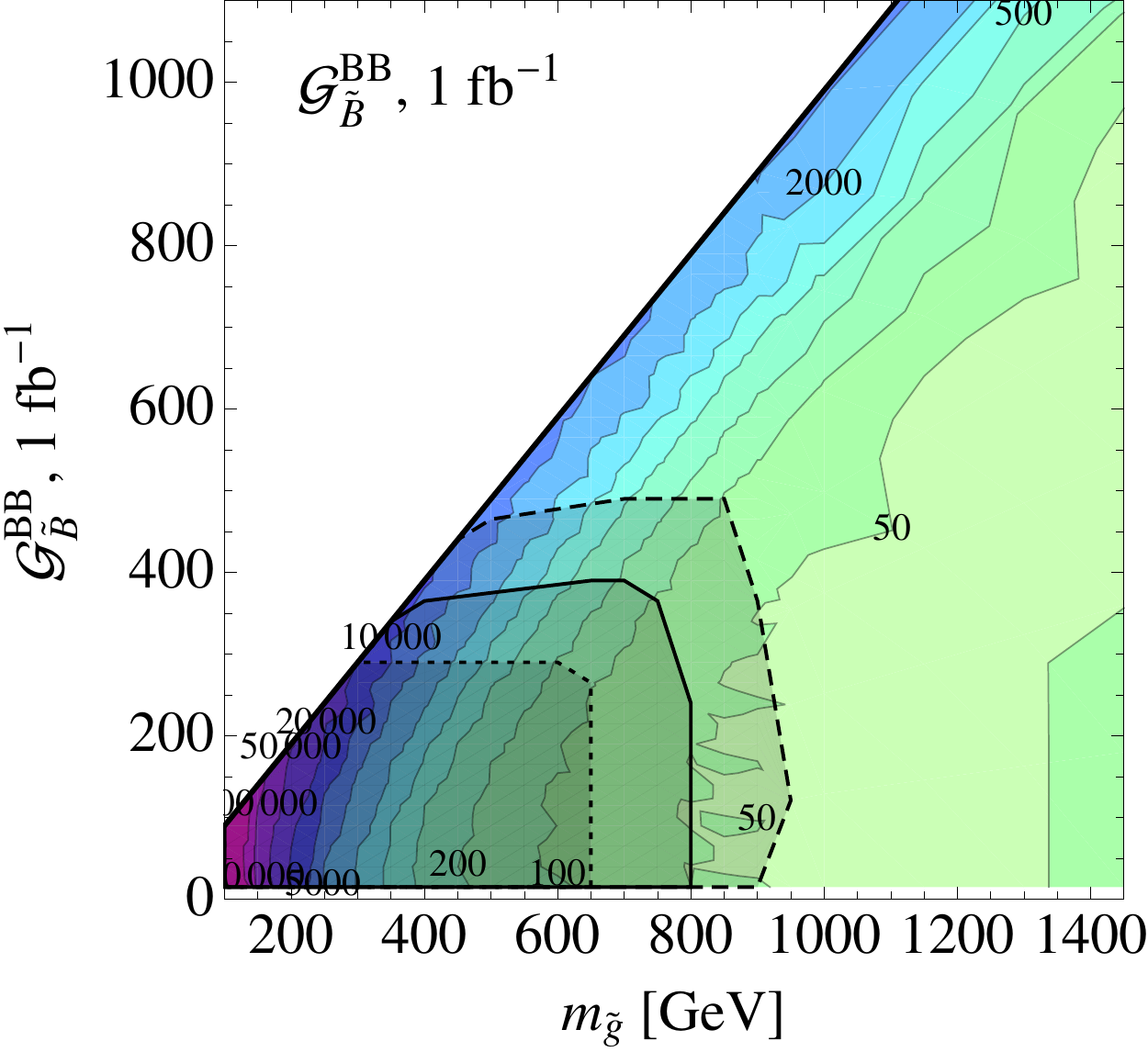}
\includegraphics[width=0.32\textwidth]{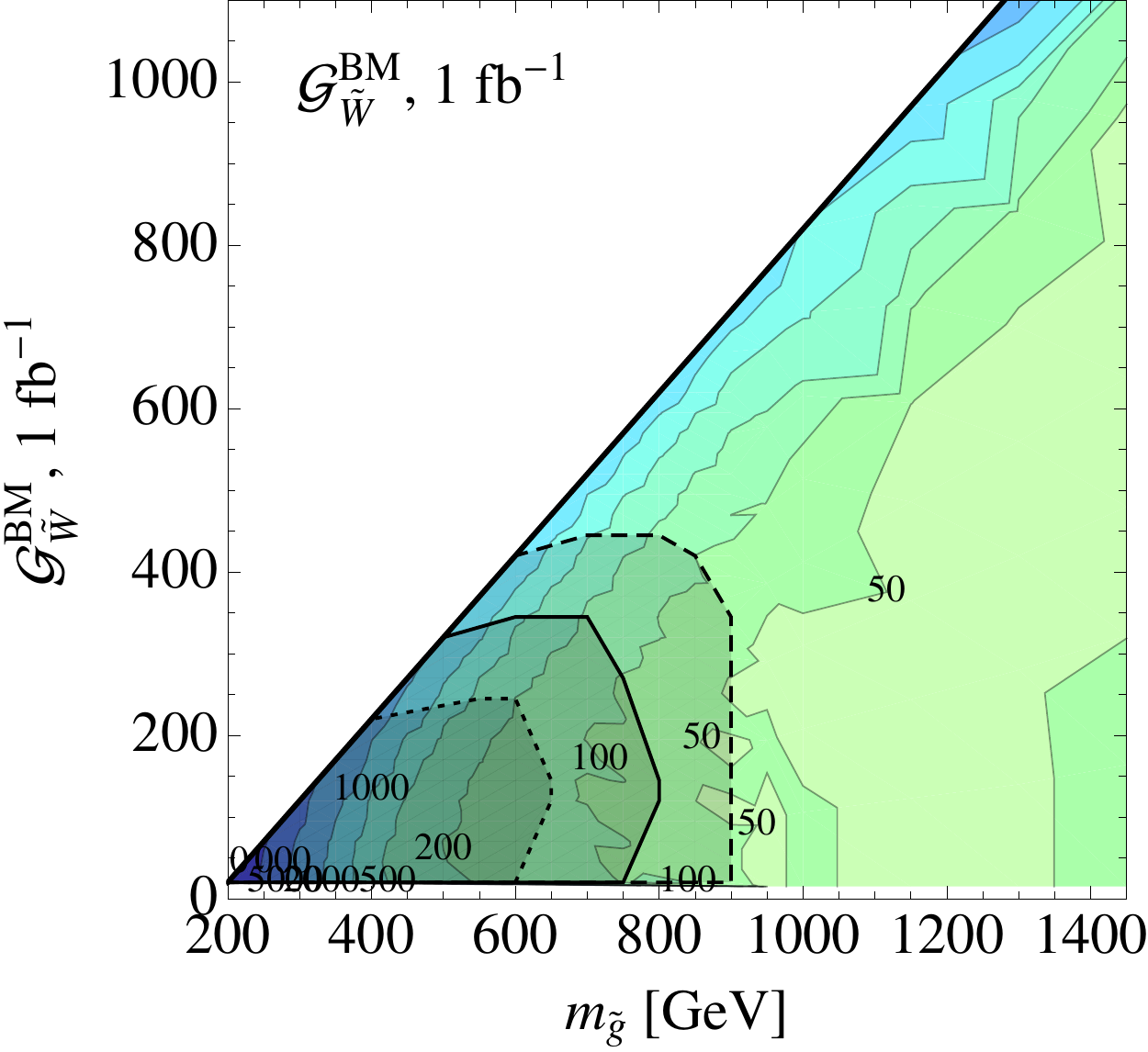}
\includegraphics[width=0.32\textwidth]{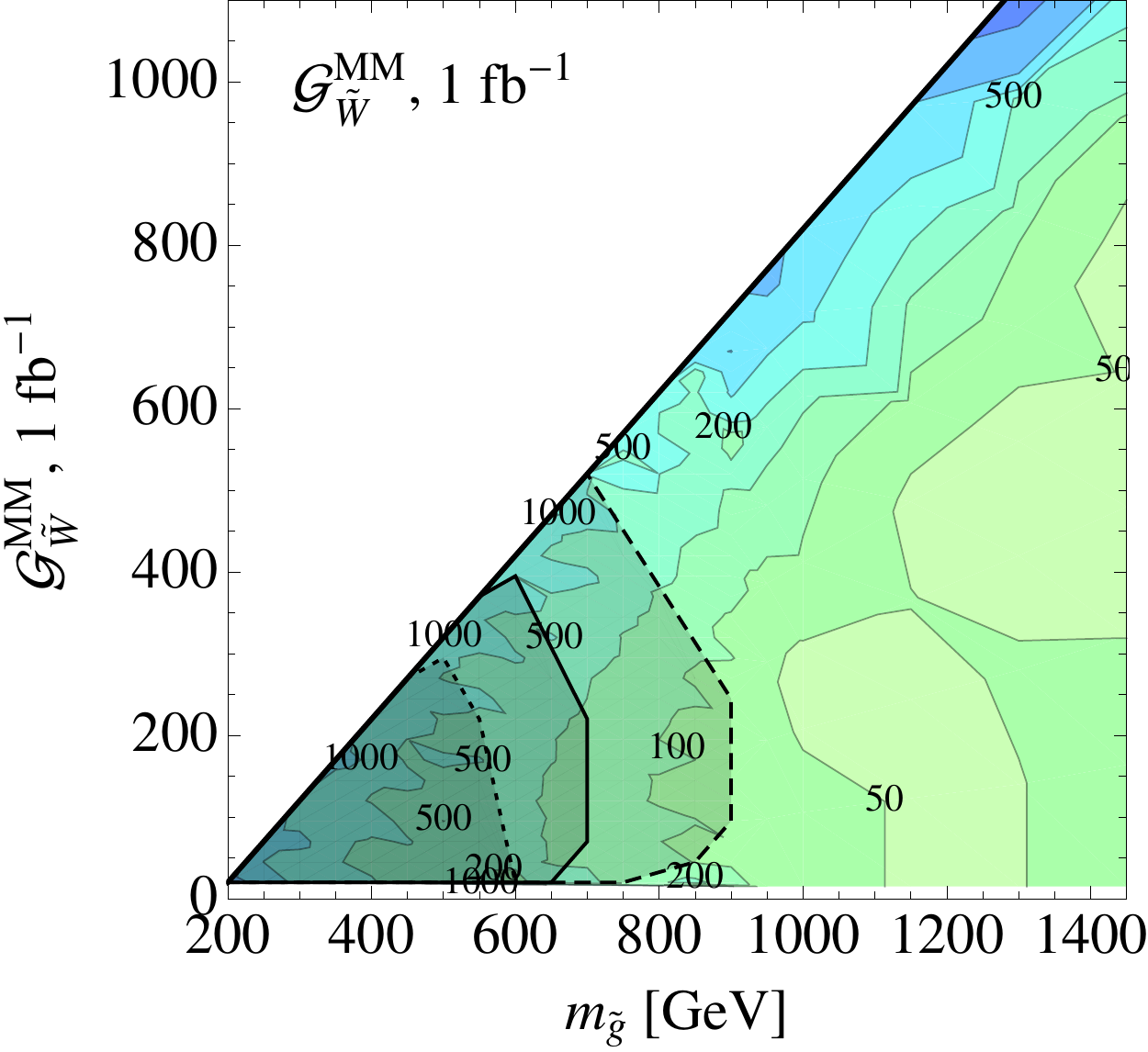}\\
\includegraphics[width=0.32\textwidth]{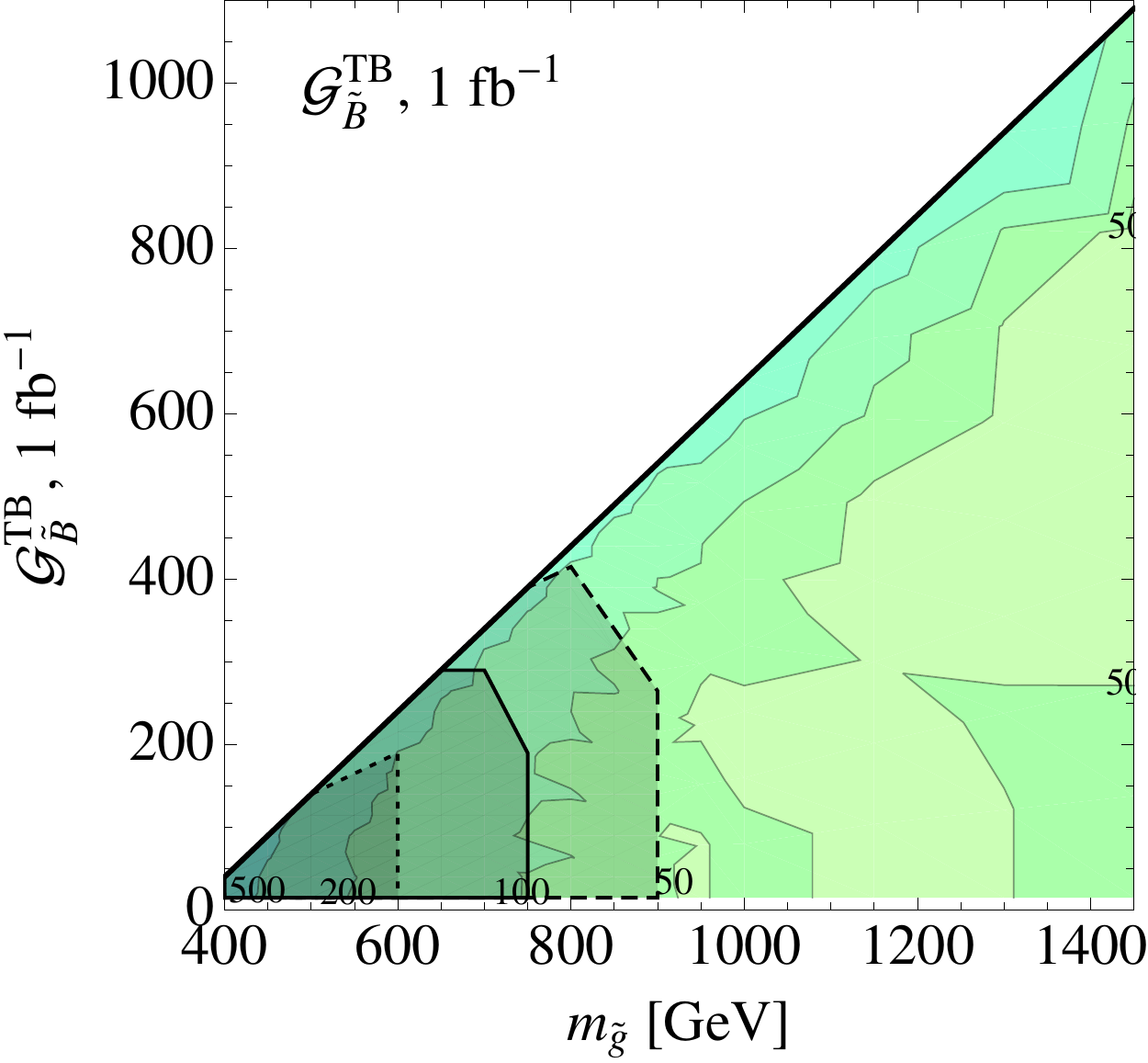}
\includegraphics[width=0.32\textwidth]{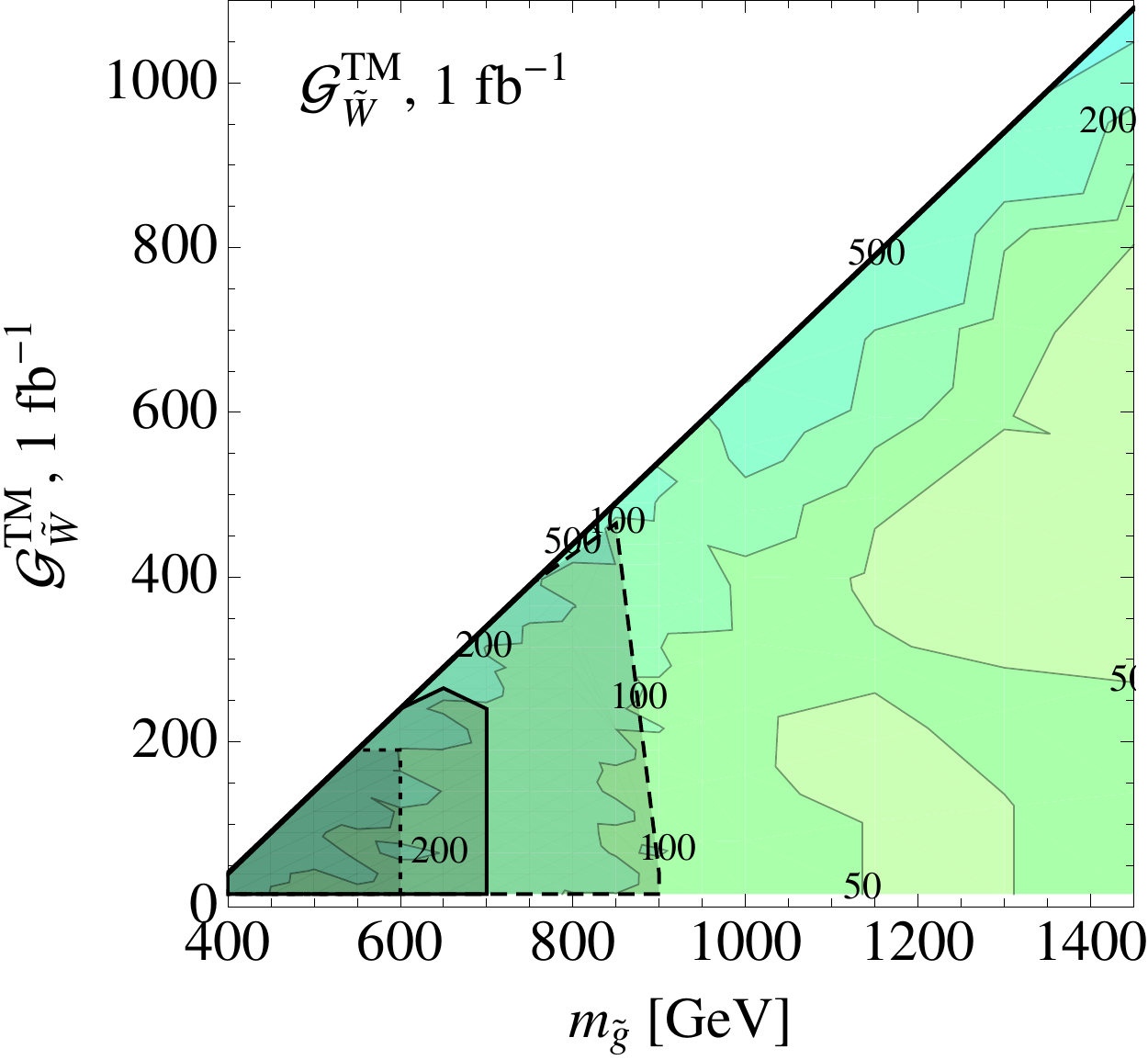}
\includegraphics[width=0.32\textwidth]{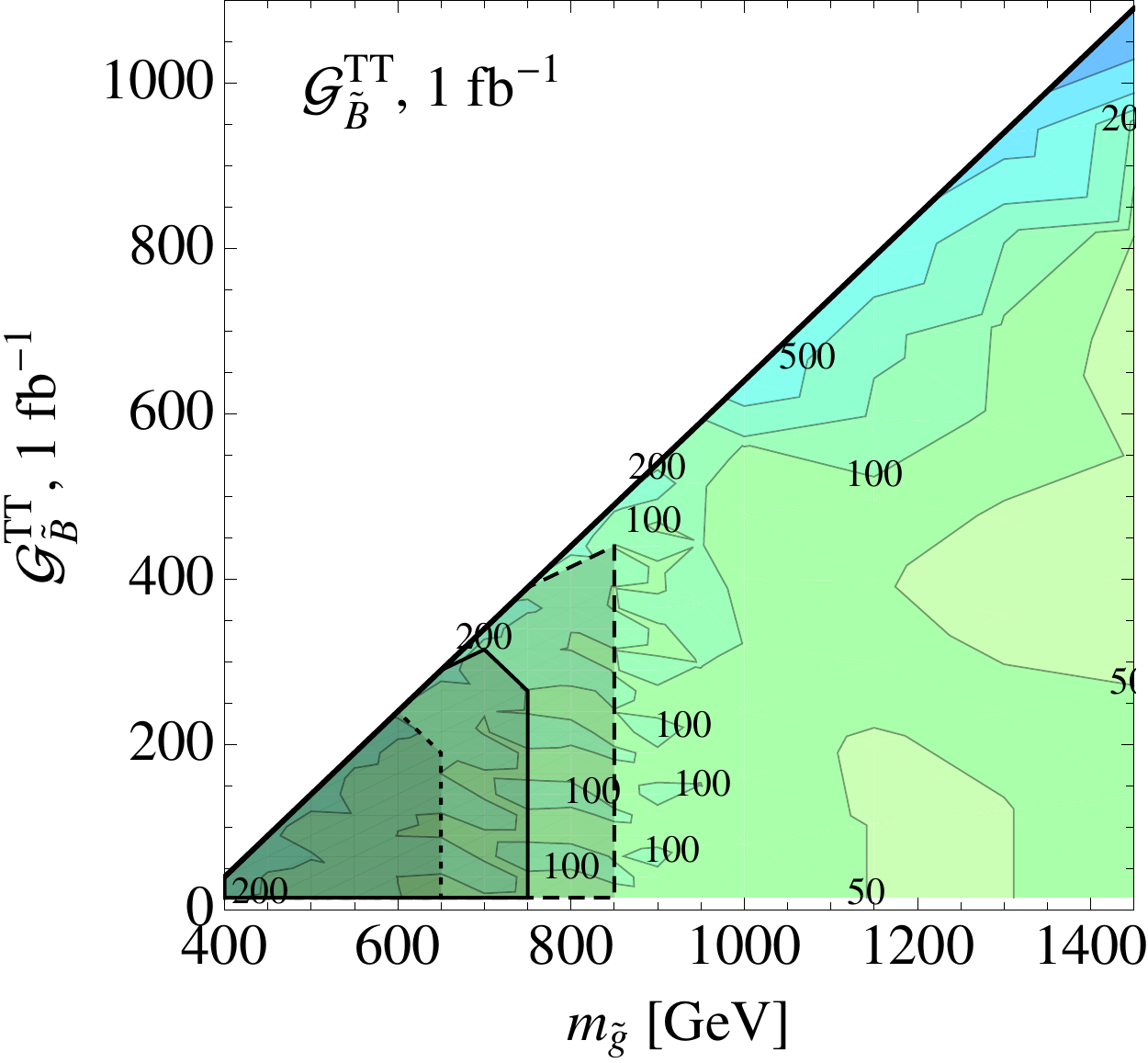}\\
\includegraphics[width=0.32\textwidth]{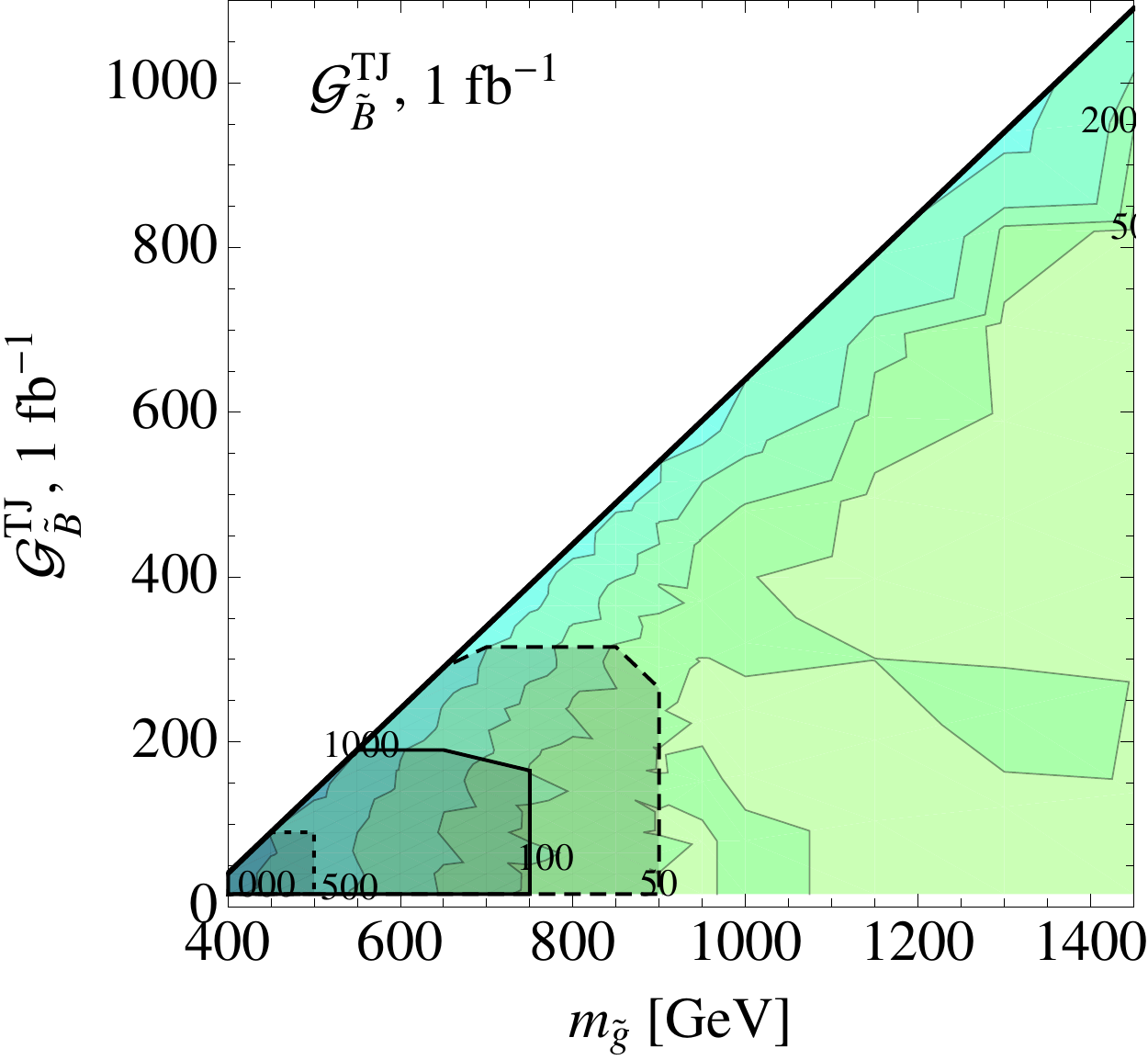}
\includegraphics[width=0.32\textwidth]{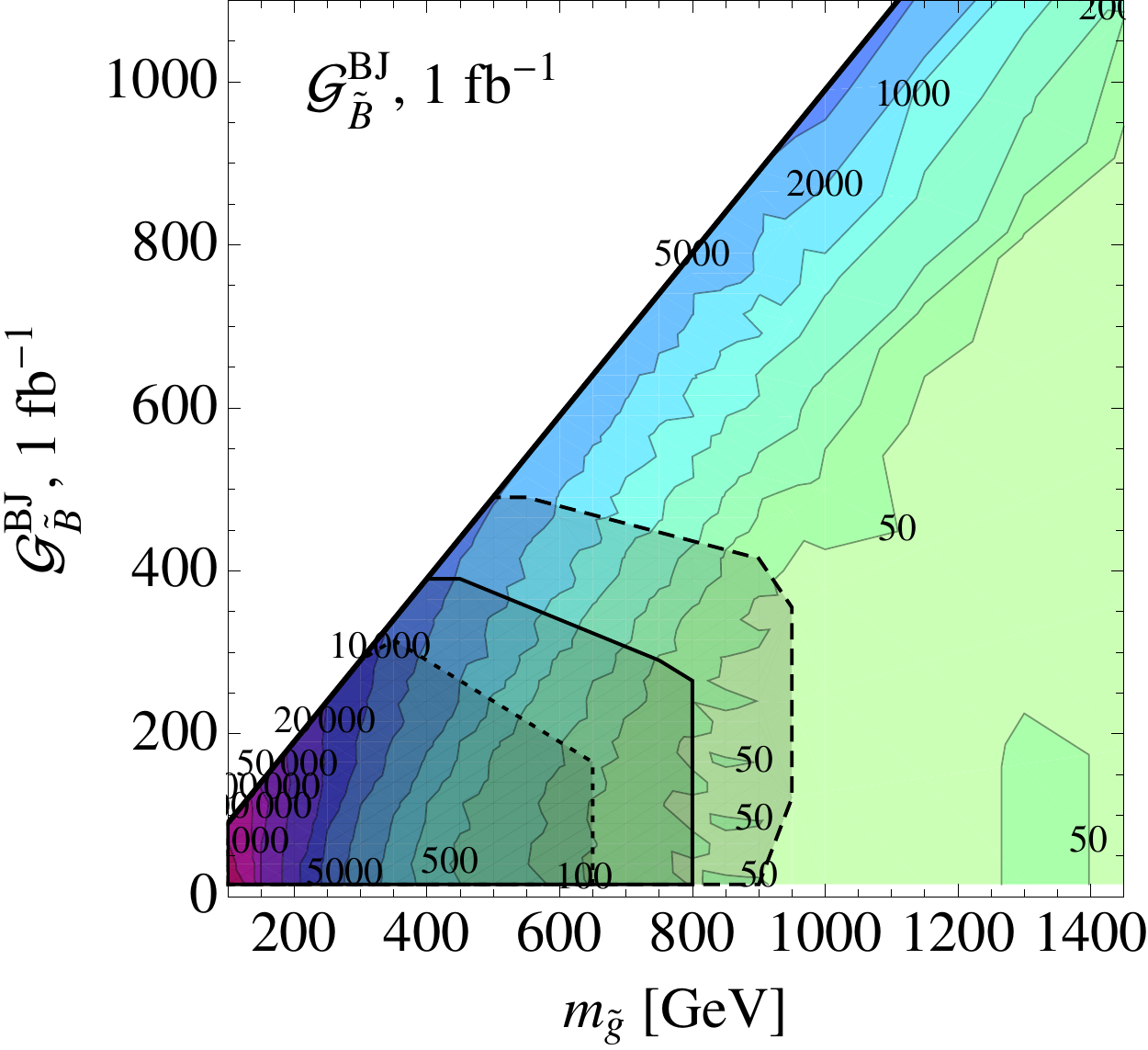}
\includegraphics[width=0.32\textwidth]{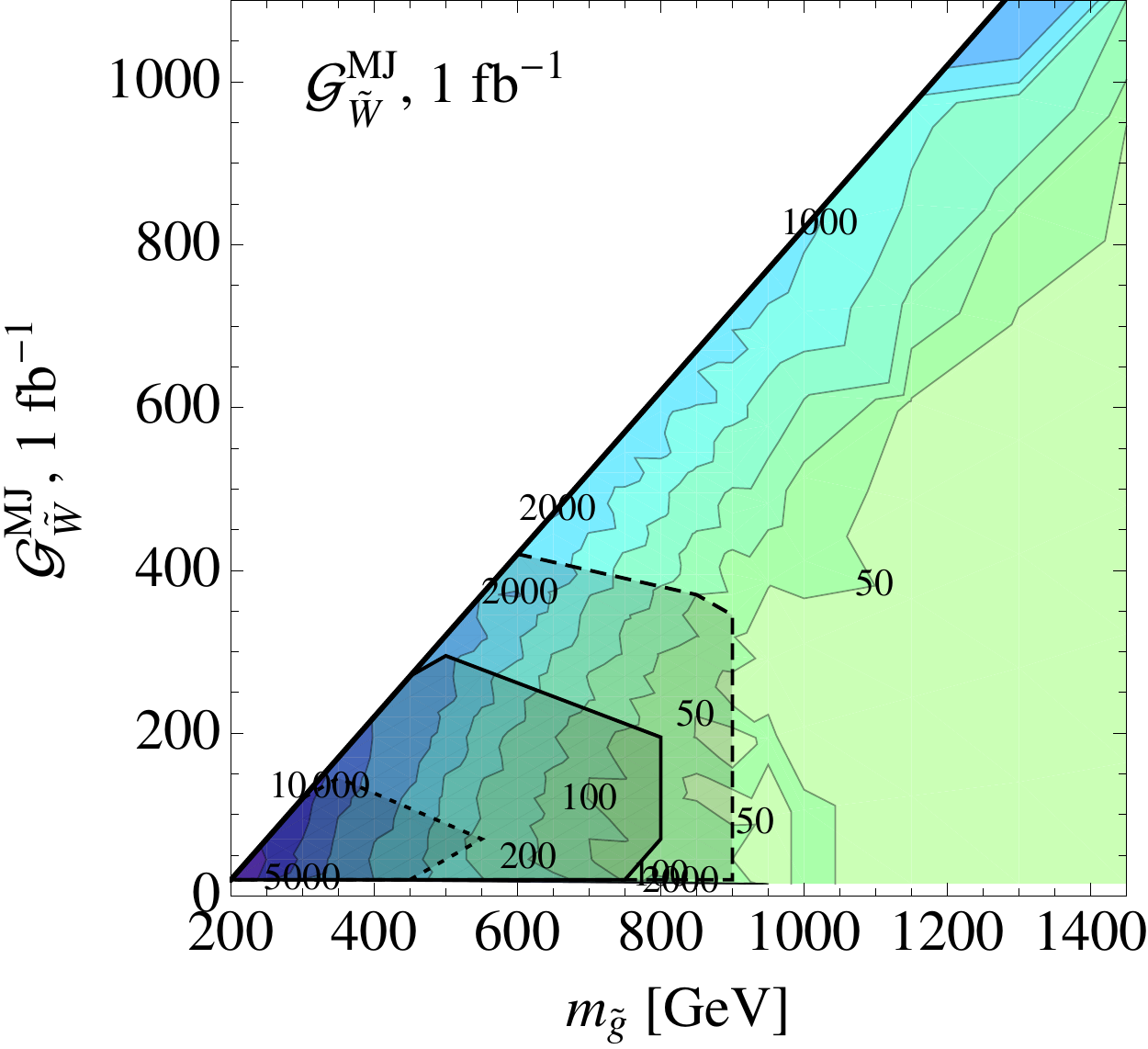}\\
\includegraphics[width=0.32\textwidth]{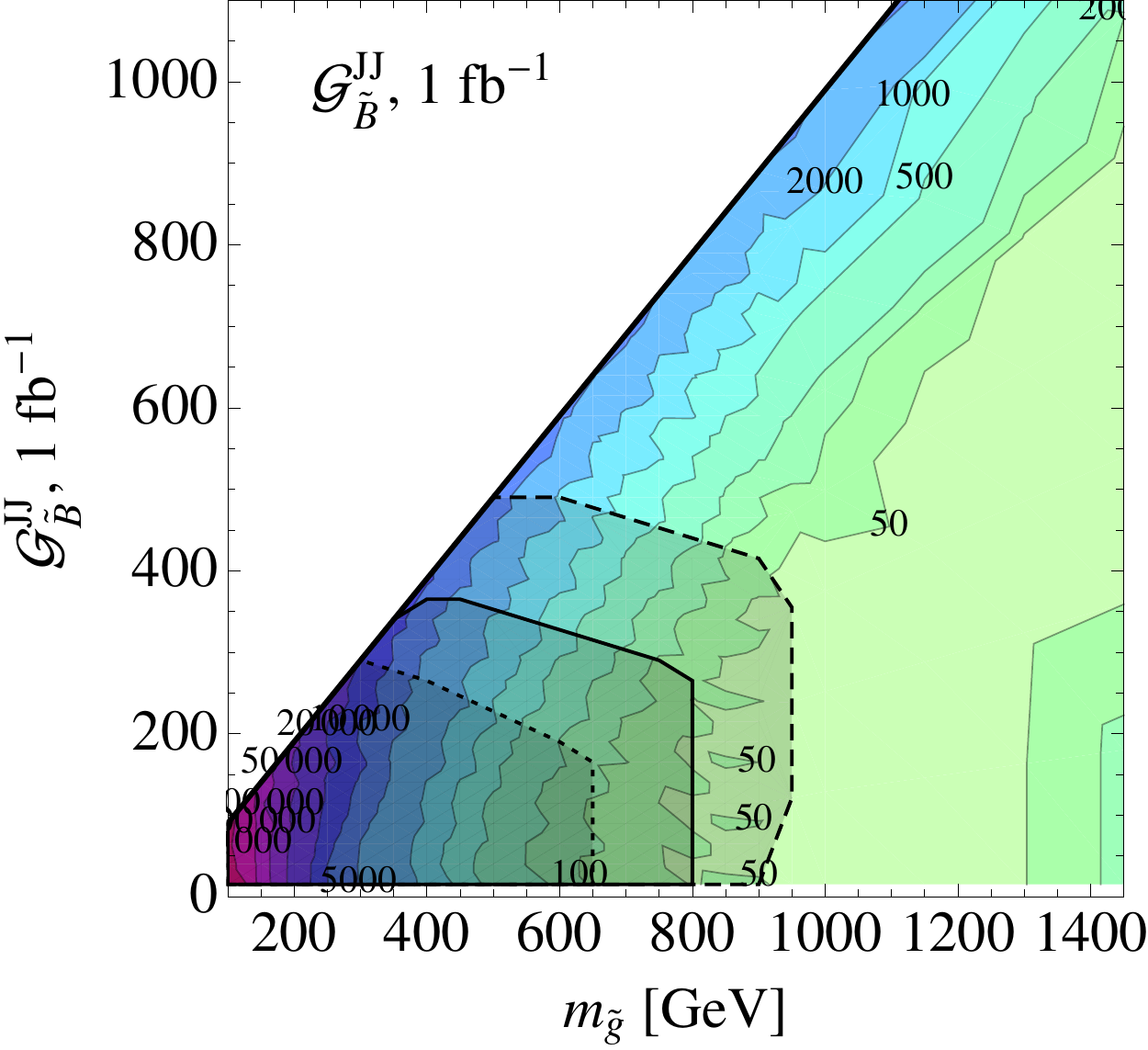}
\includegraphics[width=0.32\textwidth]{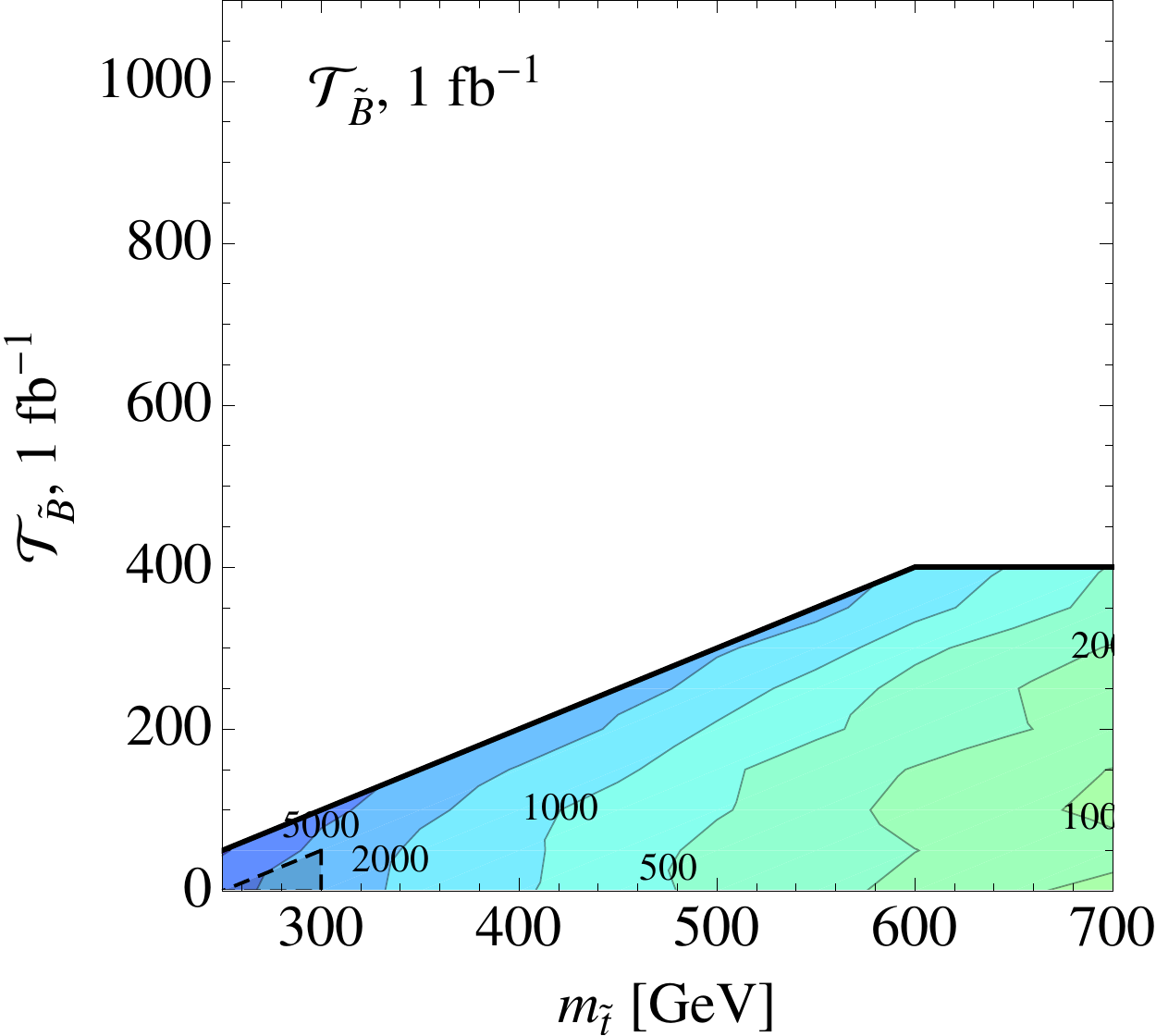}
\includegraphics[width=0.32\textwidth]{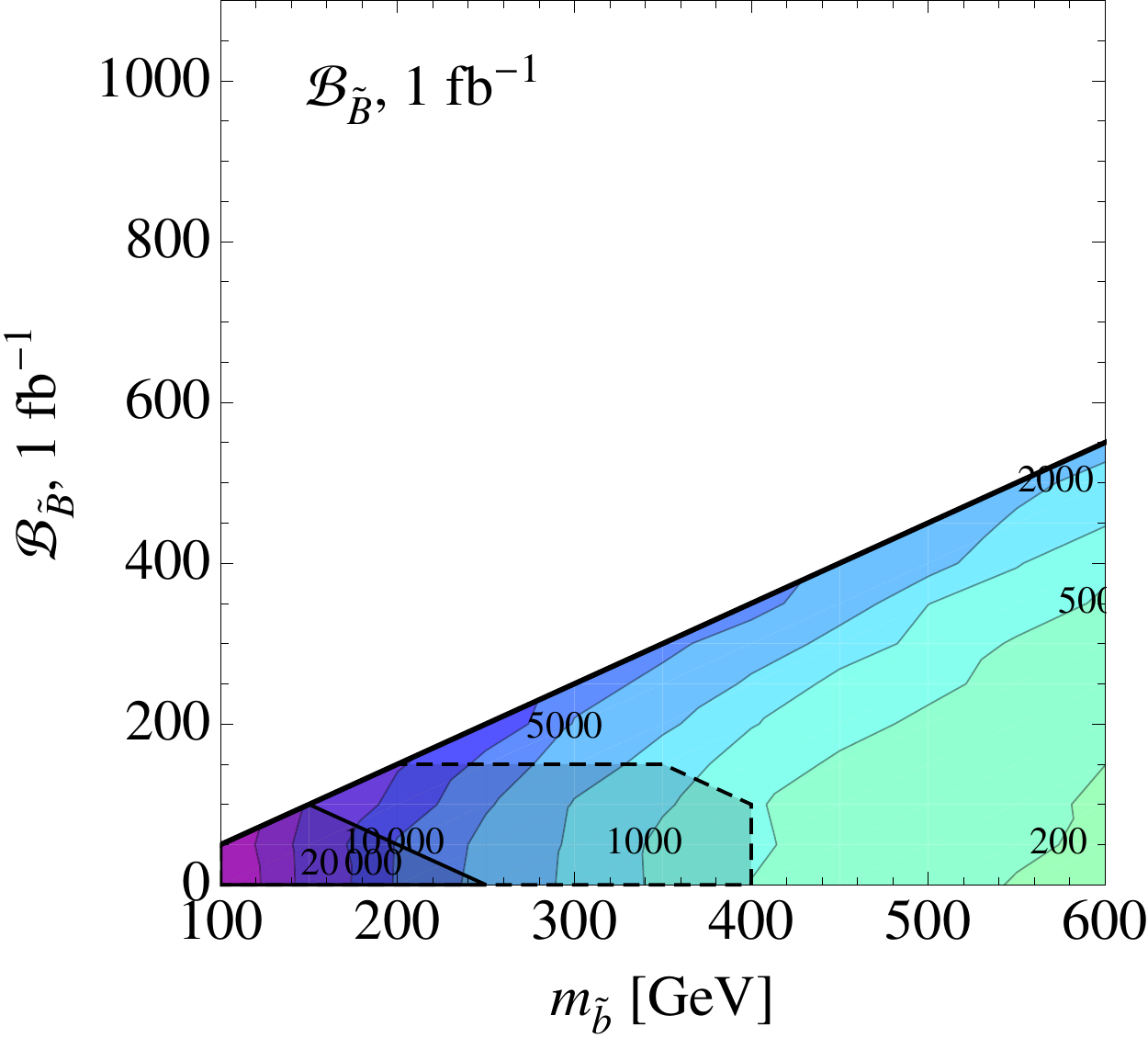}
\caption{\label{Fig:Limits} Estimated 95\% C.L.~contours for $\LL=1\ifb$ for the cross section times branching 
ratio sensitivity for various simplified models using the ATLAS jets and 
$\MET$ searches in Sec.~\ref{Sec: CurrentLimits}. 
The solid, dashed, and dotted lines give the limit for $\sigma_{pp\rightarrow XX}= 1, 3,$ and $0.3$ times $\sigma_{pp\rightarrow XX}^{\text{NLO QCD}}$, where $X=\go$ for all plots except for the middle plot in the bottom row ($X=\T$) 
and for the right plot in the bottom row ($X=\B$).}
\end{center}
\end{figure}

\section{Expected limits from optimized searches: Plots}
\label{Sec: OptLimits Plots}

In Figs.~\ref{Fig:OptLimits1}, \ref{Fig:OptLimits2}, and \ref{Fig:OptLimits3}, we show the \emph{expected} $95 \%$ C.L. 
limits cross section times branching ratio sensitivity for the simplified models in \refsec{SimplifiedModels} 
from the search regions in Table \ref{tab: SearchRegions BM} that have been optimized on the benchmarks 
in Appendix \ref{Sec: BechmarkTables}, for $\LL= 1\ifb$, $5\ifb$, and $15\ifb$.  

\begin{figure}[!t]
\begin{center}
\includegraphics[width=0.32\textwidth]{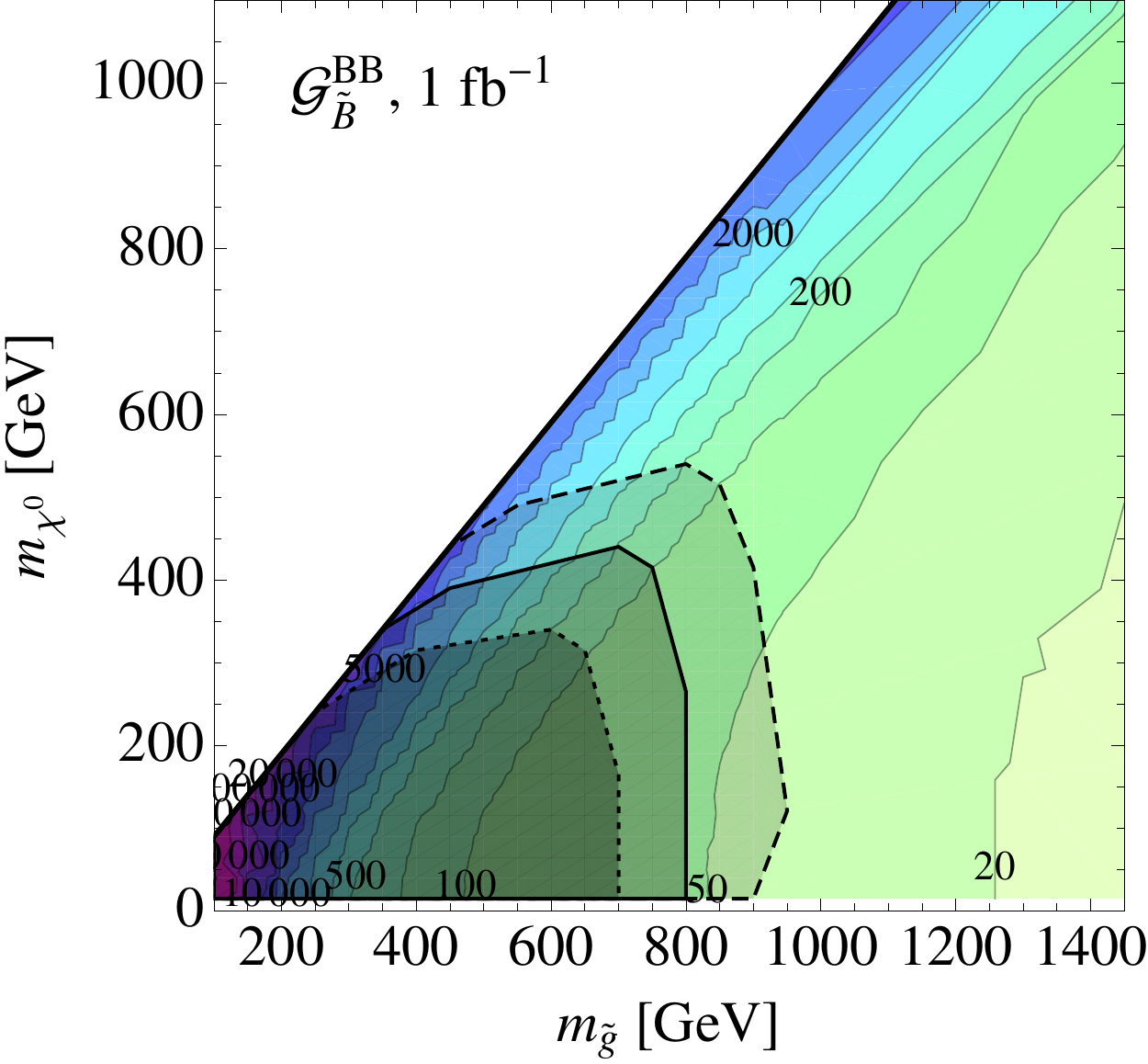}
\includegraphics[width=0.32\textwidth]{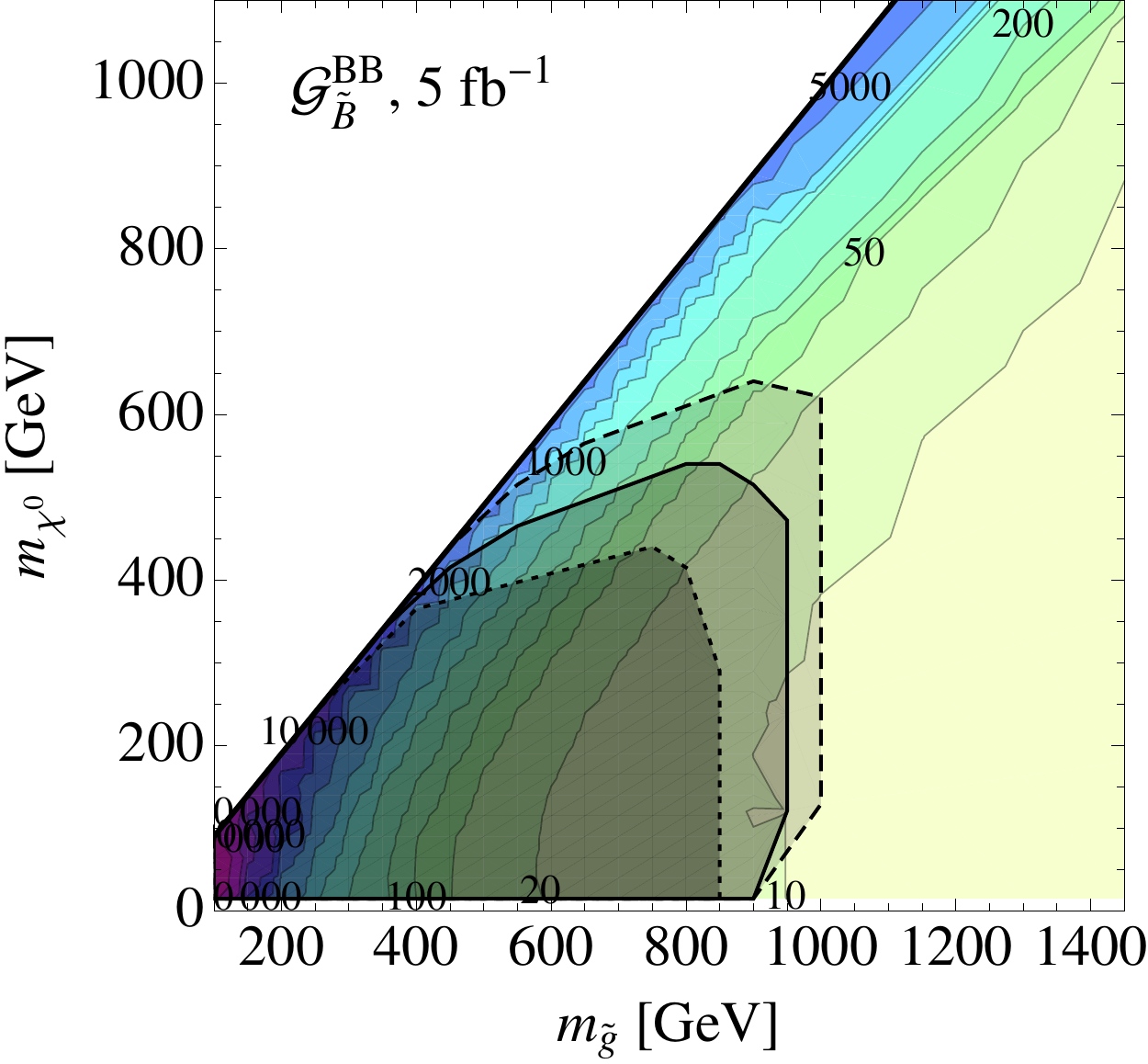}
\includegraphics[width=0.32\textwidth]{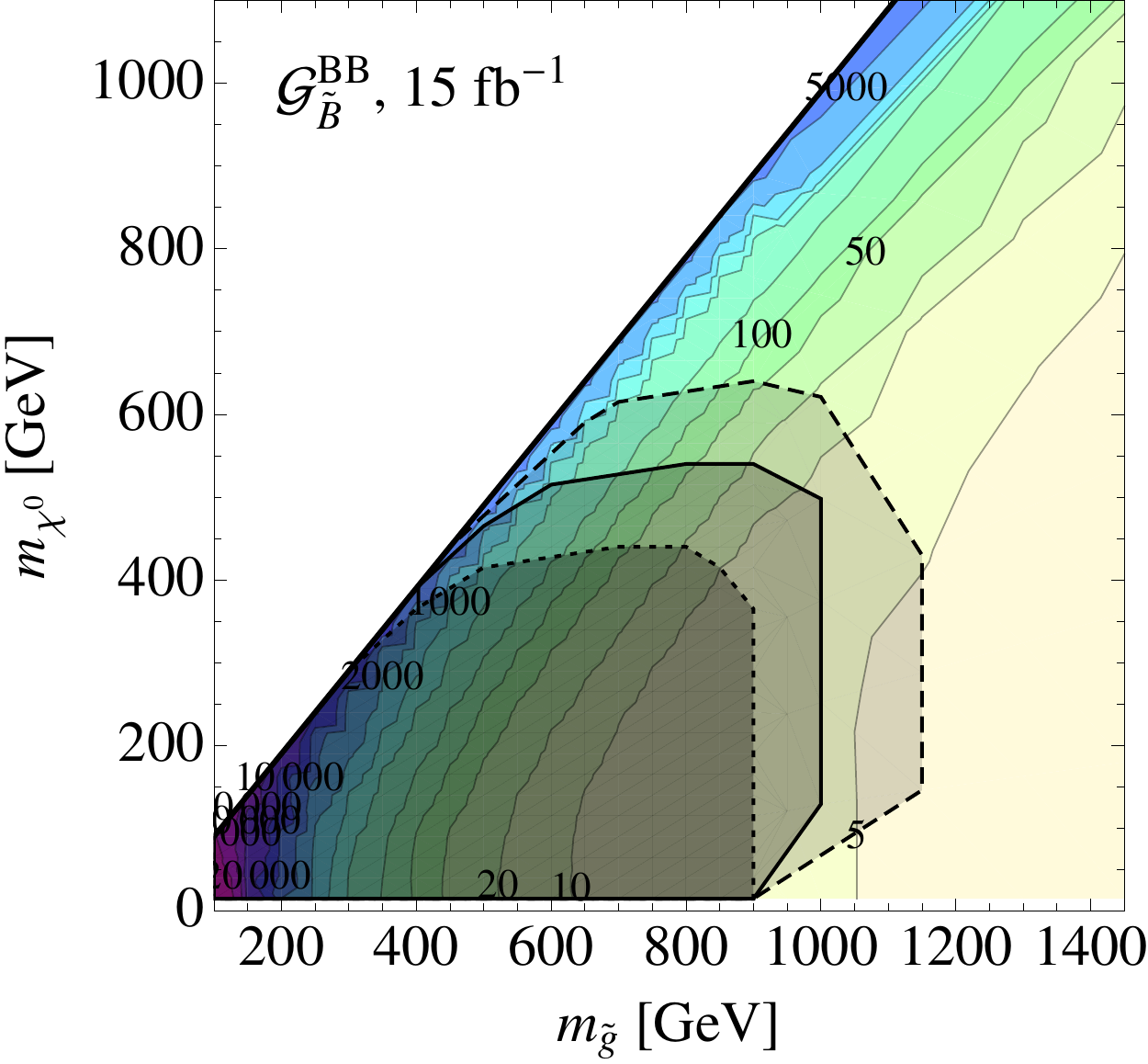}\\
\includegraphics[width=0.32\textwidth]{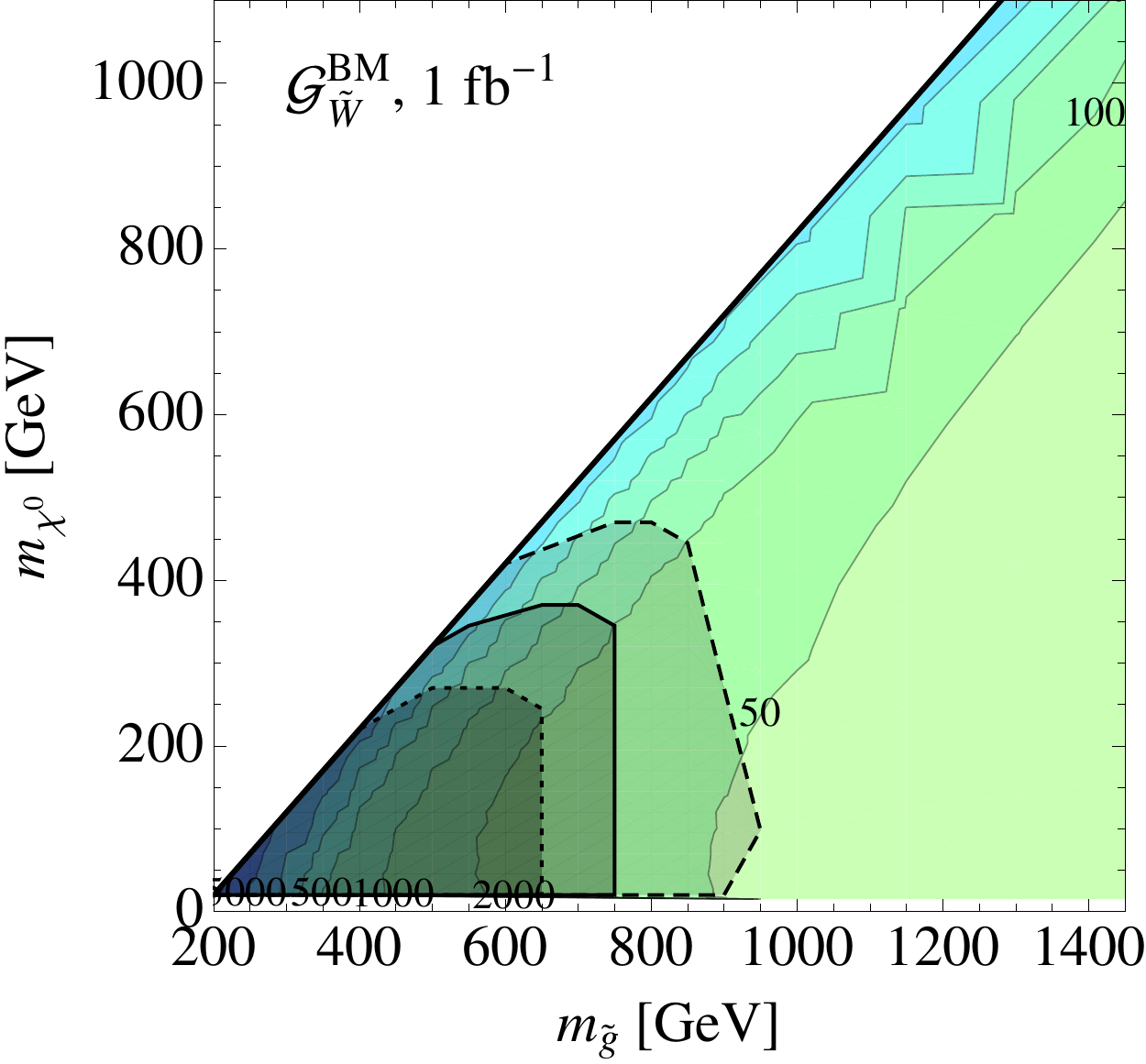}
\includegraphics[width=0.32\textwidth]{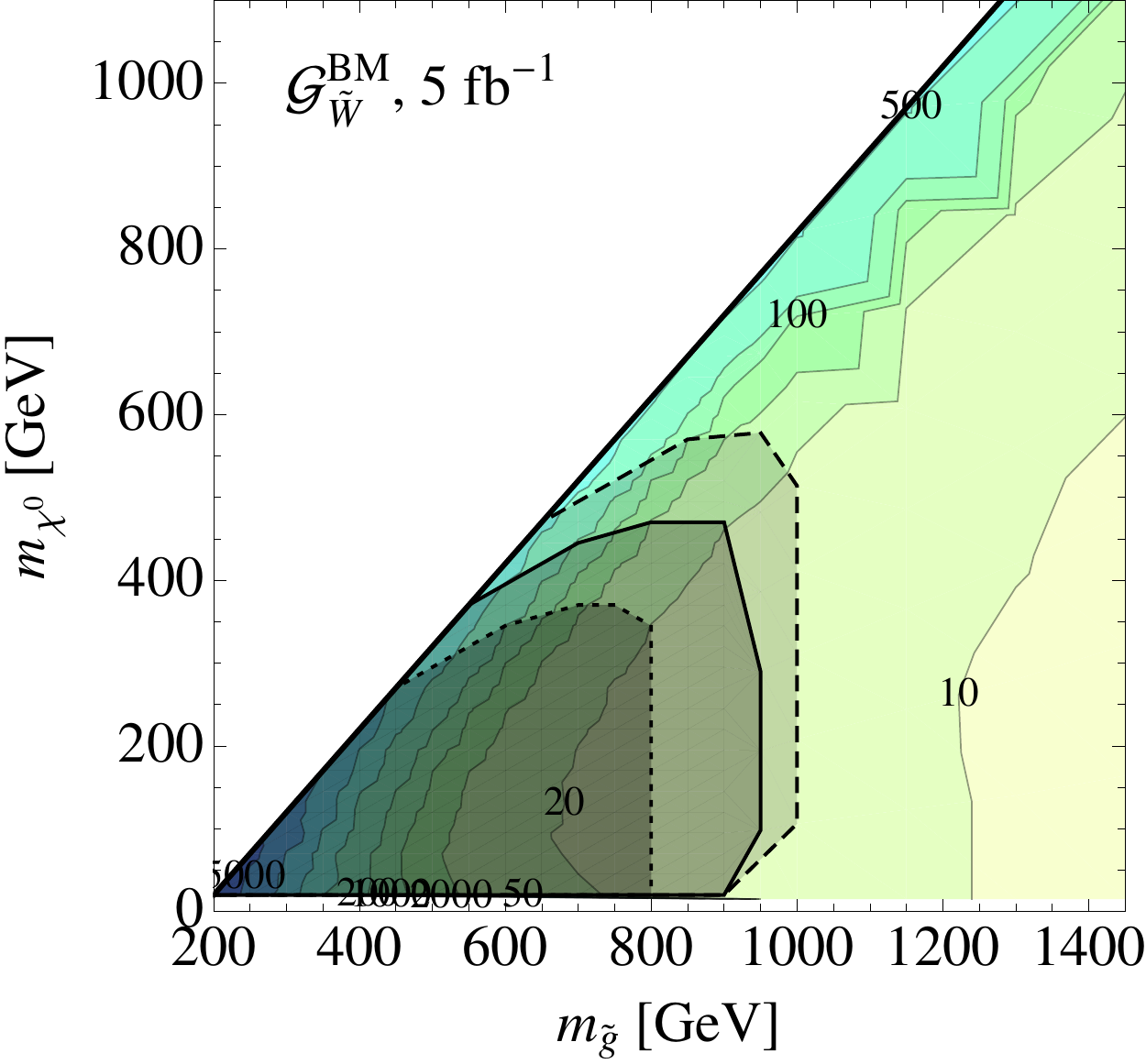}
\includegraphics[width=0.32\textwidth]{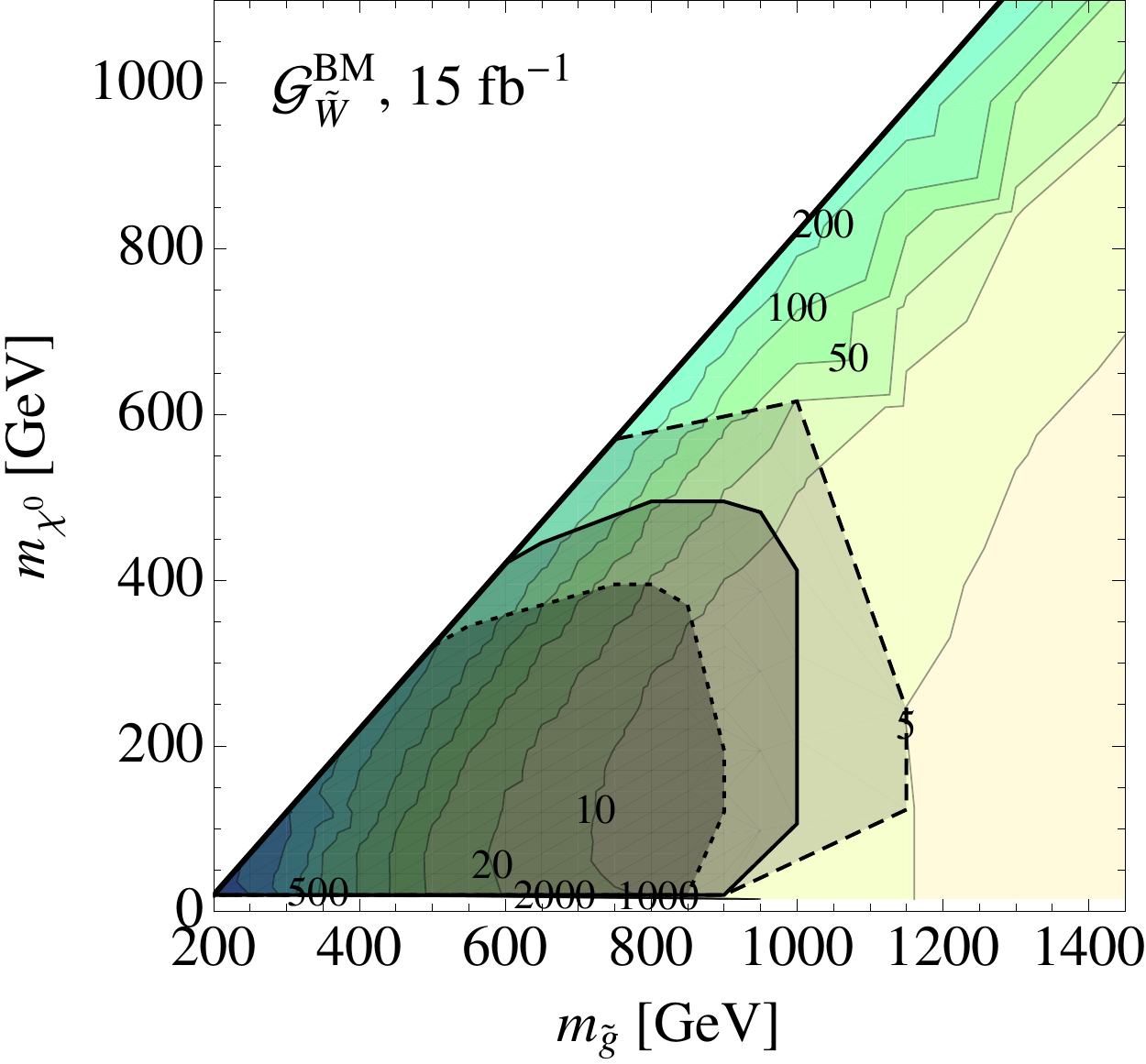}\\
\includegraphics[width=0.32\textwidth]{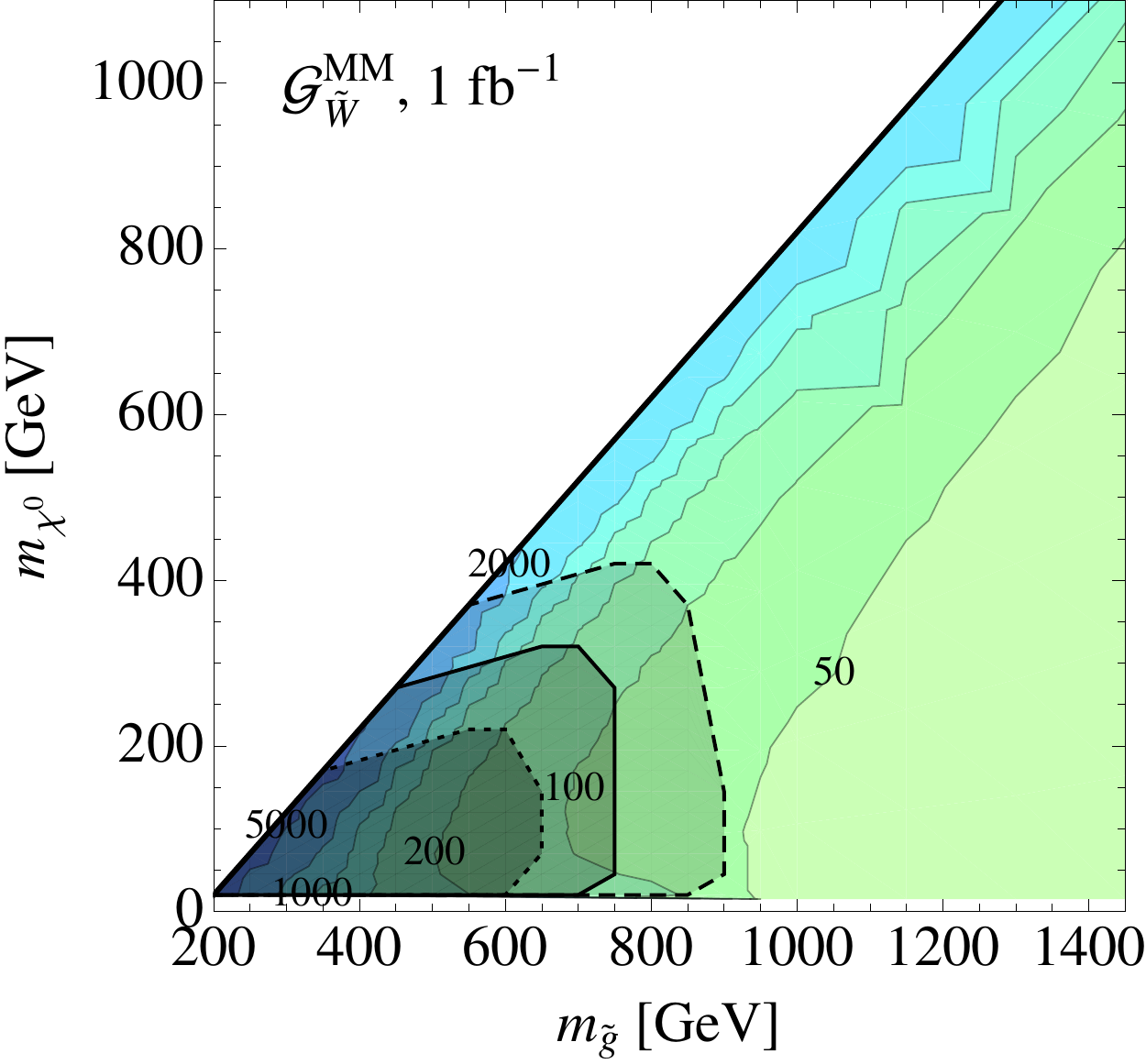}
\includegraphics[width=0.32\textwidth]{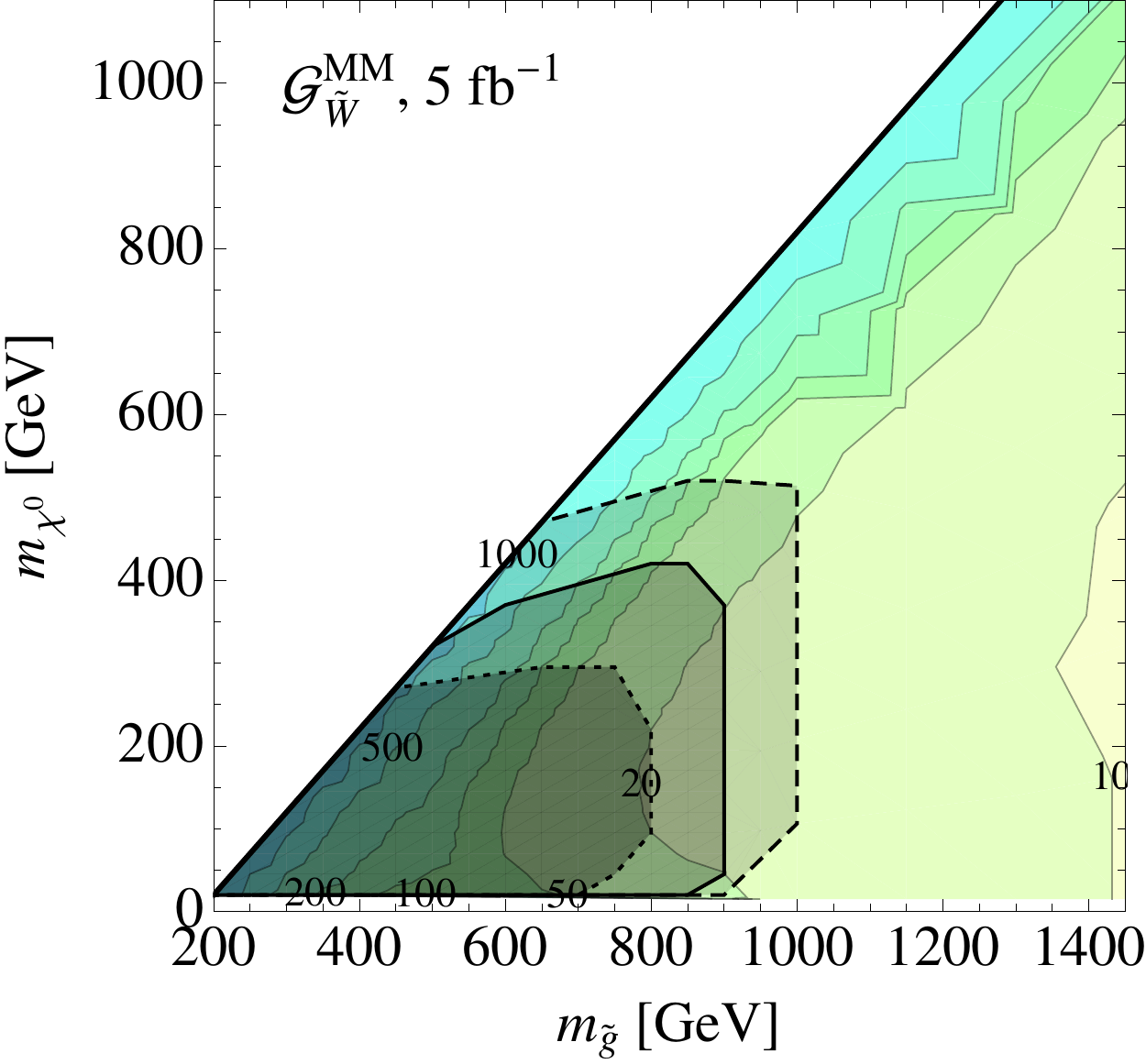}
\includegraphics[width=0.32\textwidth]{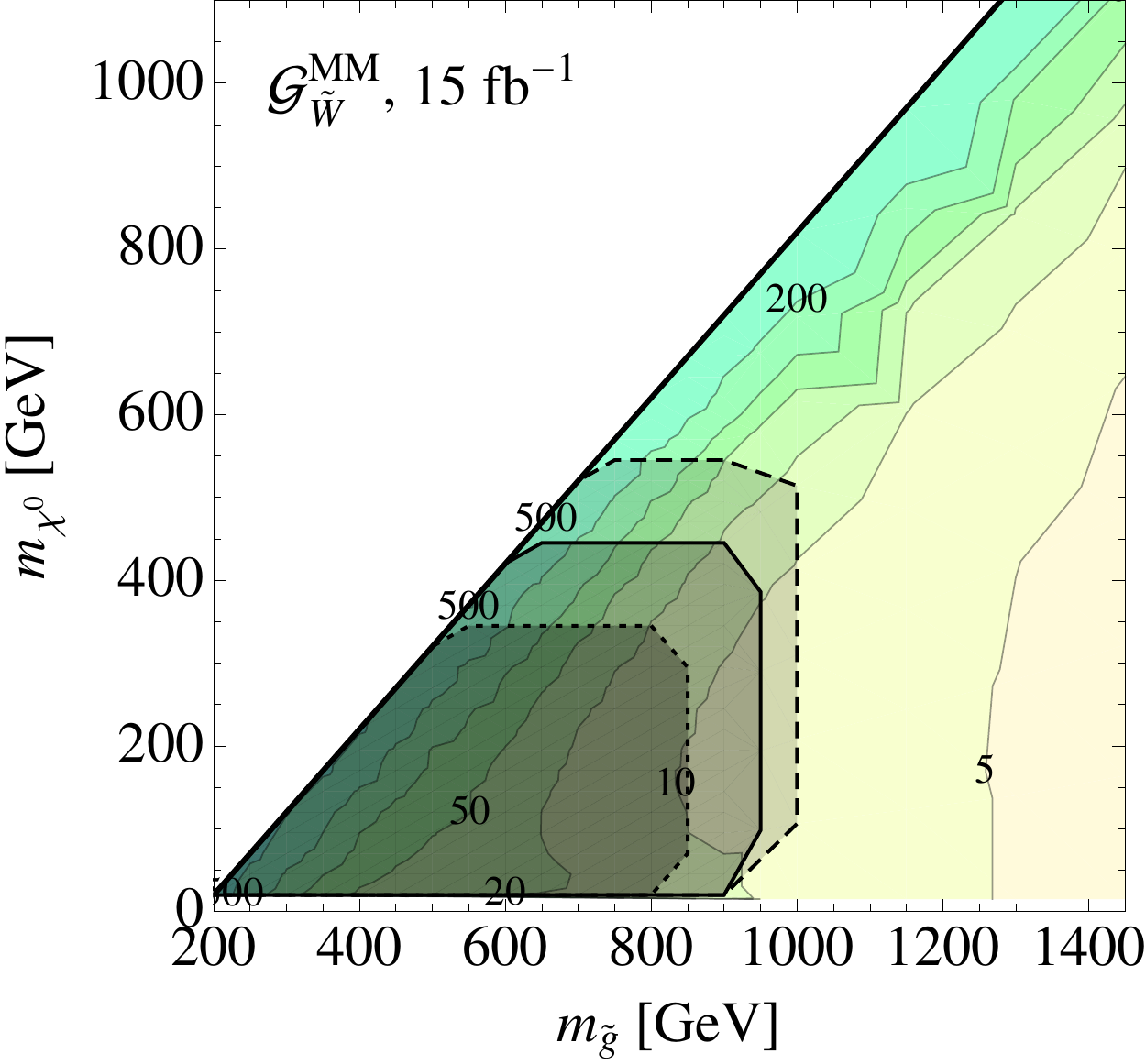}\\
\includegraphics[width=0.32\textwidth]{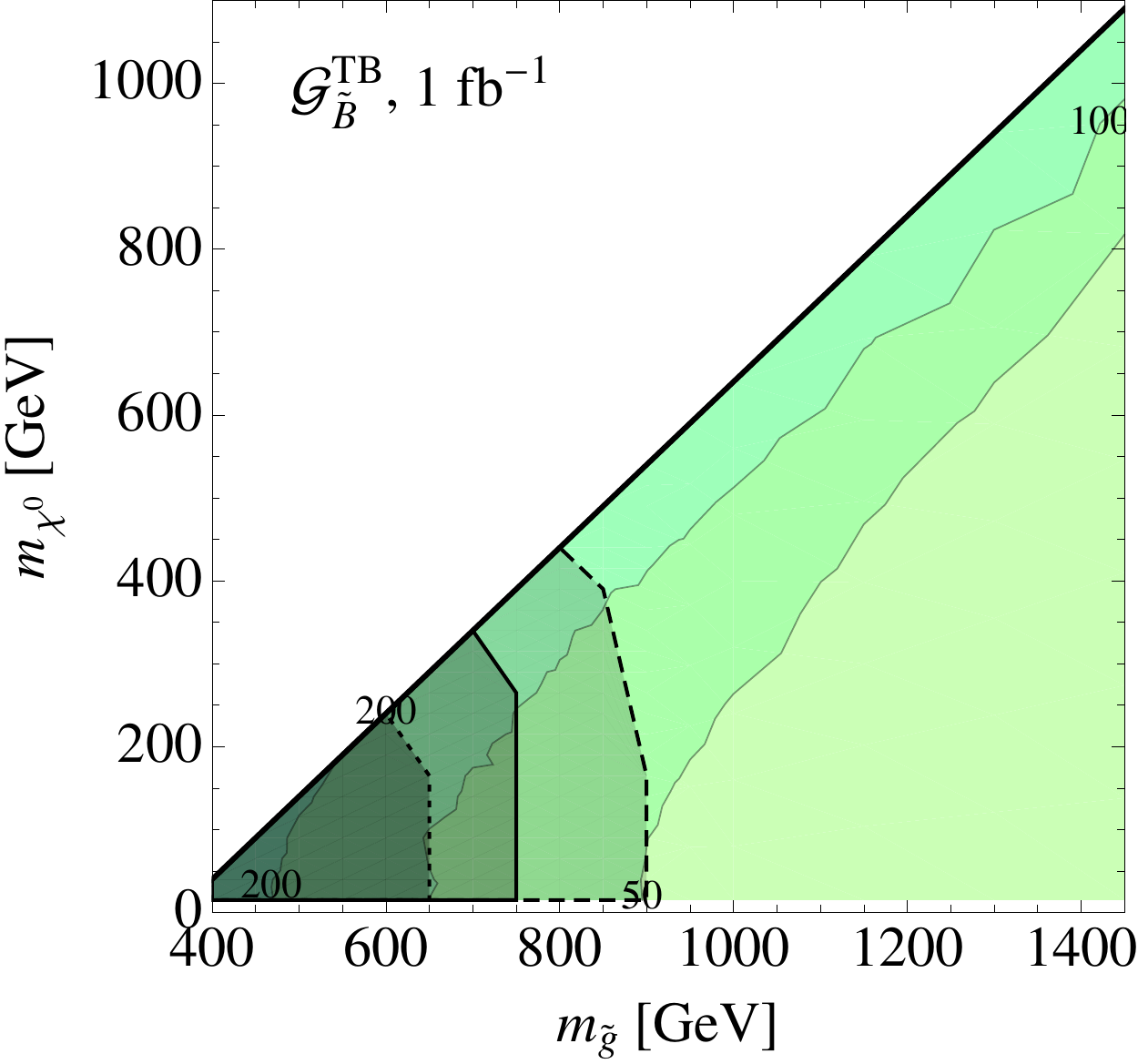}
\includegraphics[width=0.32\textwidth]{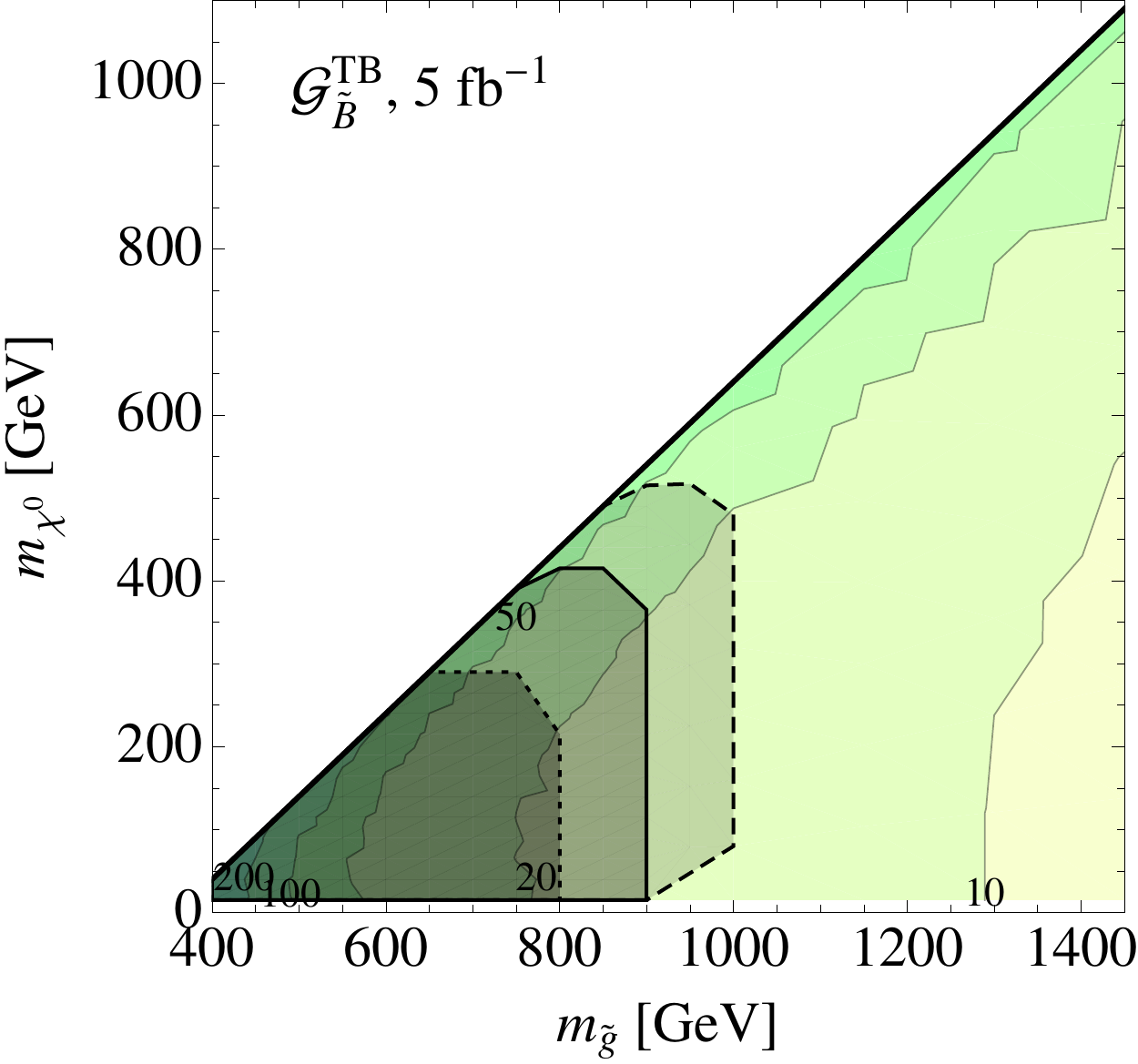}
\includegraphics[width=0.32\textwidth]{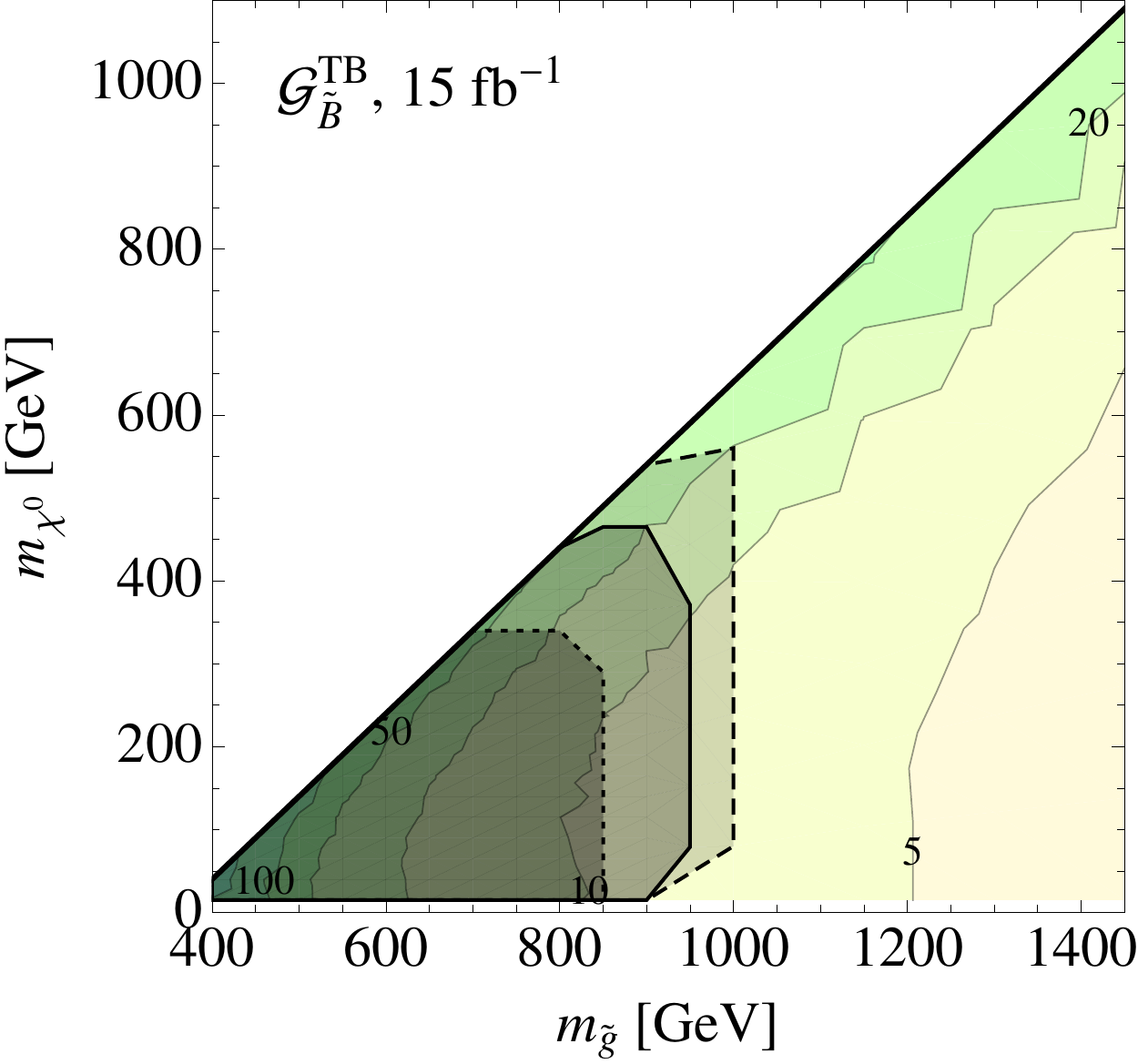}
\caption{\label{Fig:OptLimits1} Estimated 95\% C.L.~contours for the cross section times branching 
ratio sensitivity for various simplified models using the search regions in Table \ref{tab: SearchRegions BM} 
that have been optimized on the benchmarks in Appendix \ref{Sec: BechmarkTables}.  
Shown are $\LL= 1\ifb$ (left column), $5\ifb$ (middle column), 
and $15\ifb$ (right column).  
The solid, dashed, and dotted lines correspond to $\sigma_{pp\rightarrow\go\go}=$1, 3, and 0.3 times $\sigma_{pp\rightarrow \go\go}^{\text{NLO QCD}}$. 
Each row is for a different simplified model; from top to bottom, these are $\GB^{\tt BB}$, $\GW^{\tt BM}$, $\GW^{\tt MM}$, 
and $\GB^{\tt TB}$.}
\end{center}
\end{figure}

\begin{figure}[!t]
\begin{center}
\includegraphics[width=0.32\textwidth]{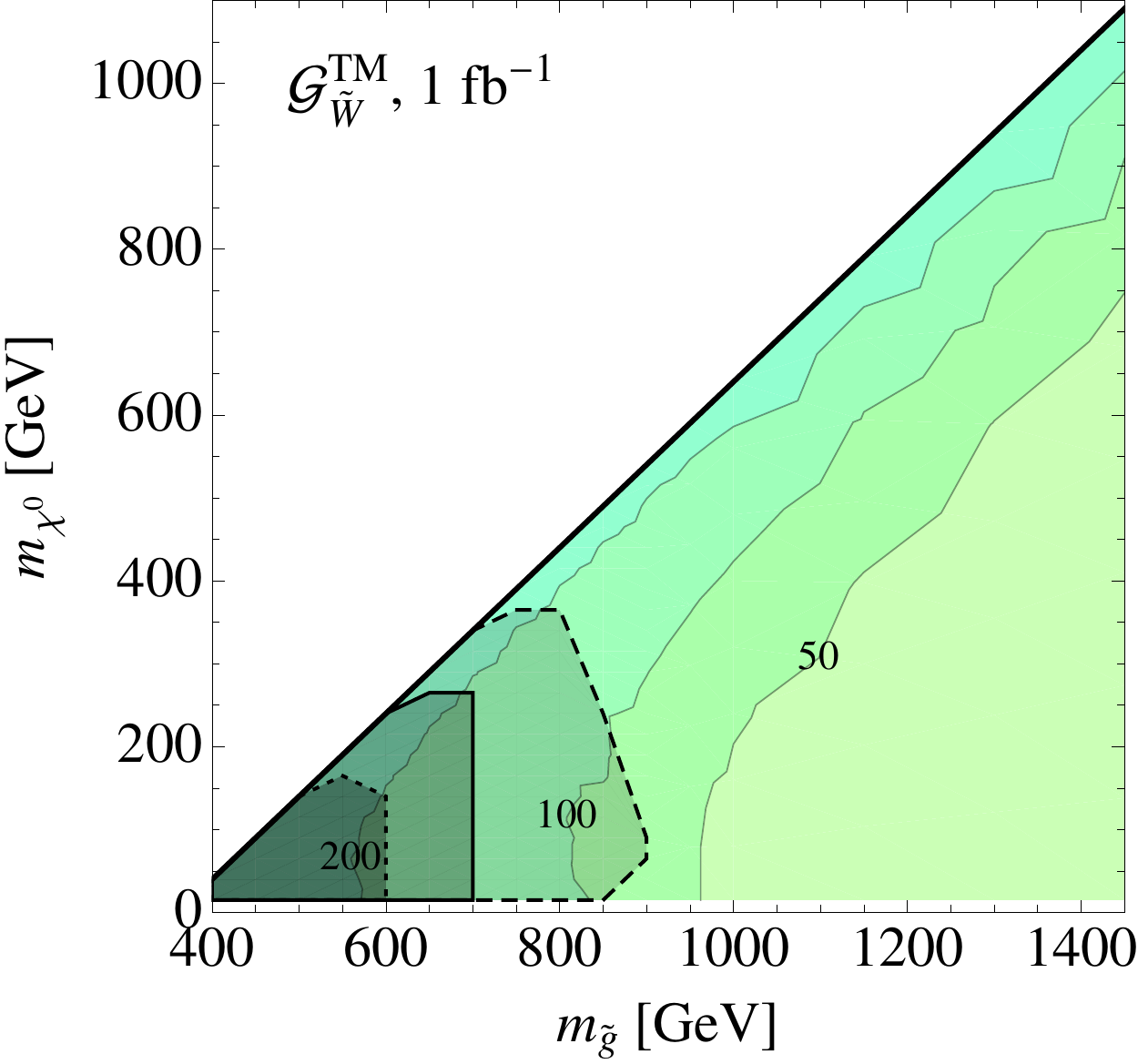}
\includegraphics[width=0.32\textwidth]{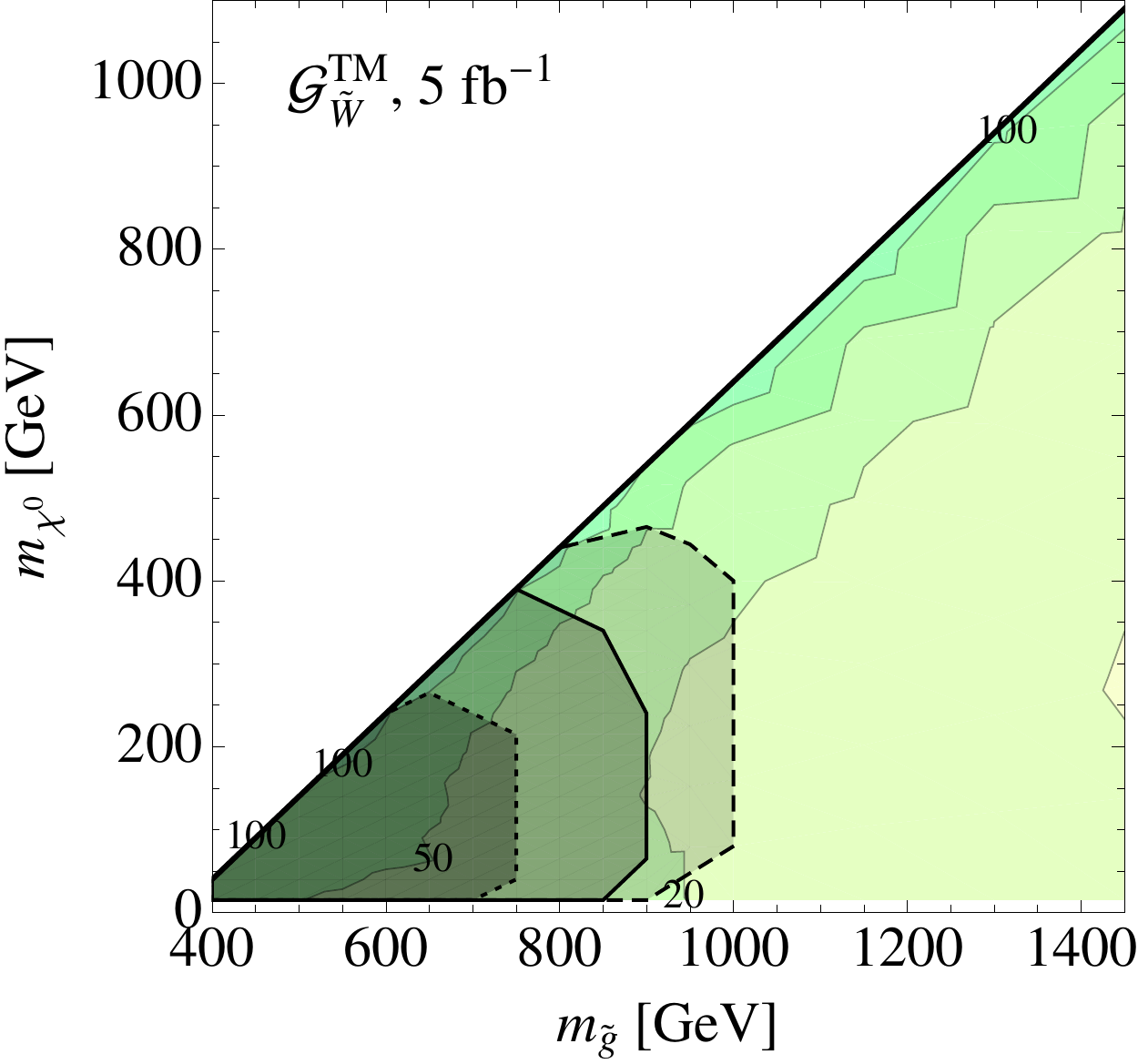}
\includegraphics[width=0.32\textwidth]{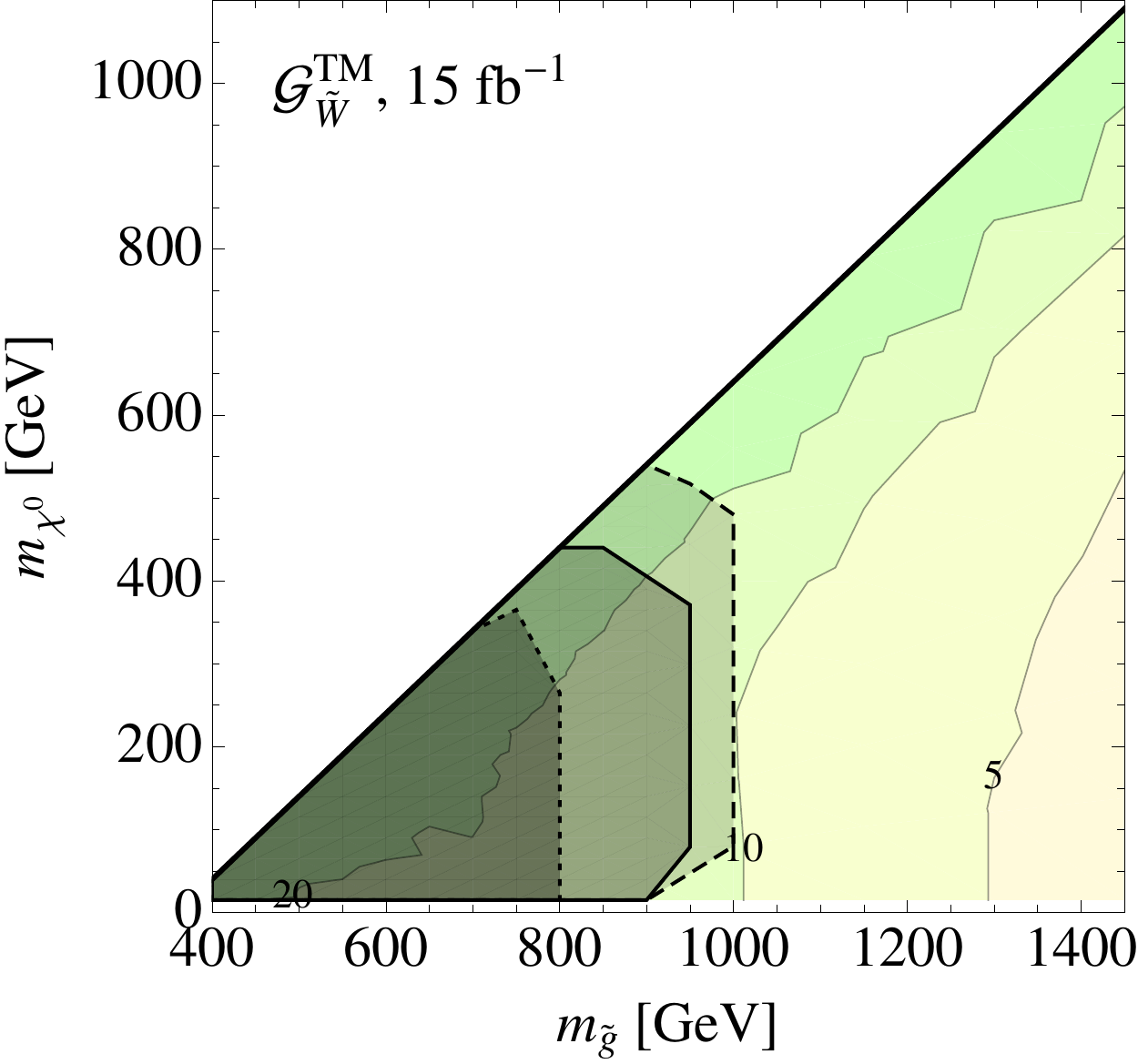}\\
\includegraphics[width=0.32\textwidth]{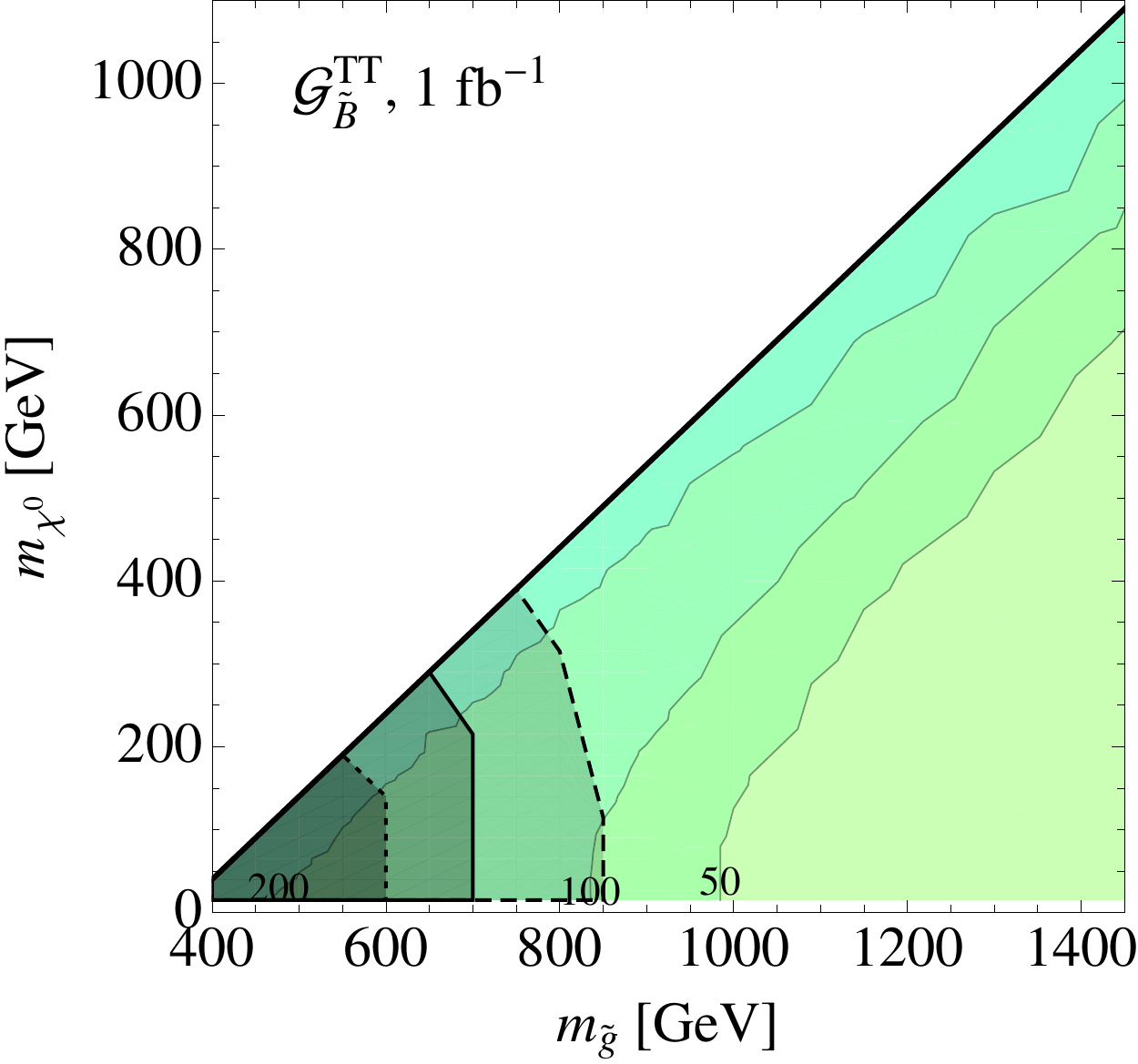}
\includegraphics[width=0.32\textwidth]{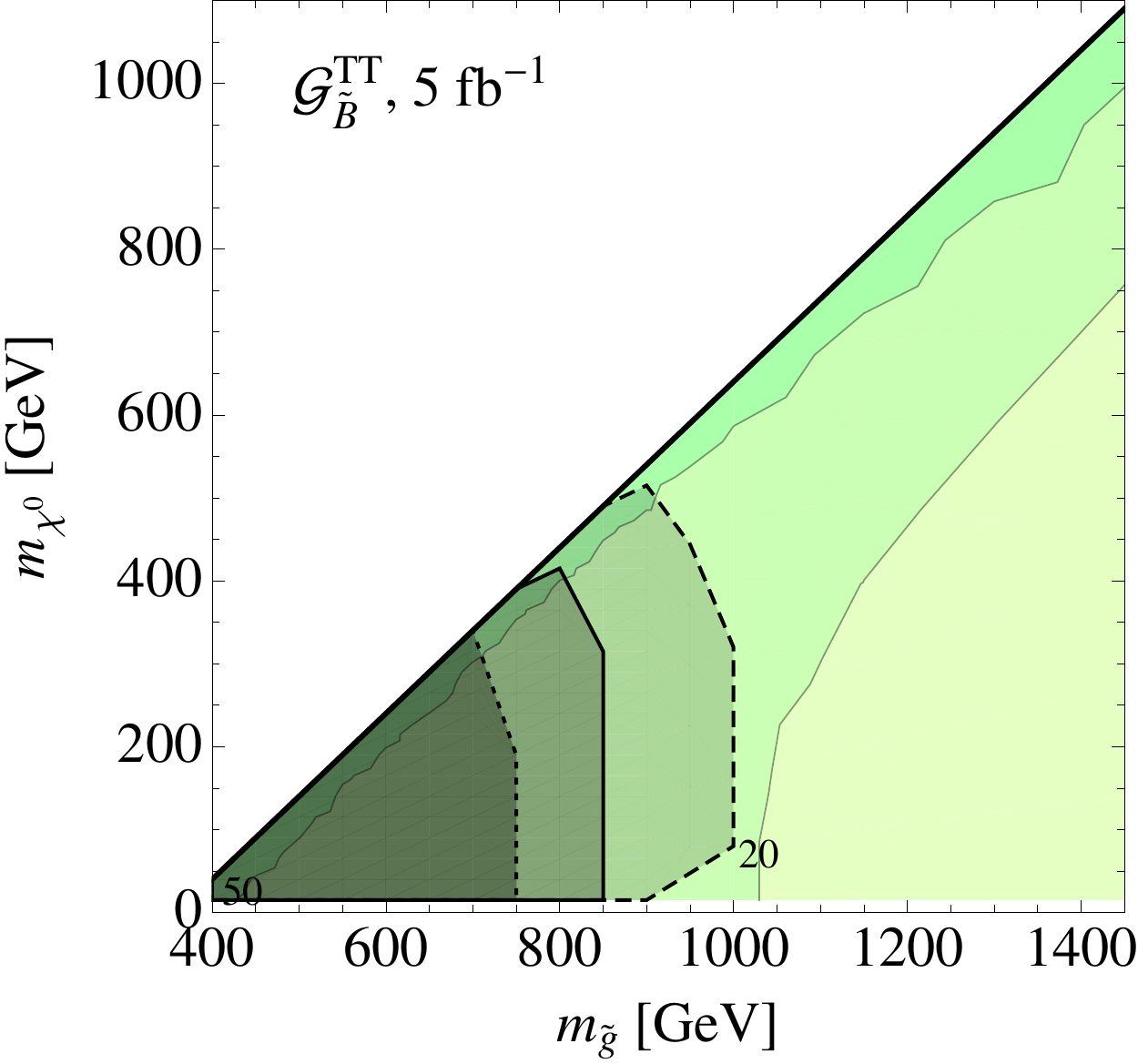}
\includegraphics[width=0.32\textwidth]{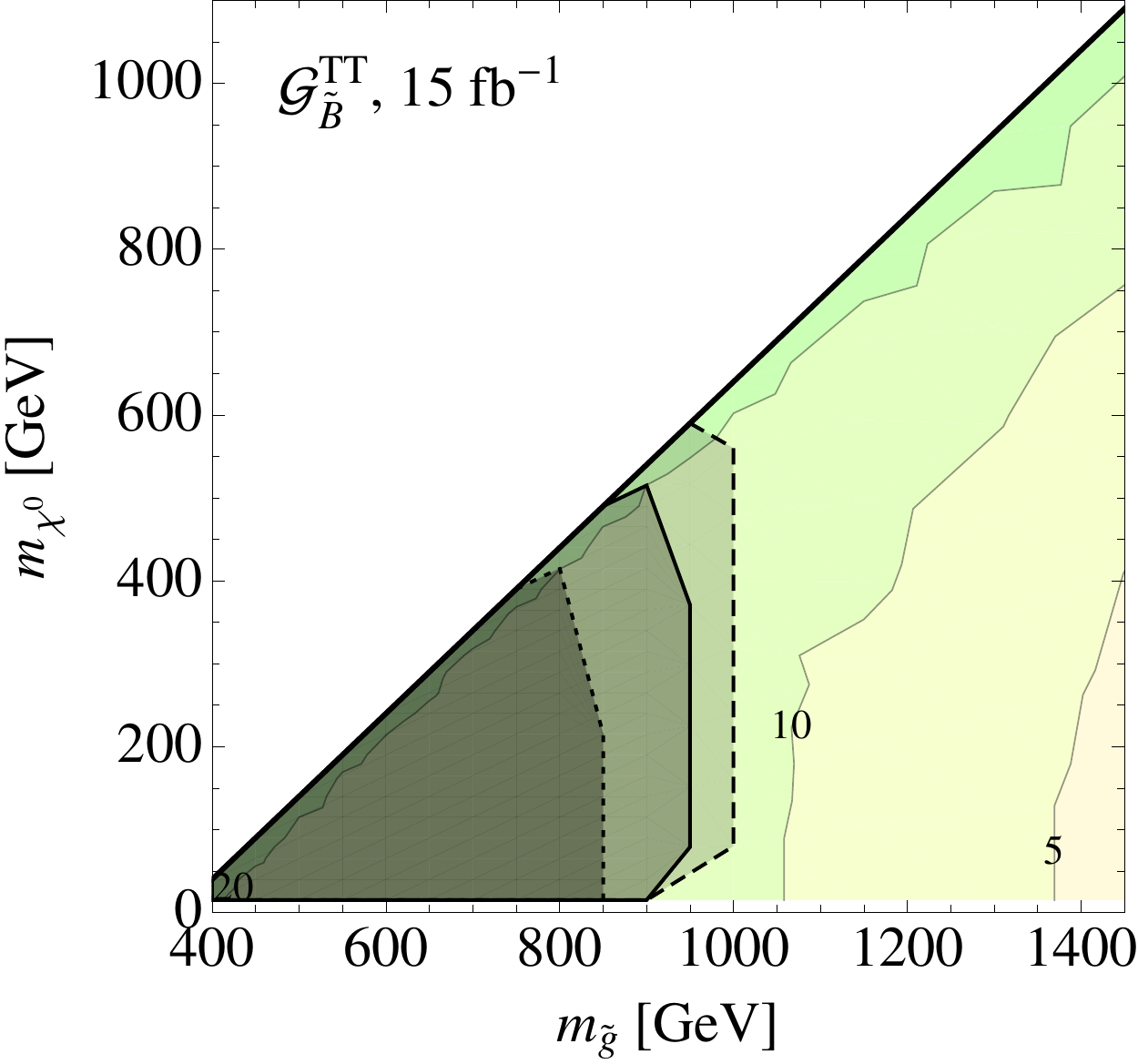}\\
\includegraphics[width=0.32\textwidth]{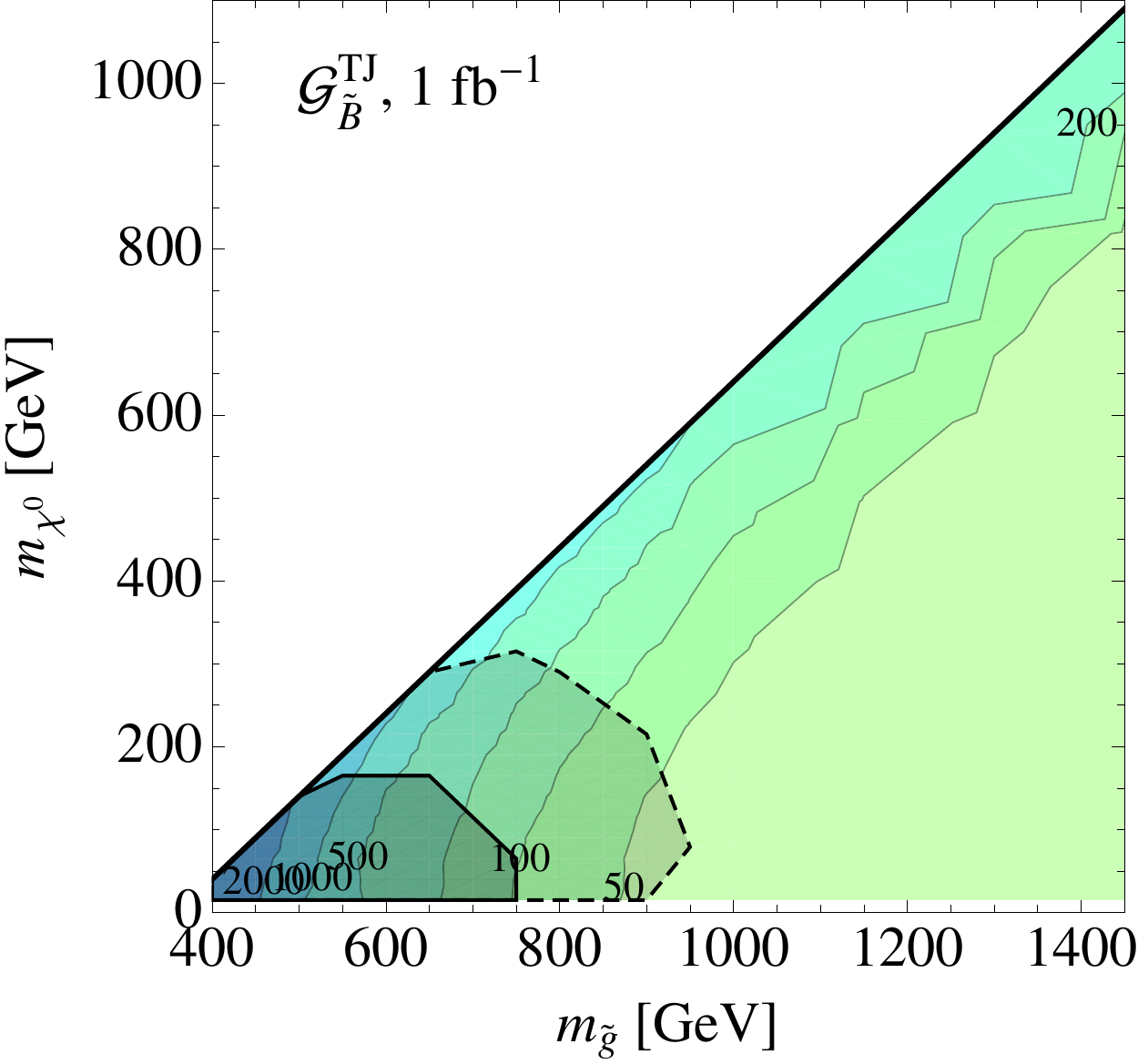}
\includegraphics[width=0.32\textwidth]{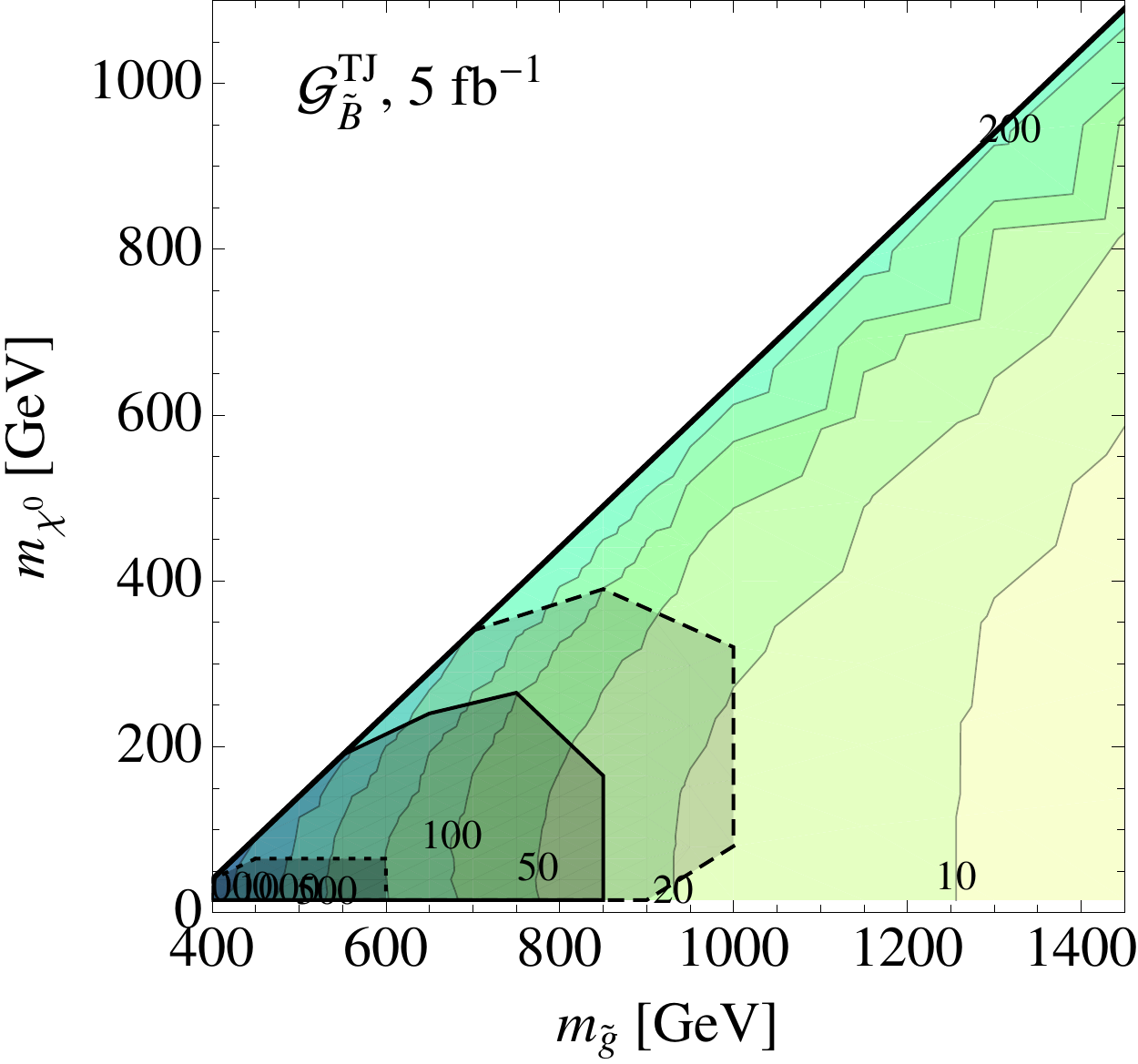}
\includegraphics[width=0.32\textwidth]{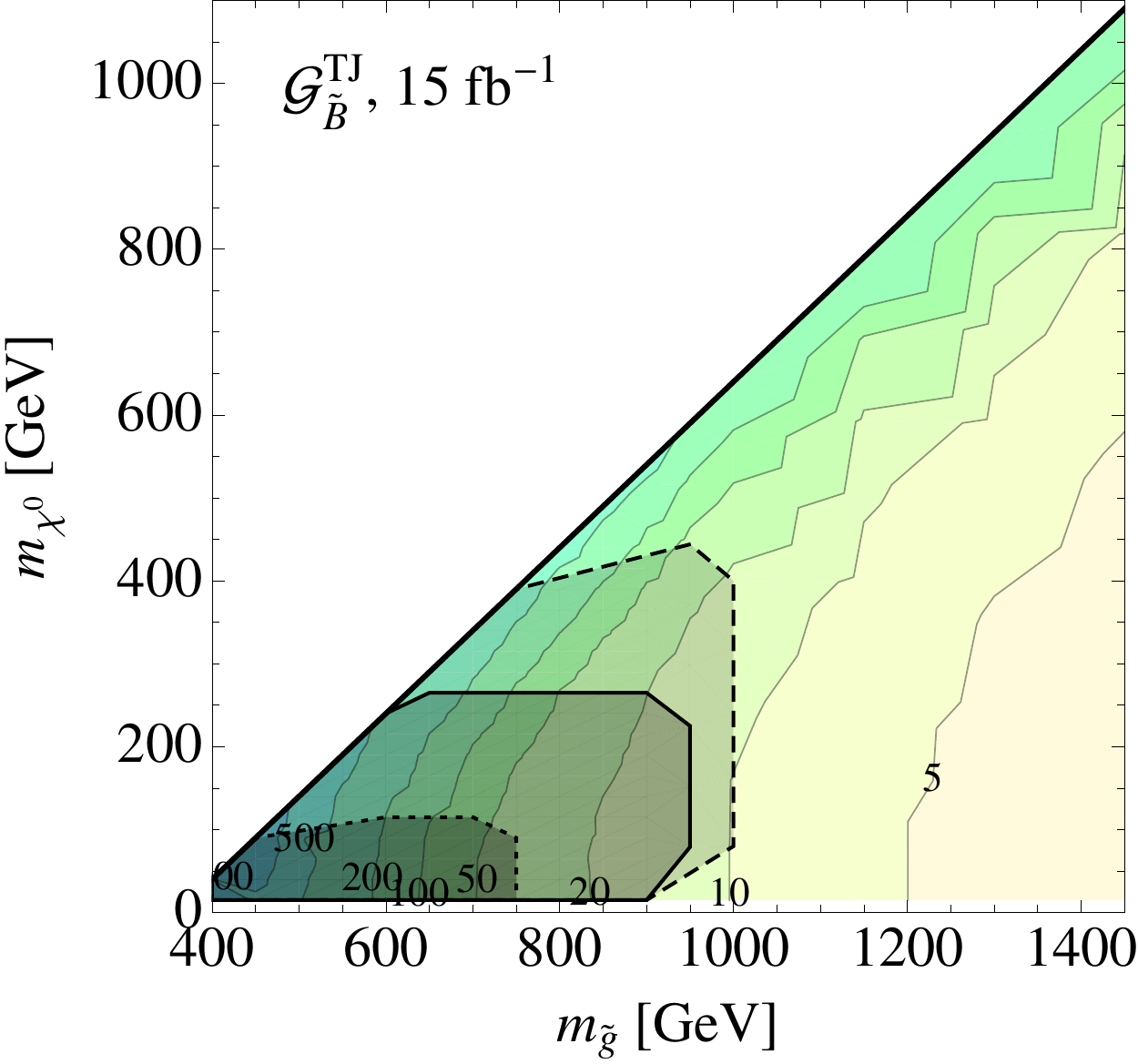}\\
\includegraphics[width=0.32\textwidth]{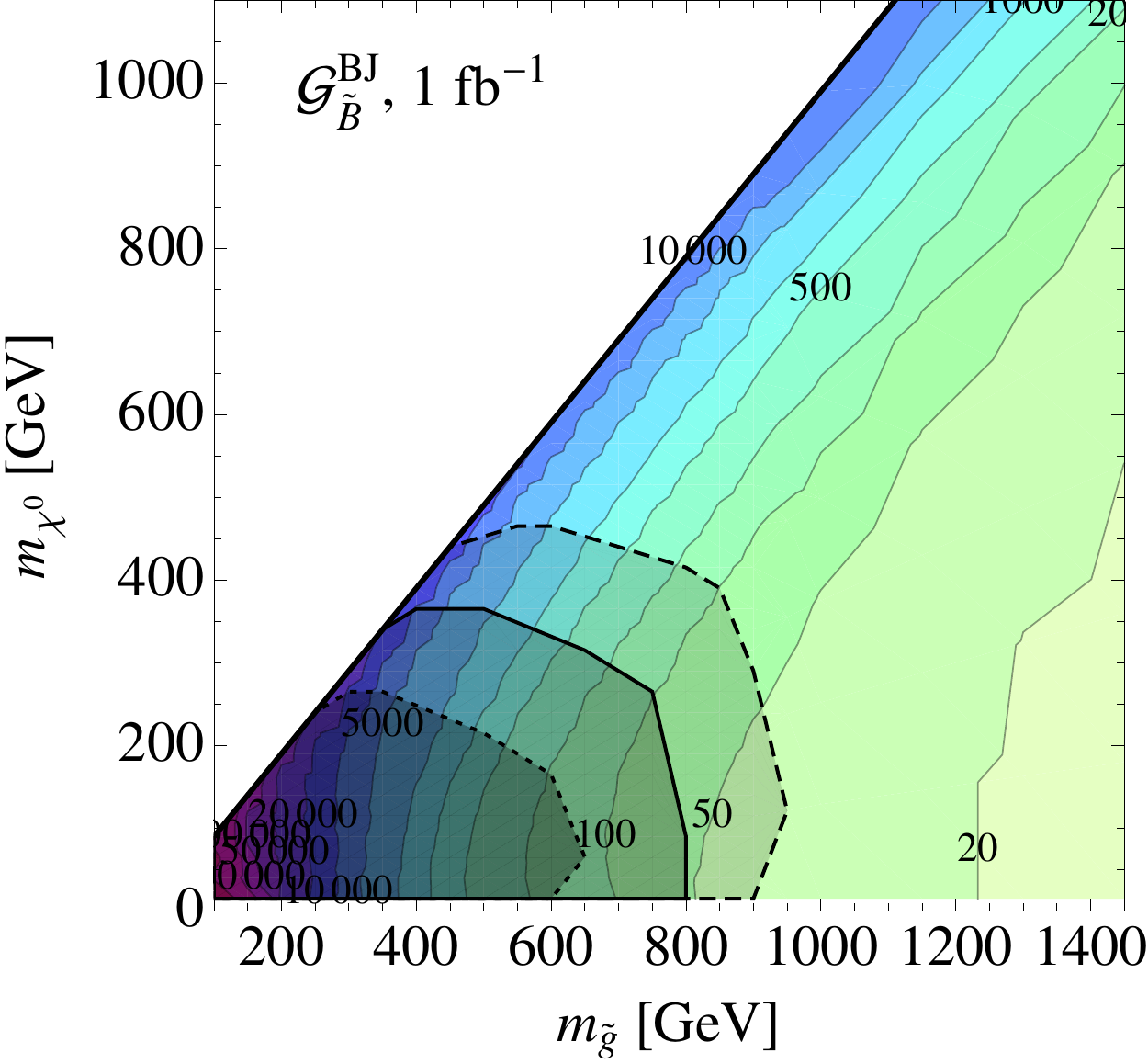}
\includegraphics[width=0.32\textwidth]{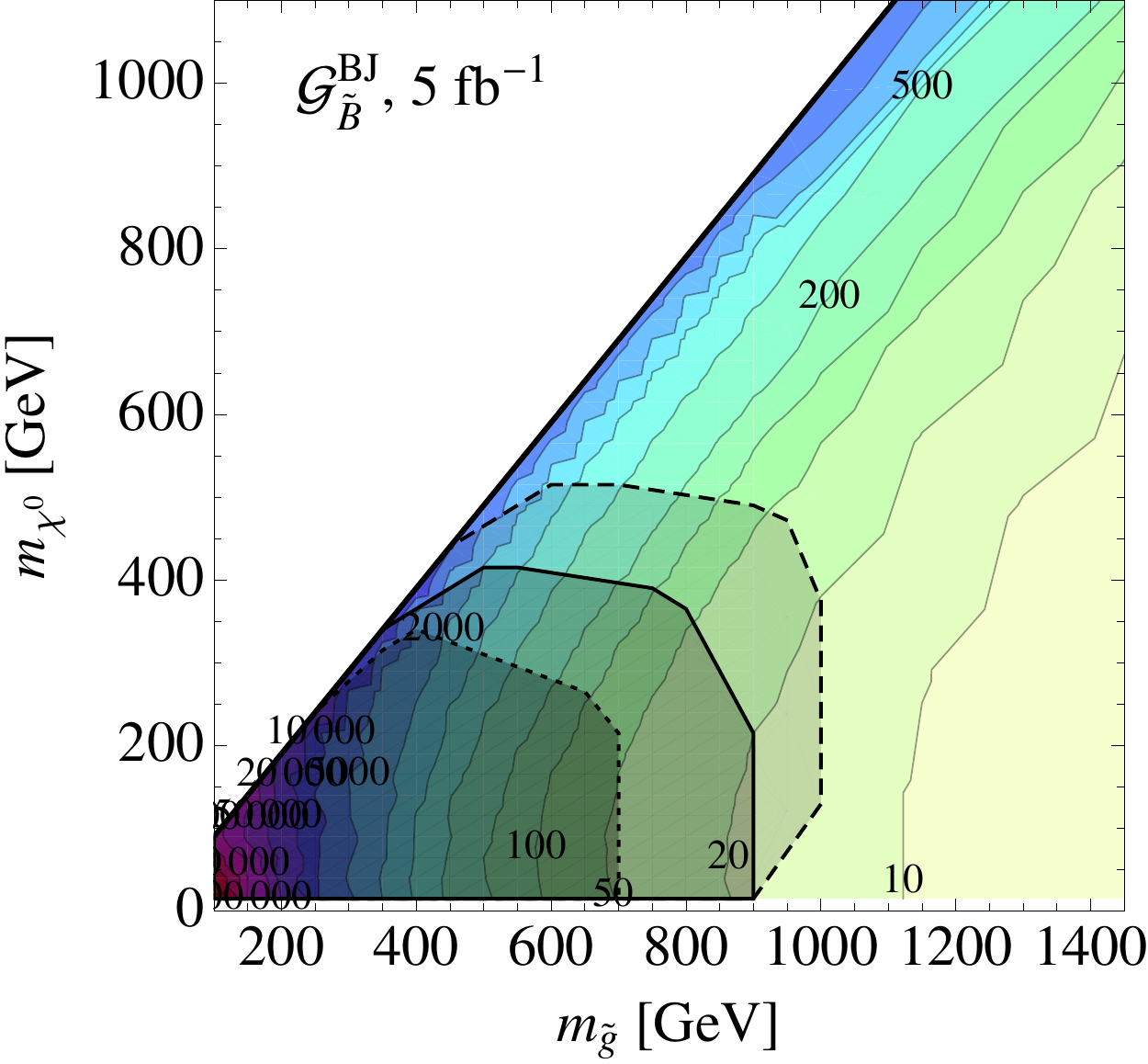}
\includegraphics[width=0.32\textwidth]{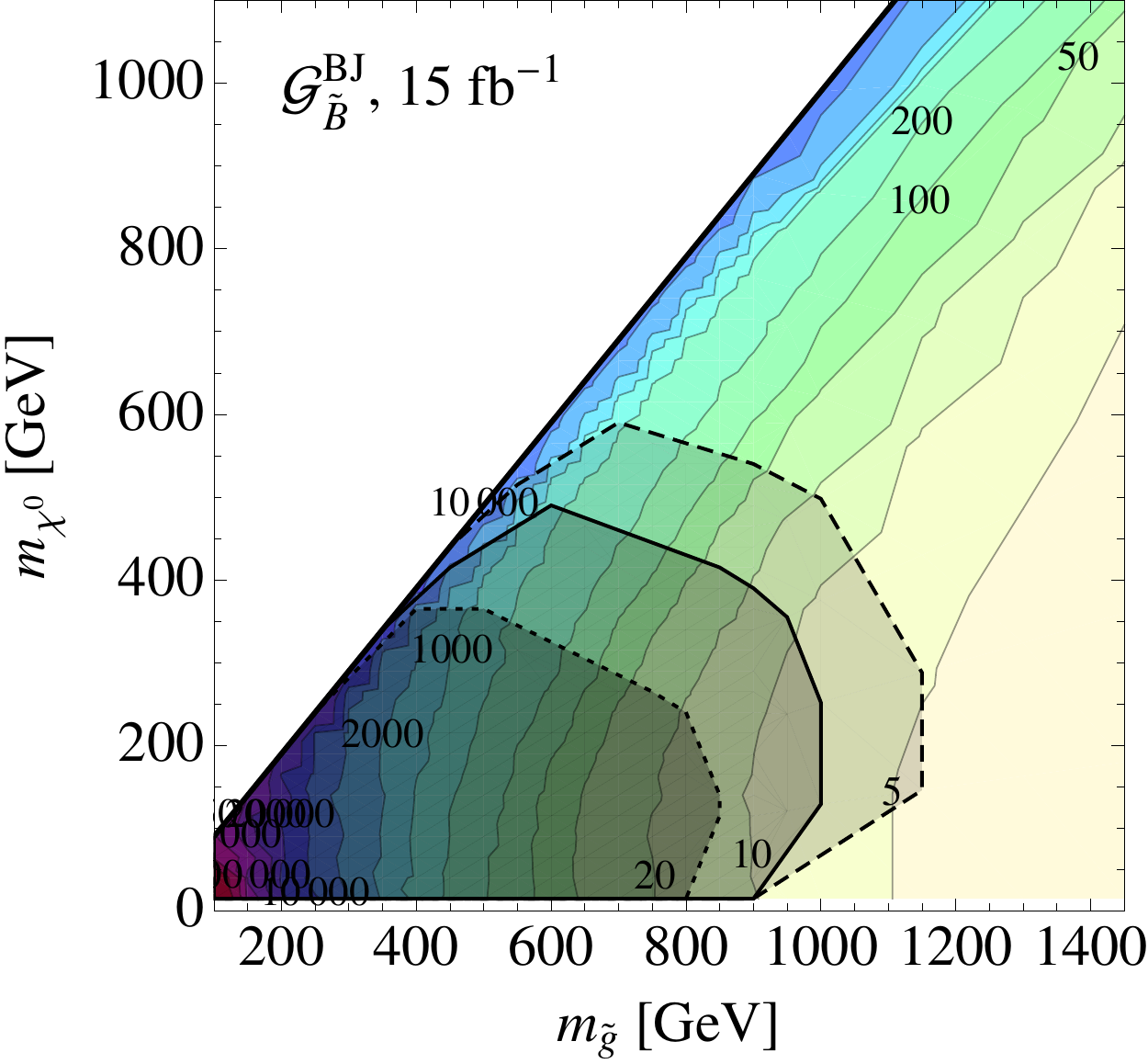}
\caption{\label{Fig:OptLimits2} Estimated 95\% C.L.~contours for the cross section times branching 
ratio sensitivity for various simplified models using the search regions in Table \ref{tab: SearchRegions BM} 
that have been optimized on the benchmarks in Appendix \ref{Sec: BechmarkTables}.  
Shown are $\LL= 1\ifb$ (left column), $5\ifb$ (middle column), 
and $15\ifb$ (right column).  
The solid, dashed, and dotted lines correspond to $\sigma_{pp\rightarrow\go\go}=$1, 3, and 0.3 times $\sigma_{pp\rightarrow \go\go}^{\text{NLO QCD}}$. 
Each row is for a different simplified model; from top to bottom, these are $\GW^{\tt TM}$, $\GB^{\tt TT}$, $\GB^{\tt TJ}$, 
and $\GB^{\tt BJ}$.}
\end{center}
\end{figure}

\begin{figure}[!t]
\begin{center}
\includegraphics[width=0.32\textwidth]{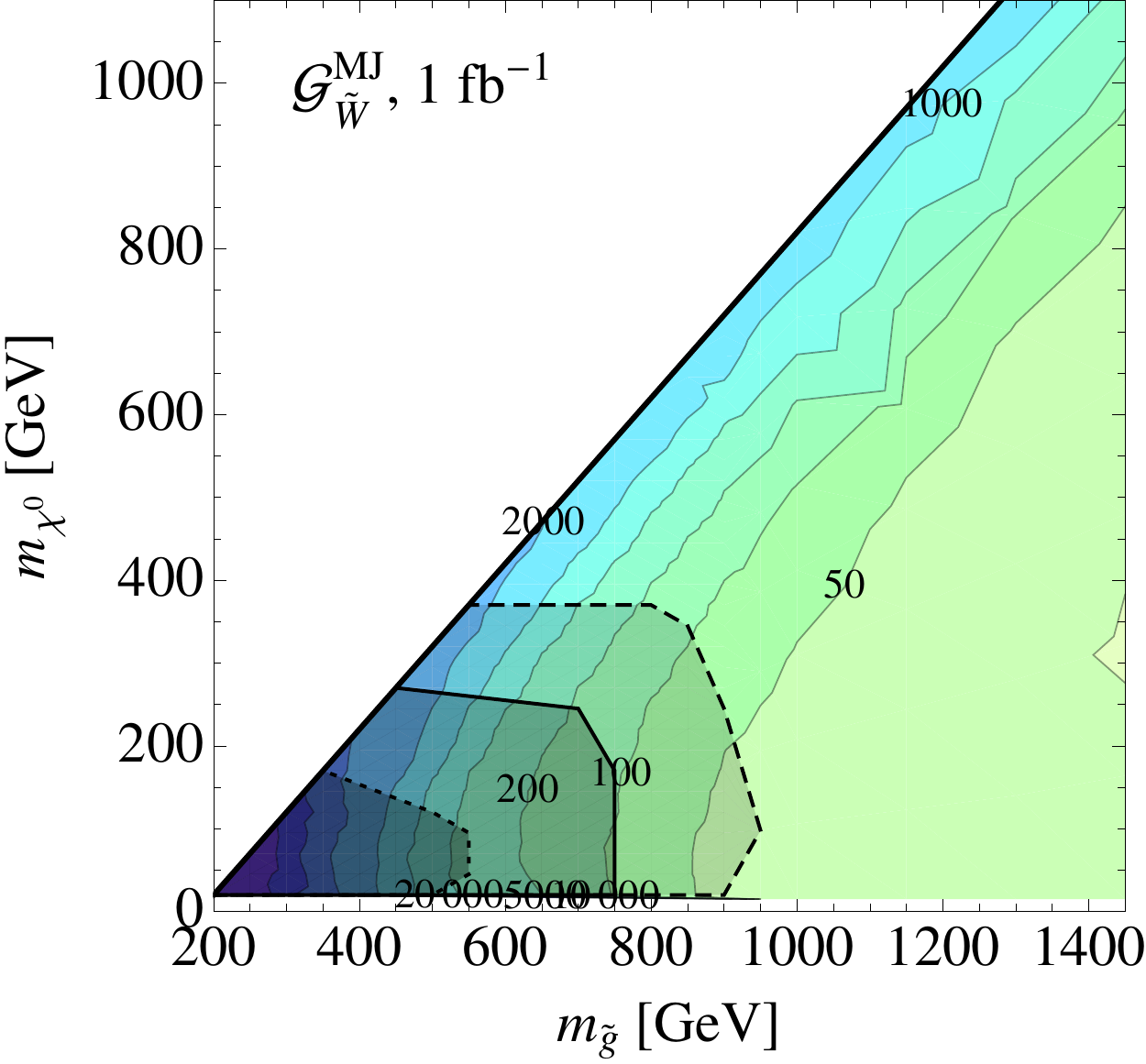}
\includegraphics[width=0.32\textwidth]{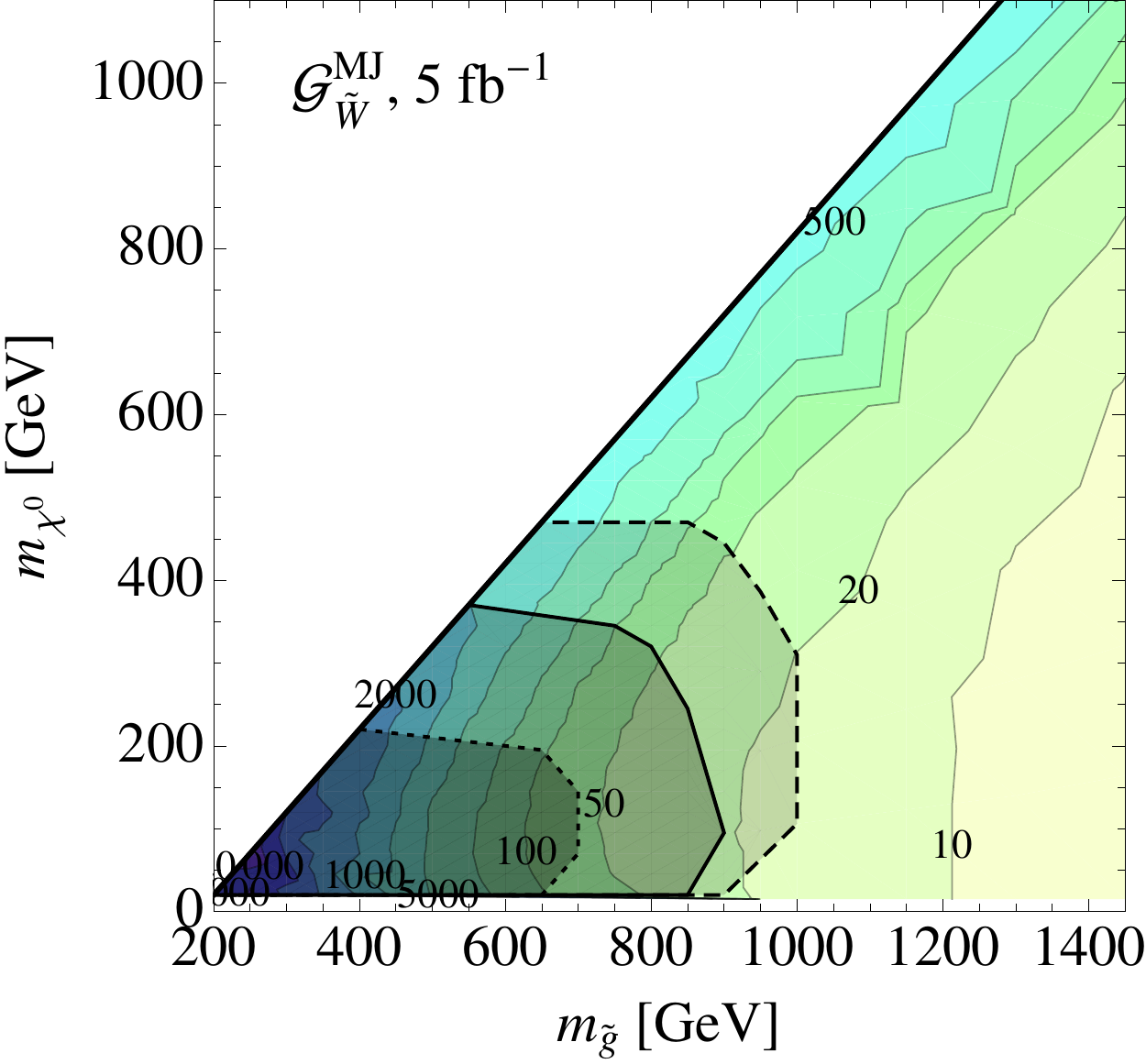}
\includegraphics[width=0.32\textwidth]{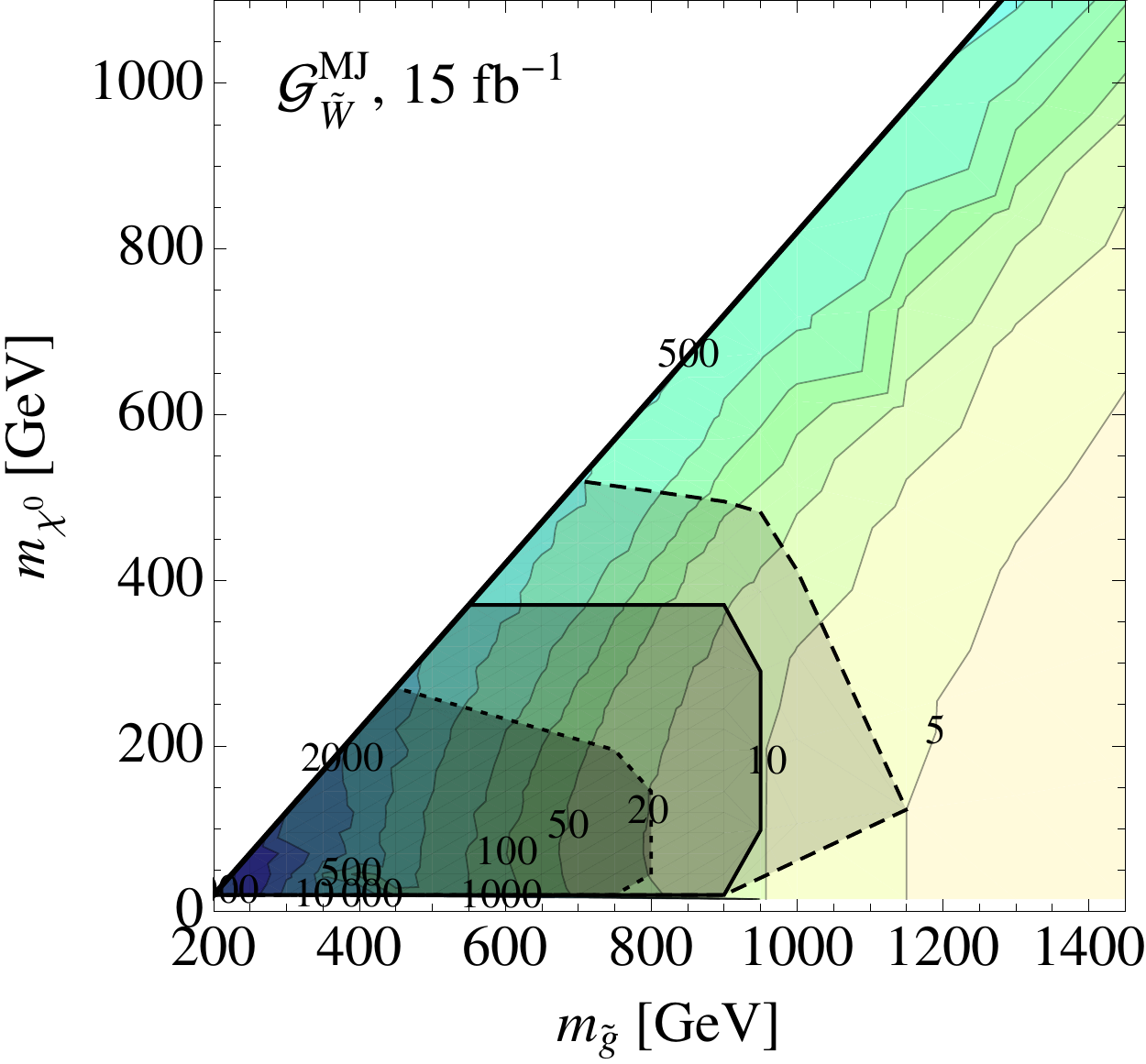}\\
\includegraphics[width=0.32\textwidth]{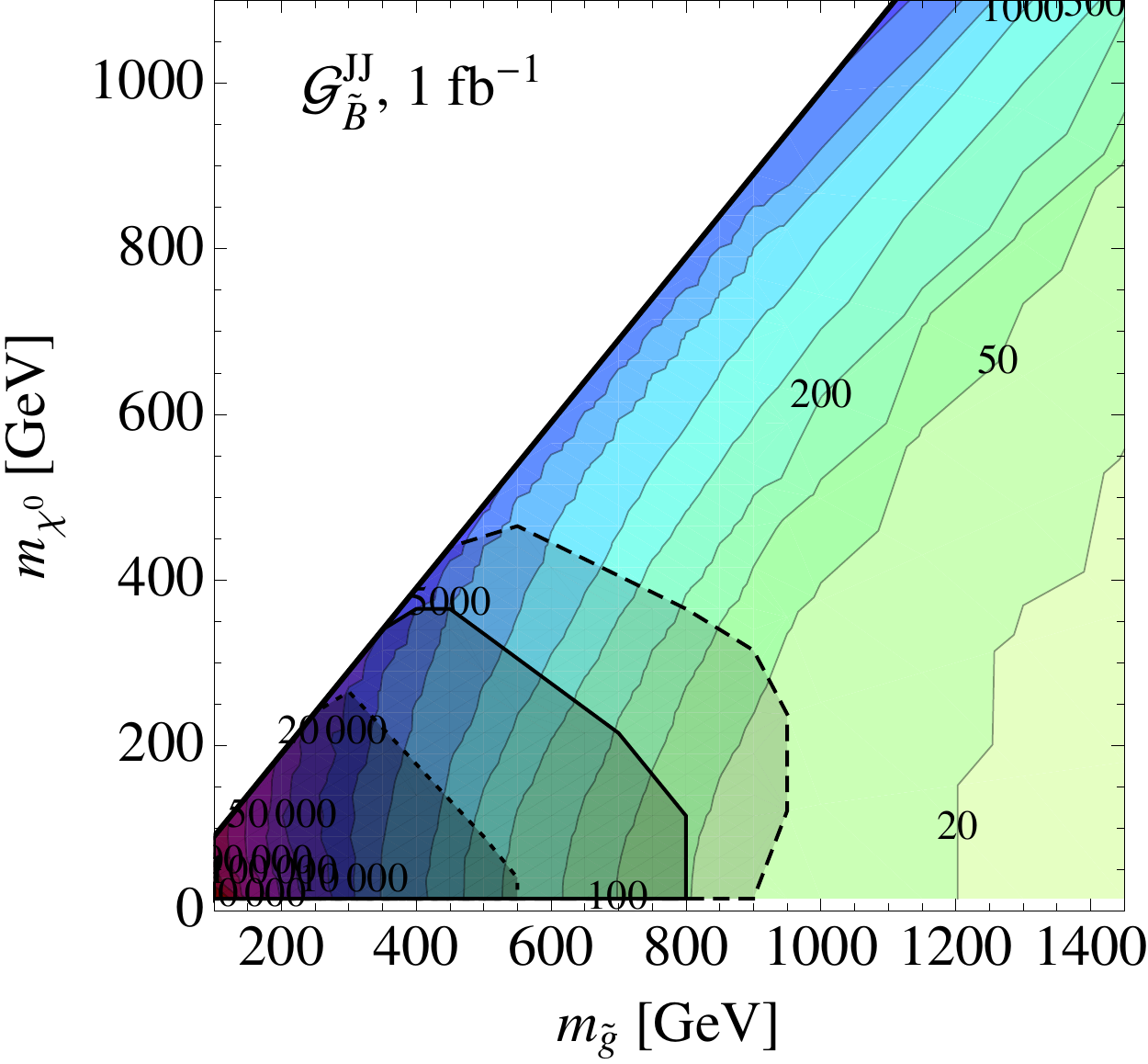}
\includegraphics[width=0.32\textwidth]{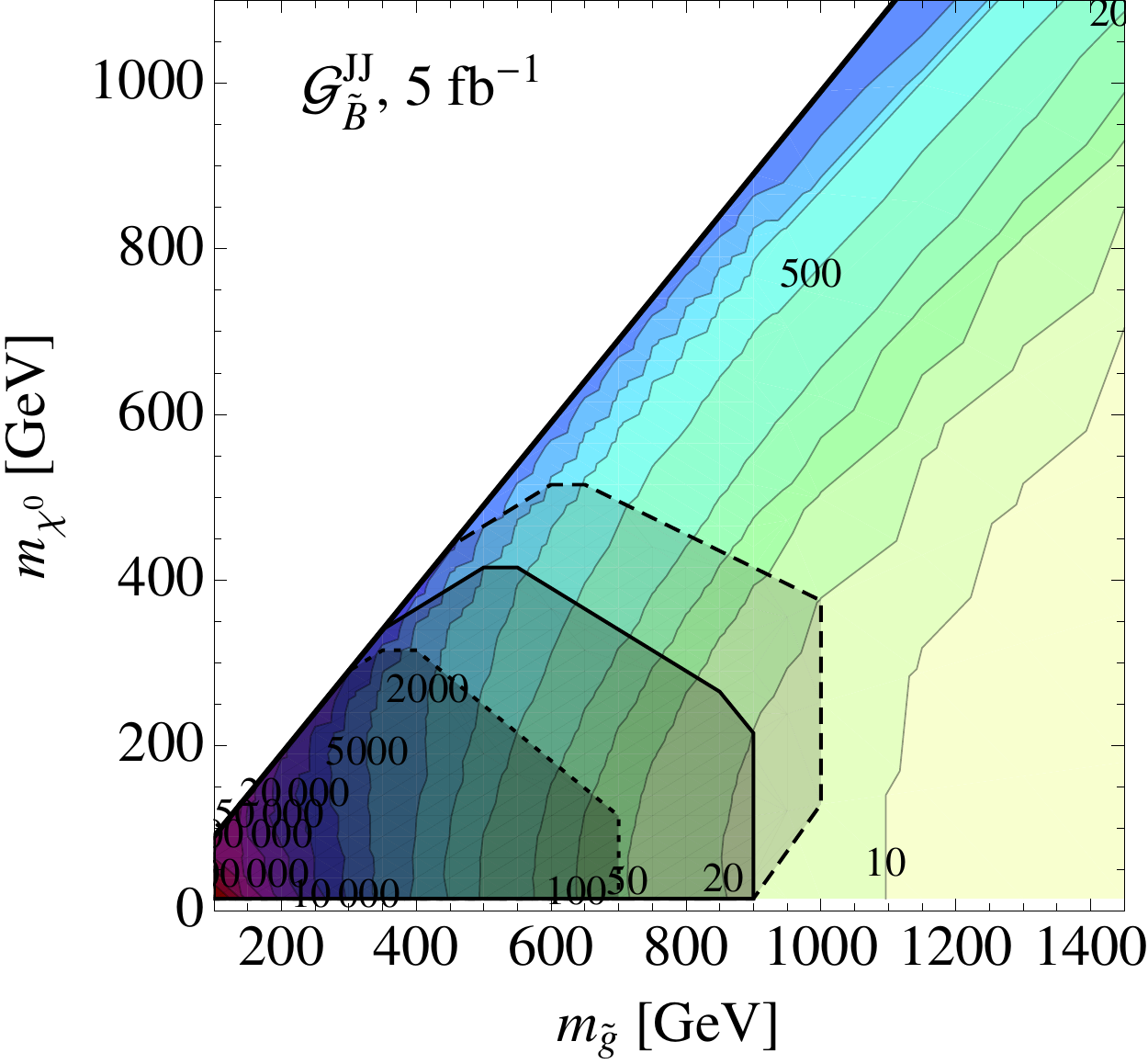}
\includegraphics[width=0.32\textwidth]{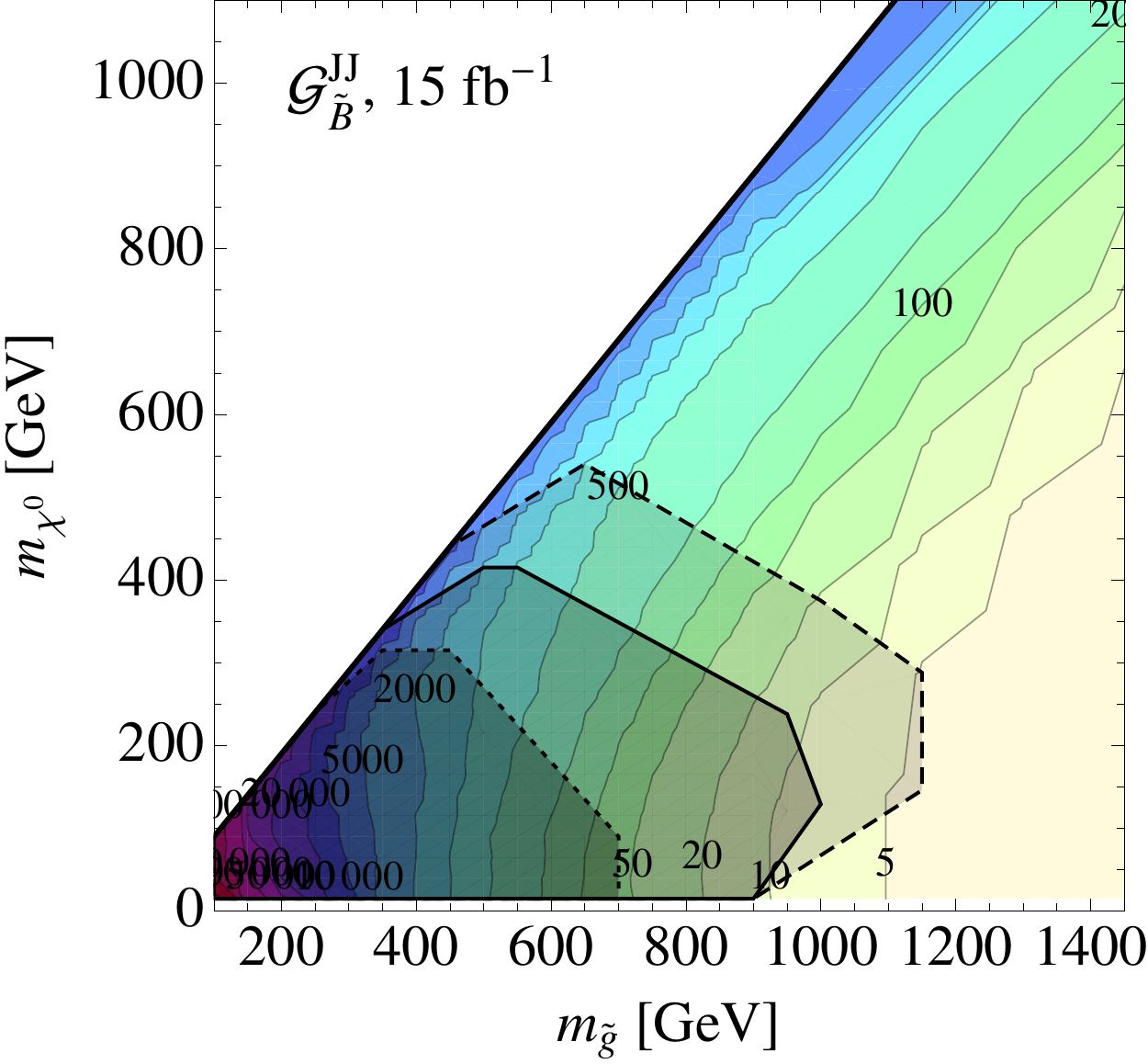}\\
\includegraphics[width=0.32\textwidth]{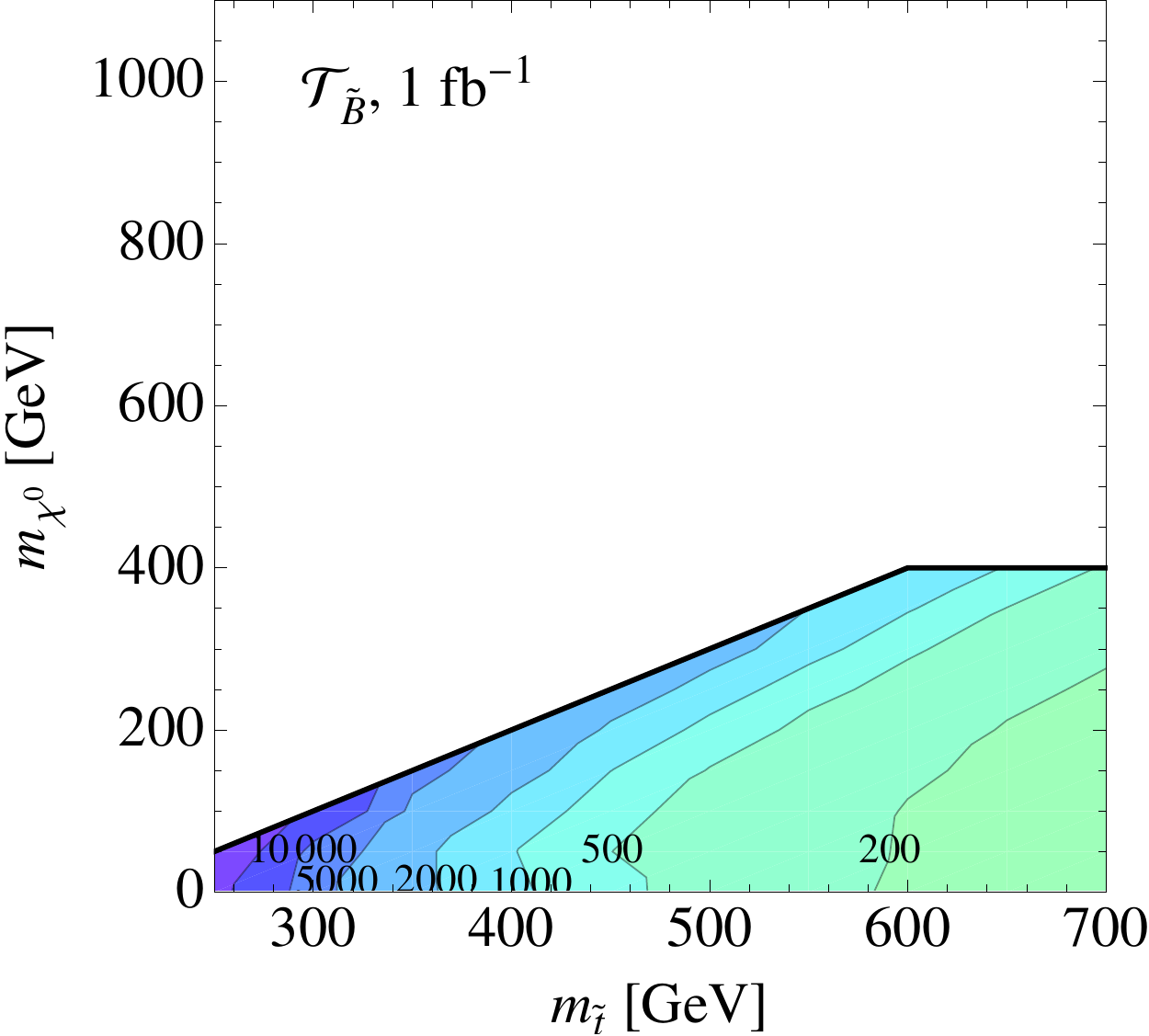}
\includegraphics[width=0.32\textwidth]{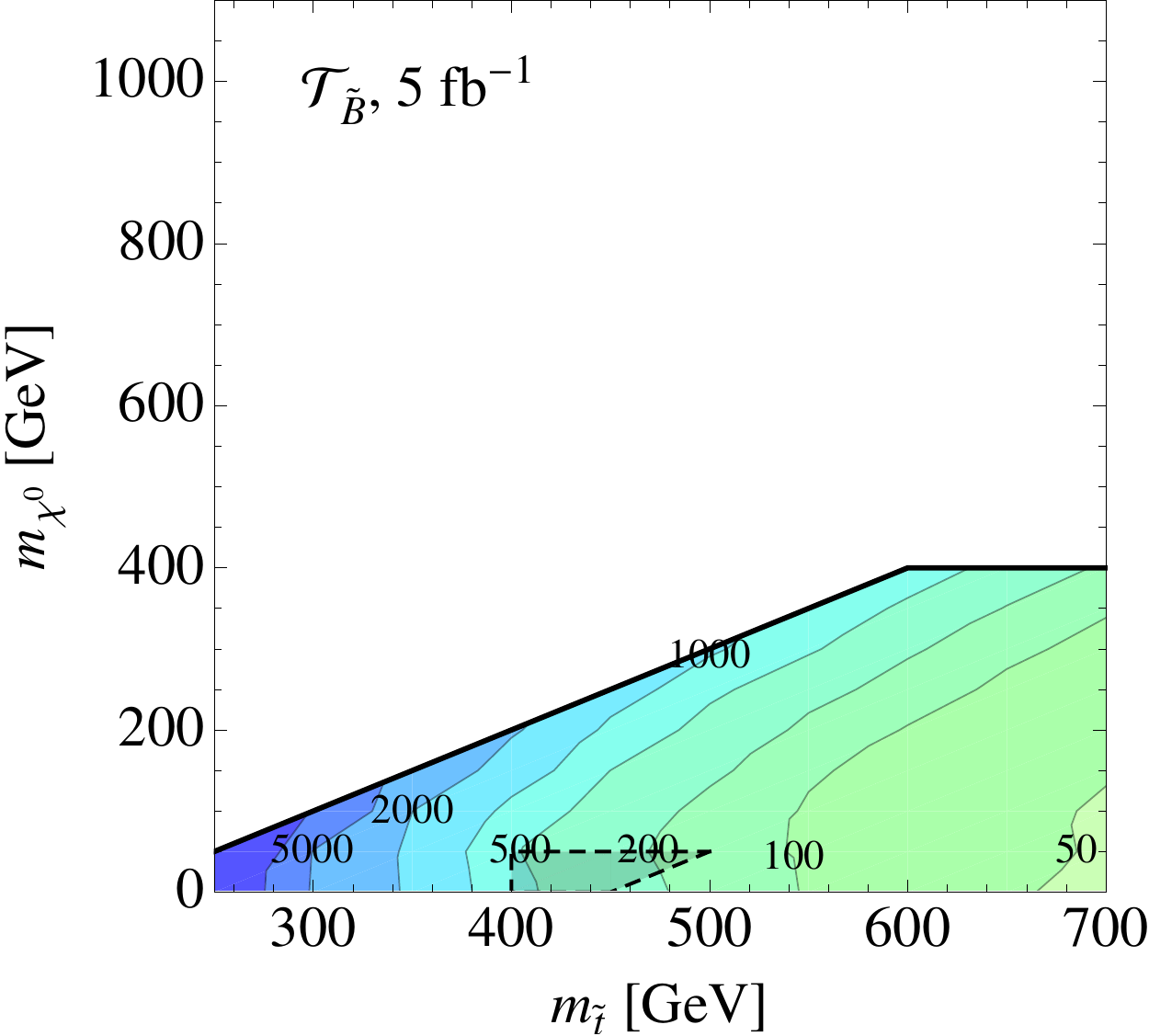}
\includegraphics[width=0.32\textwidth]{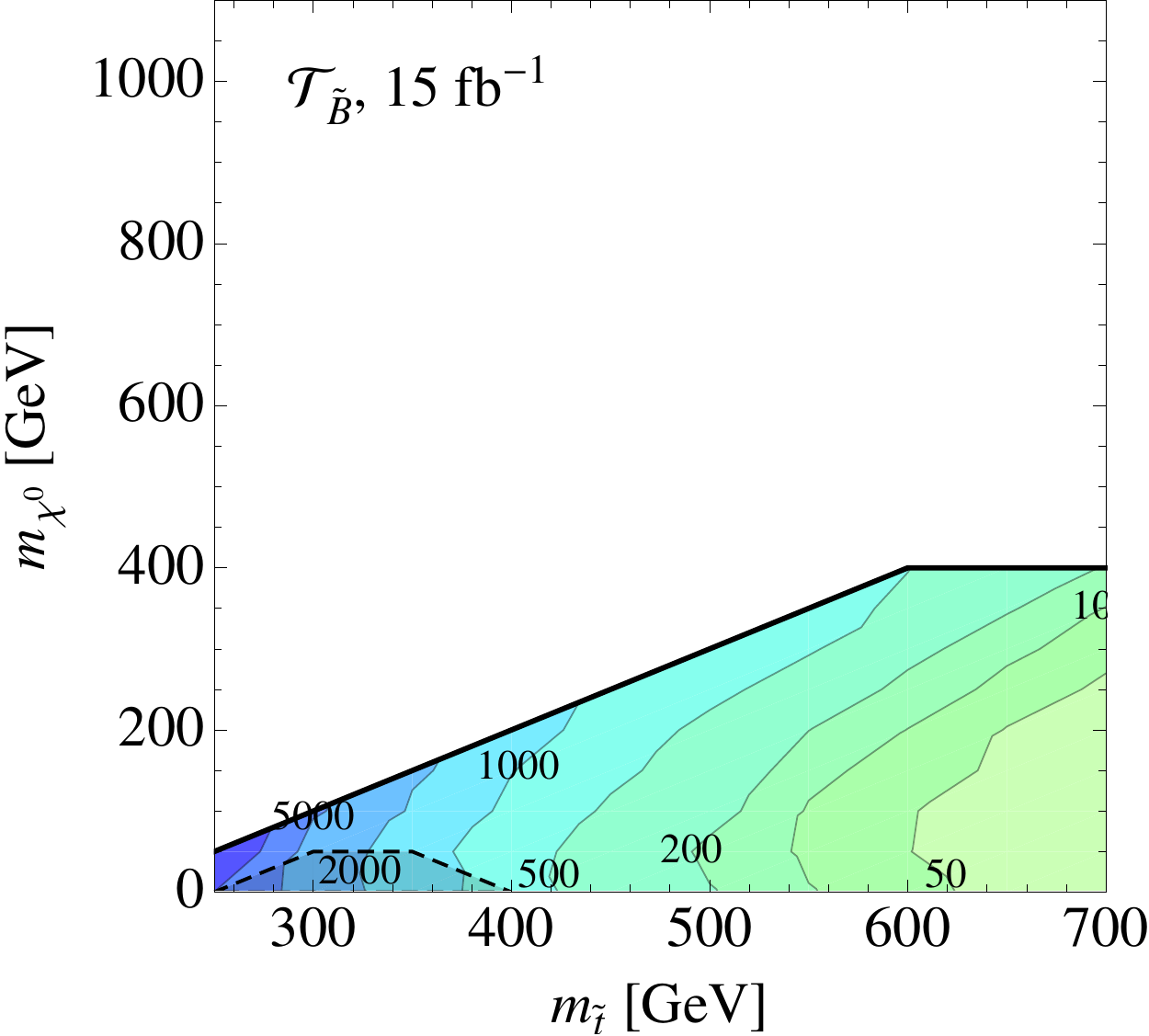}\\
\includegraphics[width=0.32\textwidth]{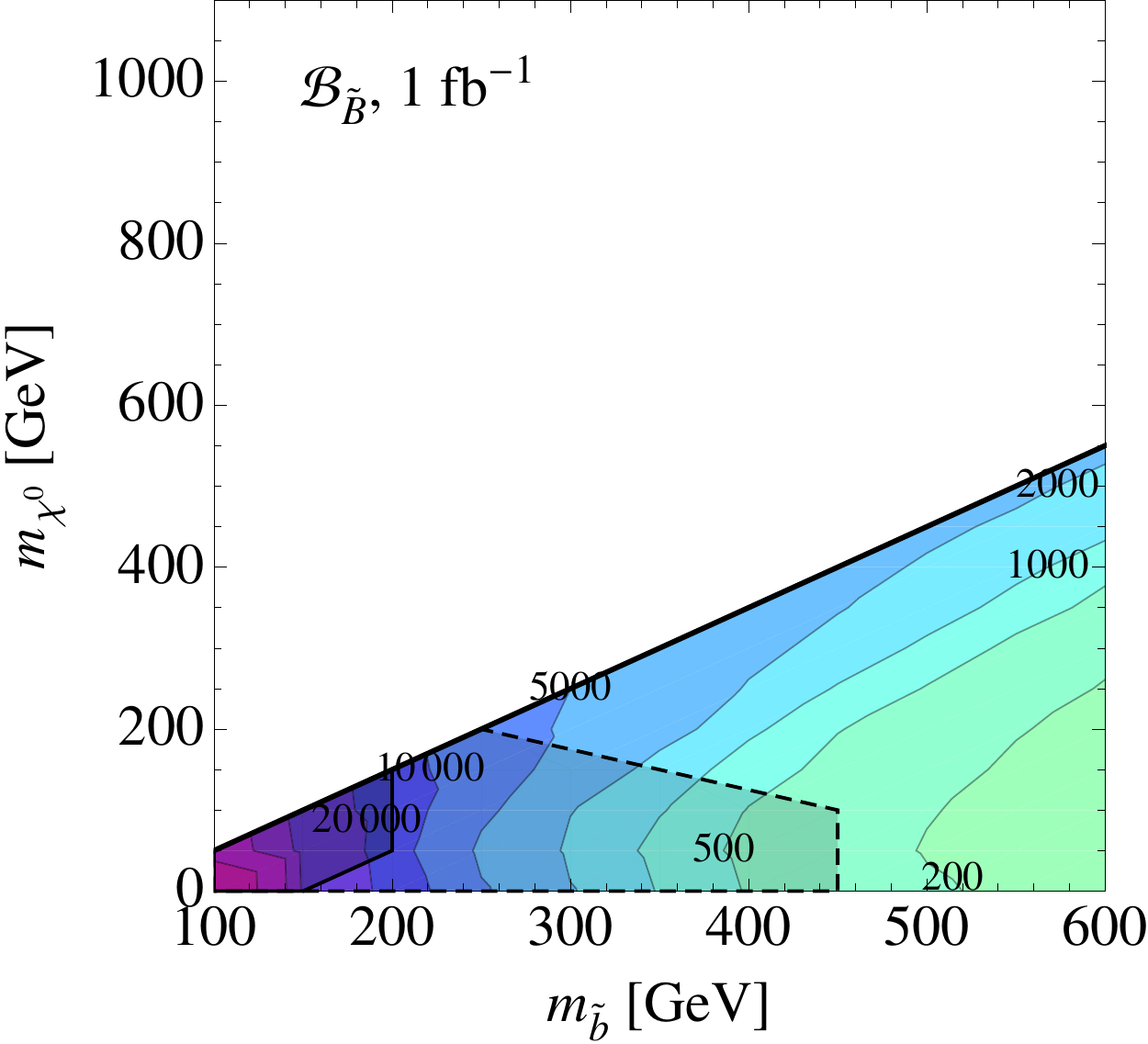}
\includegraphics[width=0.32\textwidth]{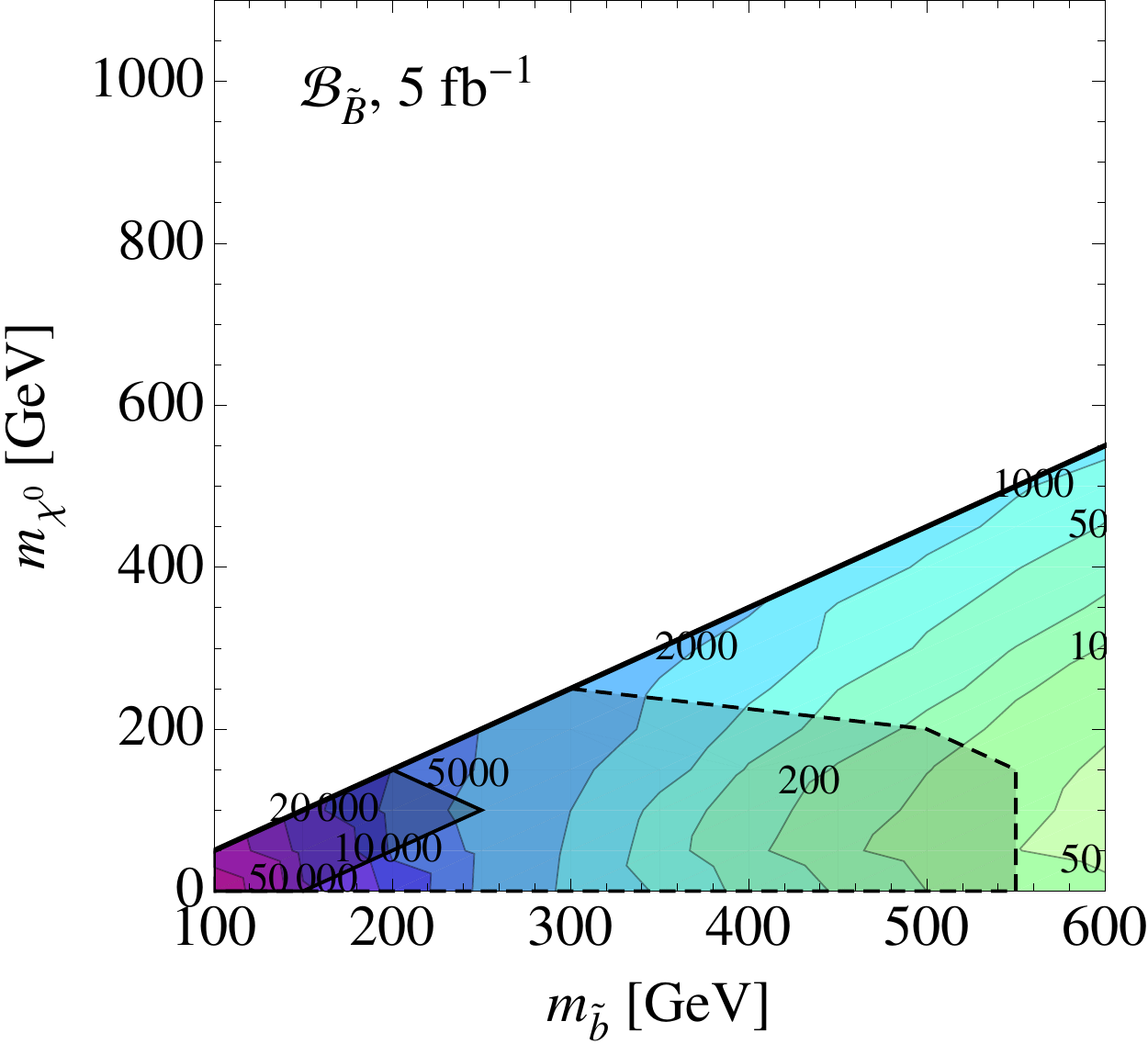}
\includegraphics[width=0.32\textwidth]{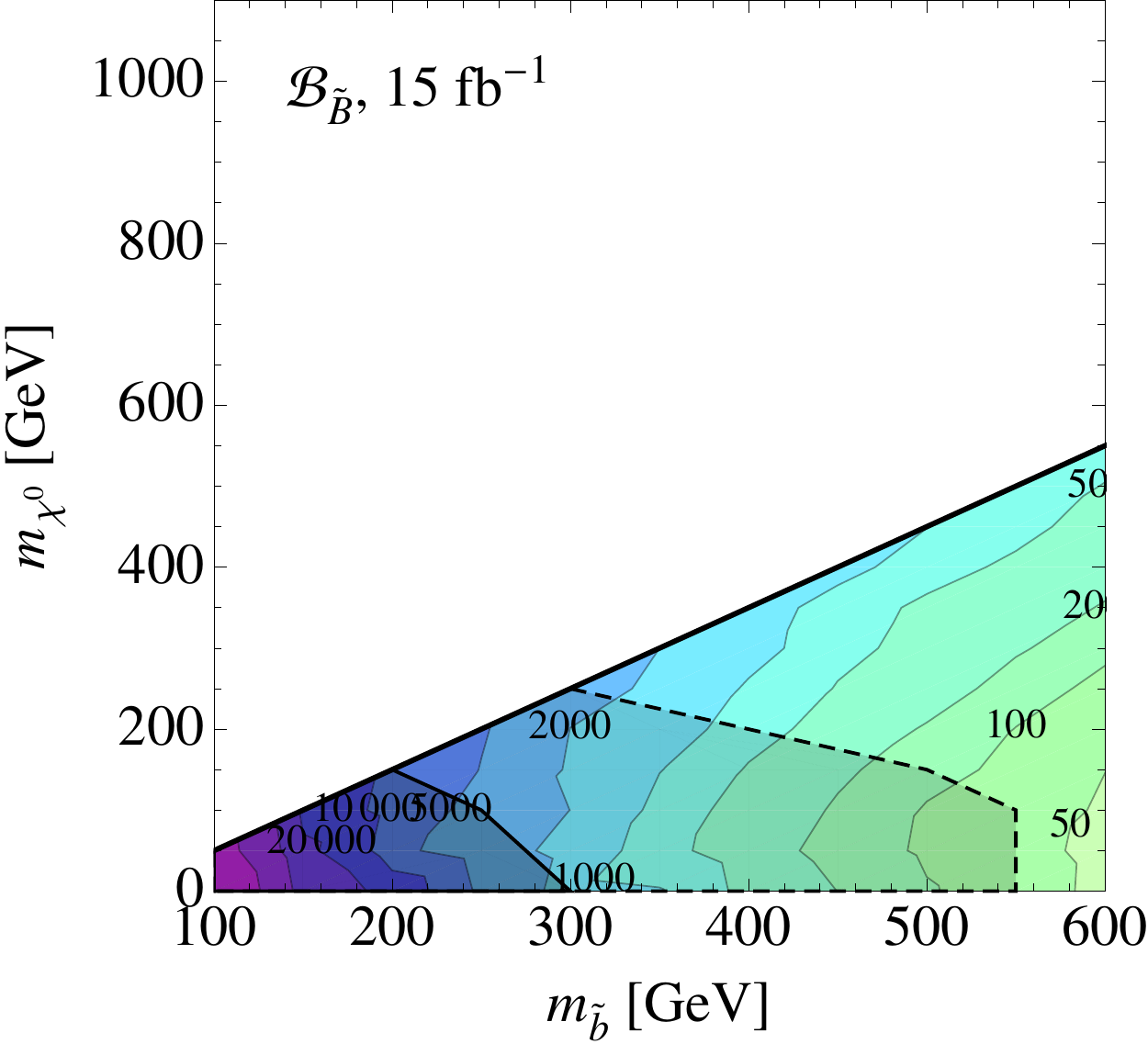}
\caption{\label{Fig:OptLimits3} Estimated 95\% C.L.~contours for the cross section times branching 
ratio sensitivity for various simplified models using the search regions in Table \ref{tab: SearchRegions BM} 
that have been optimized on the benchmarks in Appendix \ref{Sec: BechmarkTables}.  
Shown are $\LL= 1\ifb$ (left column), $5\ifb$ (middle column), 
and $15\ifb$ (right column).  
The solid, dashed, and dotted lines correspond to $\sigma_{pp\rightarrow XX}=$1, 3, and 0.3 times $\sigma_{pp\rightarrow XX}^{\text{NLO QCD}}$, where $X=\go$ (top two rows), $X=\T$ (third row), or $X=\B$ (last row). 
Each row is for a different simplified model; from top to bottom, these are $\GW^{\tt MJ}$, $\GB^{\tt JJ}$, $\TB$, 
and $\BB$.}
\end{center}
\end{figure}

\section{Benchmarks}
\label{Sec: BechmarkTables}
The list of benchmark models from the space of 12 simplified models is presented in this section. The benchmarks from the $\GB$ and $\GW$ simplified models are presented in Table \ref{Tab: GBTopologies} and Tables \ref{Tab: GBTopologiesA} and \ref{Tab: GWTopologies}, respectively. The benchmarks from the $\TB$ and $\BB$ topologies are shown in Table~\ref{Tab: TB&BBTopologies}.
\begin{table}[th!]
\begin{center}
\begin{tabular}{|c||c|c|c|c|c|c|}
\hline
Name & $m_{\go}$ (GeV) & $m_{\chi^0}$ (GeV) & $\sigma^{\text{\tiny{reach}}}_{\text{1 fb}^{-1} }$ (fb) & $\sigma^{\text{\tiny{reach}}}_{\text{5 fb}^{-1}}$ (fb) & $\sigma^{\text{\tiny{reach}}}_{\text{15 fb}^{-1} }$ (fb) & $\sigma^{\text{\tiny{QCD}}}_{\text{prod}}$ (fb)  \\
\hline\hline
$\GB^{\tt TT}$ & 500 & 115 & 592 & 129 & 44 & 2310  \\
$\GB^{\tt TT}$  & 500 & 40 & 428 & 95 & 32 & 2310 \\
$\GB^{\tt TT}$ & 650 & 40 & 139 & 65 & 26 & 335  \\
$\GB^{\tt TT}$  & 800 & 415 & 469 & 129 & 44 &  61  \\
$\GB^{\tt TT}$  & 800 & 40 & 92 & 27 & 13 &  61 \\
\hline
\hline
$\GB^{\tt BB}$ & 100 & 40 & 353000 & 265000 & 226000 & 21.2x10$^6$  \\
$\GB^{\tt BB}$ & 200 & 15 & 17800 & 11400 & 10400 & 625000  \\
$\GB^{\tt BB}$ & 200 & 165 & 3360 & 3230 & 3210 & 625000  \\
$\GB^{\tt BB}$ & 350 & 165 & 875 & 591 & 373 & 24200  \\
$\GB^{\tt BB}$ & 500 & 40 & 94 & 37 & 24 & 2310  \\
$\GB^{\tt BB}$ & 600 & 365 & 236 & 112 & 70 & 617  \\
$\GB^{\tt BB}$ & 700 & 265 & 57 & 20 & 11 & 186  \\
$\GB^{\tt BB}$ & 750 & 490 & 153 & 62 & 41 & 106  \\
$\GB^{\tt BB}$ & 800 & 765 & 4056 & 1840 & 1490 &  61  \\
$\GB^{\tt BB}$ & 800 & 40 & 42 & 11 & 5.2 &  61  \\
$\GB^{\tt BB}$ & 900 & 540 & 65 & 23 & 13 & 21  \\
\hline
\hline
$\GB^{\tt JJ}$ & 150 & 15 & 12900 & 128000 & 115000 & 2.86x10$^6$  \\
$\GB^{\tt JJ}$ & 200 & 165 & 39300 & 25700 & 19900 & 625000  \\
$\GB^{\tt JJ}$ & 300 & 115 & 6450 & 4970 & 4300 & 62100  \\
$\GB^{\tt JJ}$ & 500 & 40 & 406 & 306 & 278 & 2310  \\
$\GB^{\tt JJ}$ & 600 & 515 & 2590 & 1440 & 939 & 617  \\
$\GB^{\tt JJ}$ & 650 & 115 & 129 & 82 & 67 & 335  \\
$\GB^{\tt JJ}$ & 750 & 215 & 90 & 52 & 41 & 106  \\
$\GB^{\tt JJ}$ & 800 & 765 & 3700 & 2750 & 2250 &  61  \\
$\GB^{\tt JJ}$ & 850 & 40 & 517 & 351 & 244 & 36  \\
$\GB^{\tt JJ}$ & 850 & 590 & 39 & 19 & 12 & 36  \\
\hline
\end{tabular}
\caption{Benchmark models from the pure $\GB$ simplified models. Also shown are the estimated cross section reach for $\mathcal{L}=1,5,15$ fb$^{-1}$ in addition to the NLO production cross section.  }
\label{Tab: GBTopologies}
\end{center}
\end{table}

\begin{table}[th!]
\begin{center}
\begin{tabular}{|c||c|c|c|c|c|c|}
\hline
Name & $m_{\go}$ (GeV) & $m_{\chi^0}$ (GeV) & $\sigma^{\text{\tiny{reach}}}_{\text{1 fb}^{-1} }$ (fb) & $\sigma^{\text{\tiny{reach}}}_{\text{5 fb}^{-1}}$ (fb) & $\sigma^{\text{\tiny{reach}}}_{\text{15 fb}^{-1} }$ (fb) & $\sigma^{\text{\tiny{QCD}}}_{\text{prod}}$ (fb)  \\
\hline\hline
$\GB^{\tt TB}$  & 500 & 115 & 239 & 146 & 92 & 2310  \\
$\GB^{\tt TB}$ & 500 & 40 & 175 & 100 & 63 & 2310 \\
$\GB^{\tt TB}$ & 650 & 40 & 88 & 29 & 14 & 335  \\
$\GB^{\tt TB}$ & 800 & 415 & 152 & 59 & 37 &  61  \\
$\GB^{\tt TB}$ & 800 & 40 & 66 & 17 & 8.3 &  61 \\
\hline
\hline
$\GB^{\tt TJ}$ & 450 & 65 & 1680 & 1320 & 1080 & 4760  \\
$\GB^{\tt TJ}$ & 550 & 140 & 653 & 470 & 354 & 1170  \\
$\GB^{\tt TJ}$ & 650 & 40 & 177 & 102 & 83 & 335  \\
$\GB^{\tt TJ}$ & 800 & 415 & 349 & 234 & 183 & 61  \\
$\GB^{\tt TJ}$ & 800 & 40 & 79 & 39 & 24 &  61 \\
\hline
\hline
$\GB^{\tt BJ}$ & 200 & 165 & 25000 & 17900 & 13000 & 625000  \\
$\GB^{\tt BJ}$ & 200 & 40 & 35100 & 25400 & 11800 & 625000 \\
$\GB^{\tt BJ}$ & 500 & 40 & 311 & 197 & 179 & 2310  \\
$\GB^{\tt BJ}$ & 800 & 765 & 4120 & 2960 & 2510 &  61  \\
$\GB^{\tt BJ}$ & 800 & 40 & 58 & 29 & 17 &  61 \\
\hline
\hline
\end{tabular}
\caption{Benchmark models from the hybrid $\GB$ simplified models. Also shown are the estimated cross section reach for $\mathcal{L}=1,5,15$ fb$^{-1}$ in addition to the NLO production cross section.  }
\label{Tab: GBTopologiesA}
\end{center}
\end{table}

\begin{table}[th!]
\begin{center}
\begin{tabular}{|c||c|c|c|c|c|c|}
\hline
Name & $m_{\go}$ (GeV) & $m_{\chi^0}$ (GeV) & $\sigma^{\text{\tiny{reach}}}_{\text{1 fb}^{-1} }$ (fb) & $\sigma^{\text{\tiny{reach}}}_{\text{5 fb}^{-1}}$ (fb) & $\sigma^{\text{\tiny{reach}}}_{\text{15 fb}^{-1} }$ (fb) & $\sigma^{\text{\tiny{QCD}}}_{\text{prod}}$ (fb)  \\
\hline\hline
$\GW^{\tt TM}$ & 500 & 115 & 422 & 184 & 63 & 2310  \\
$\GW^{\tt TM}$ & 500 & 40 & 324 & 126 & 44 & 2310 \\
$\GW^{\tt TM}$ & 650 & 40 & 115 & 52 & 25 & 335  \\
$\GW^{\tt TM}$ & 800 & 415 & 243 & 130 & 66 &  61  \\
$\GW^{\tt TM}$ & 800 & 40 & 81 & 25 & 12 &  61  \\
\hline
\hline
$\GW^{\tt BM}$ & 300 & 45 & 1370 & 1180 & 1010 & 62100  \\
$\GW^{\tt BM}$ & 400 & 220 & 2660 & 1300 & 619 & 10400  \\
$\GW^{\tt BM}$ & 600 & 170 & 113 & 40 & 25 & 617  \\
$\GW^{\tt BM}$ & 800 & 595 & 1160 & 452 & 240 &  61  \\
$\GW^{\tt BM}$ & 800 & 45 & 55 & 15 & 6.9 &  61 \\
\hline
\hline
$\GW^{\tt MM}$ & 300 & 45 & 3230 & 695 & 272 & 62100  \\
$\GW^{\tt MM}$ & 450 & 270 & 3190 & 1530 & 674 & 4760  \\
$\GW^{\tt MM}$ & 550 & 45 & 150 & 86 & 51 & 1170  \\
$\GW^{\tt MM}$ & 800 & 595 & 1290 & 727 & 413 &  61  \\
$\GW^{\tt MM}$ & 800 & 45 & 69 & 21 & 10 &  61 \\
\hline
\end{tabular}
\caption{Benchmark models from the $\GW$ simplified models. Also shown are the estimated cross section reach for $\mathcal{L}=1,5,15$ fb$^{-1}$ in addition to the NLO production cross section.}
\label{Tab: GWTopologies}
\end{center}
\end{table}

\begin{table}[th!]
\begin{center}
\begin{tabular}{|c||c|c|c|c|c|c|}
\hline
Name & $m_{\go}$ (GeV) & $m_{\chi^0}$ (GeV) & $\sigma^{\text{\tiny{reach}}}_{\text{1 fb}^{-1} }$ (fb) & $\sigma^{\text{\tiny{reach}}}_{\text{5 fb}^{-1}}$ (fb) & $\sigma^{\text{\tiny{reach}}}_{\text{15 fb}^{-1} }$ (fb) & $\sigma^{\text{\tiny{QCD}}}_{\text{prod}}$ (fb)  \\
\hline\hline
$\TB$  & 250 & 0 & 15100 & 9960 & 5980 & 180000  \\
$\TB$  & 350 & 50 & 1970 & 1500 & 1104 & 24200  \\
$\TB$  & 500 & 200 & 536 & 349 & 289 & 2310  \\
$\TB$  & 500 & 50 & 240 & 124 & 104 & 2310 \\
$\TB$ & 650 & 350 & 321 & 178 & 144 & 335  \\
$\TB$ & 650 & 50 & 96 & 49 & 32 & 335  \\
\hline
\hline
$\BB$ & 100 & 0 & 219000 & 203000 & 124000 & 21.2x10$^6$  \\
$\BB$ & 200 & 50 & 11200 & 8620 & 5370 & 625000  \\
$\BB$ & 350 & 200 & 2260 & 1680 & 1260 & 24200  \\
$\BB$ & 350 & 50 & 481 & 438 & 427 & 24200  \\
$\BB$ & 400 & 50 & 263 & 209 & 171 & 10400  \\
$\BB$ & 450 & 150 & 230 & 168 & 133 & 4760  \\
$\BB$ & 500 & 350 & 989 & 586 & 348 & 2310  \\
$\BB$ & 500 & 50 & 142 & 71 & 54 & 2310 \\
$\BB$ & 550 & 0 & 121 & 65 & 45 & 1170  \\
$\BB$ & 600 & 350 & 233 & 153 & 120 & 617  \\
\hline
\end{tabular}
\caption{Benchmark models from the hybrid $\TB$ and $\BB$ simplified models. Also shown are the estimated cross section reach for $\mathcal{L}=1,5,15$ fb$^{-1}$ in addition to the NLO production cross section.}
\label{Tab: TB&BBTopologies}
\end{center}
\end{table}

\end{document}